\documentclass{IEEEtaes}

\jvol{XX}
\jnum{XX}
\jmonth{XXXXX}
\paper{XXXXX}
\pubyear{XXXX}
\doiinfo{XXX}

\usepackage{cite}
\usepackage{amssymb,amsmath,latexsym,amsfonts,amsthm,mathtools}

\usepackage{accents}

\usepackage{algorithmic}
\usepackage{graphicx}
\usepackage{float,subcaption}
\usepackage{textcomp}
\usepackage{xcolor}
\definecolor{darkpastelpurple}{rgb}{0.59, 0.44, 0.84}
\usepackage{hyperref}
\hypersetup{colorlinks, breaklinks, citecolor=blue, linkcolor=blue, urlcolor=blue}
\usepackage[nameinlink]{cleveref}
\Crefformat{figure}{#2Fig.~#1#3}
\Crefmultiformat{figure}{Figs.~#2#1#3}{ and~#2#1#3}{, #2#1#3}{ and~#2#1#3}

\usepackage{tikz}
\usetikzlibrary{automata, shapes, arrows, calc, arrows.meta, fit, positioning}
\usepackage{pgfplots}

\theoremstyle{plain}
\newtheorem{theorem}{Theorem}
\newtheorem{lemma}{Lemma}
\newtheorem{corollary}{Corollary}

\newtheorem*{problem*}{Problem}
\theoremstyle{remark}
\newtheorem{remark}{Remark}

\newtheorem{definition}{Definition}
\theoremstyle{definition}

\newcommand{\sign}{\mathrm{sign}}
\usepackage{mathrsfs}

\usepackage{newtxmath}
\usepackage{booktabs}

\IEEEoverridecommandlockouts

\begin{document}
	\title{Cooperative Guidance for Aerial Defense in Multiagent Systems} 
	
	\author{Shivam Bajpai}
	\affil{Department of Aerospace Engineering and Engineering Mechanics, University of Cincinnati, OH} 
	\author{Abhinav Sinha}
	\member{Senior Member, IEEE}
	\affil{Department of Aerospace Engineering and Engineering Mechanics, University of Cincinnati, OH} 
	\author{Shashi Ranjan Kumar}
	\member{Senior Member, IEEE}
	\affil{Department of Aerospace Engineering, Indian Institute of Technology Bombay, Powai, India} 
	
	\receiveddate{Manuscript received XXXXX 00, 0000; revised XXXXX 00, 0000; accepted XXXXX 00, 0000.\\
		This paragraph of the first footnote will contain the date on which you submitted your paper for review, which is populated by IEEE. It is IEEE style to display support information, including sponsor and financial support acknowledgment, here and not in an acknowledgment section at the end of the article. For example, ``This work was supported in part by the some project.'' }
	\authoraddress{S. Bajpai and A. Sinha are with the Guidance, Autonomy, Learning, and Control for Intelligent Systems (GALACxIS) Lab, Department of Aerospace Engineering and Engineering Mechanics, University of Cincinnati, OH 45221, USA. (e-mails: bajpaism@mail.uc.edu, abhinav.sinha@uc.edu).\newline S. R. Kumar is with the Intelligent Systems \& Control (ISaC) Lab, Department of Aerospace Engineering, Indian Institute of Technology Bombay, Mumbai 400076, India. (e-mail: srk@aero.iitb.ac.in).}

	\maketitle
	
	\begin{abstract}
		This paper addresses a critical aerial defense challenge in contested airspace, involving three autonomous aerial vehicles-- a hostile drone (the pursuer), a high-value drone (the evader), and a protective drone (the defender). We present a cooperative guidance framework for the evader-defender team that guarantees interception of the pursuer before it can capture the evader, even under highly dynamic and uncertain engagement conditions. Unlike traditional heuristic, optimal control, or differential game-based methods, we approach the problem within a time-constrained guidance framework, leveraging true proportional navigation based approach that ensures robust and guaranteed solutions to the aerial defense problem. The proposed strategy is computationally lightweight, scalable to a large number of agent configurations, and does not require knowledge of the pursuer's strategy or control laws. From arbitrary initial geometries, our method guarantees that key engagement errors are driven to zero within a fixed time, leading to a successful mission. Extensive simulations across diverse and adversarial scenarios confirm the effectiveness of the proposed strategy and its relevance for real-time autonomous defense in contested airspace environments.
	\end{abstract}
	
	
	\section{Introduction}\label{sec:introduction}
	\IEEEPARstart{I}{n} modern contested airspace, the rapid increase in the presence of fast, agile, and networked unmanned aerial vehicles (UAVs) has intensified the need for reliable aerial defense strategies. Small, inexpensive UAVs can be deployed in large numbers to overwhelm defenses, execute coordinated attacks, or serve as reconnaissance assets to enable subsequent strikes. In such environments, traditional two-body point-defense systems may be inadequate, as engagements often occur at short ranges with limited reaction time. In contrast, cooperative multiagent defense, where defenders actively protect high-value UAVs (evaders) from hostile pursuers, offers a critical layer of protection. This results in a three-body engagement necessitating greater autonomy and strategic decision-making within stricter constraints on engagement duration.
	
	Research on the kinematics of three-body engagements can be traced to \cite{4101686}, where a closed-form solution was obtained for constant-bearing collision courses, and to \cite{4102335}, which determined the intercept point in the evader-centered reference frame. Subsequent studies have adopted optimal control formulations for three-agent engagements with specific performance objectives, such as minimizing energy or interception cost \cite{doi:10.2514/1.G001083,doi:10.2514/1.51765,doi:10.2514/1.49515,doi:10.2514/1.58531,doi:10.2514/1.61832}. For example, the work in \cite{doi:10.2514/1.G001083} integrated a differential game approach into cooperative optimal guidance to maximize pursuer–evader separation, while that in \cite{doi:10.2514/1.51765} addressed cooperative pursuit–evasion strategies for defender–evader teams with arbitrary-order linearized agent dynamics, assuming the pursuer’s guidance law was known. A multiple-model adaptive estimator enabling cooperative information sharing to predict the pursuer’s likely linear strategy was proposed in \cite{doi:10.2514/1.49515}. Three-layer cooperation with explicit defender–evader communication was examined in \cite{doi:10.2514/1.58531}, whereas results in \cite{doi:10.2514/1.61832} provided algebraic capture conditions under which a pursuer could bypass the defender. While these approaches benefit from analytical tractability, their reliance on linearized dynamics can reduce robustness under large initial heading errors or highly nonlinear kinematics.
	
	Nonlinear guidance strategies address these shortcomings by avoiding small-error assumptions and explicitly incorporating turn-rate constraints. Representative works include \cite{6315051,doi:10.2514/1.G000659,9274339,doi:10.1007/s10846-022-01570-y,doi:10.1016/j.ast.2020.105787}. In \cite{6315051}, a sliding-mode control–based terminal intercept guidance and autopilot design was developed for a defender to shield the evader from an incoming pursuer. A related sliding-mode cooperative defense law was presented in \cite{doi:10.2514/1.G000659}, while the work in \cite{9274339} extended this to multi-defender engagements to enable simultaneous interception of the pursuer before it reaches the evader. In \cite{doi:10.2514/1.G003059}, the authors proposed a hybrid cooperative guidance law for defenders that combined inertial delay control, prescribed performance control, and sliding mode control. This fusion offered greater flexibility and performance but relied on an intermediate switching mechanism to transition between the individual guidance schemes.
	
	In this paper, we develop a cooperative guidance strategy for the evader and the defender that is inherently robust to arbitrary and potentially aggressive maneuvers of the pursuer, eliminating the need to model or predict its specific guidance law. Moreover, our approach includes different levels of cooperation between the evader and the defender based on whether the defender can access the evader's guidance law. 
	
	The proposed approach is formulated entirely within a nonlinear engagement framework, removing restrictive assumptions on small heading errors or linearized dynamics and thereby broadening applicability to diverse operational conditions. The evader's and the defender's control laws are designed to make the relevant error variables vanish within a fixed time, which helps ensure that the defender is able to protect the evader starting from any feasible and arbitrary engagement geometry. 
	
	A key novelty is that the proposed guidance strategy for the defender is developed on the true proportional navigation principle. This allows rapid course adjustments and reliable interception even in engagements with fast-changing relative motion. Unlike other variants of the popular proportional navigation guidance, which approximate the guidance command based solely on the line-of-sight rate, the current method is based on computing acceleration directly perpendicular to the instantaneous line-of-sight, utilizing both radial and tangential control components. This grants the defender significantly enhanced agility, precise trajectory shaping, and improved interception capability against maneuvering threats.
	
	\section{Problem Formulation}\label{sec:problem}
\begin{figure}[h!]
		\centering
		\includegraphics[width=.7\linewidth]{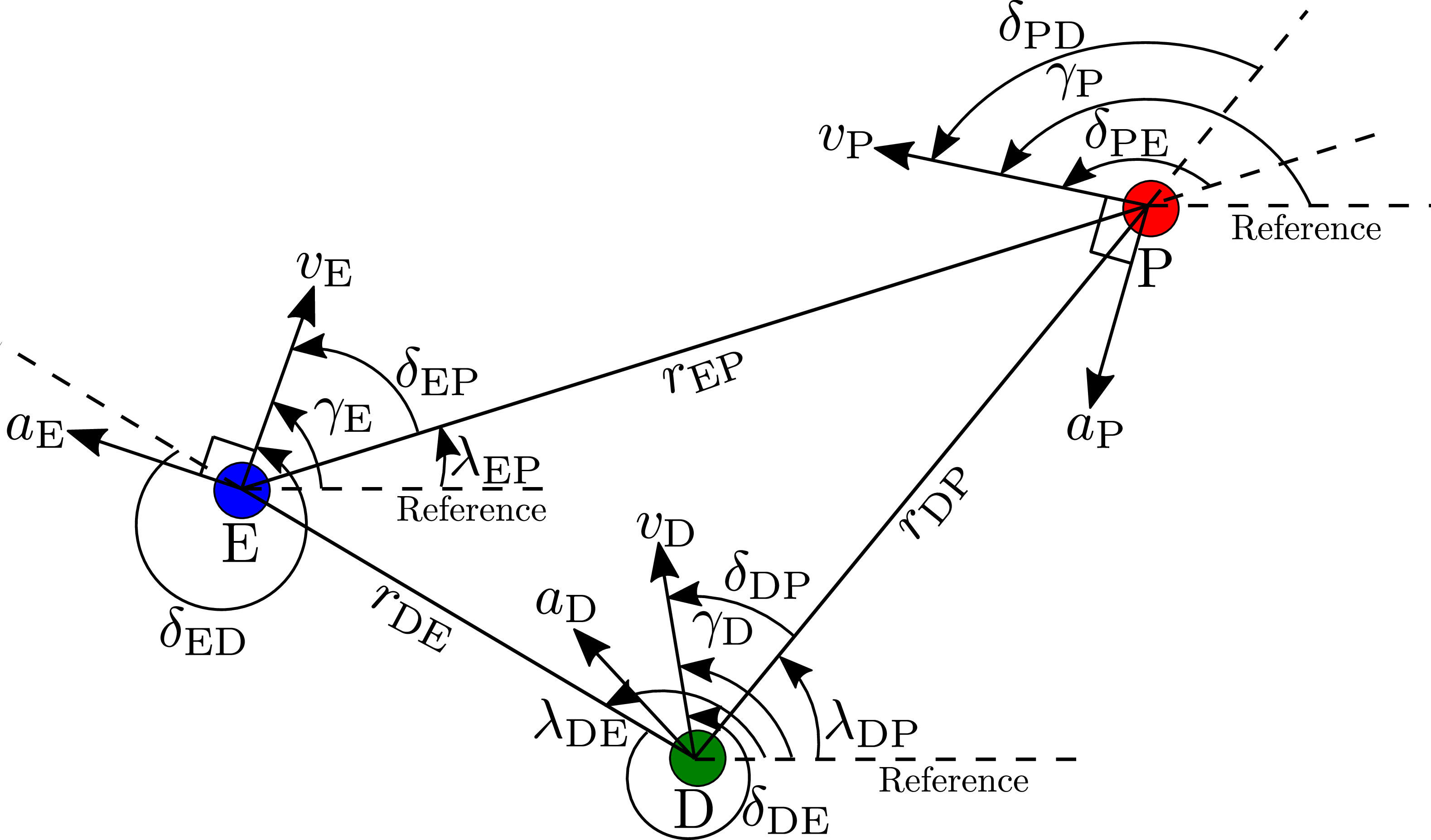}
		\caption{Aerial defense scenario in 2D.}
		\label{fig:enggeo}
	\end{figure}

	We study a cooperative aerial defense problem in contested airspace involving three autonomous nonholonomic agents-- a hostile pursuer (P), a high-value evader (E), and a protective defender (D), as shown in \Cref{fig:enggeo}. The pursuer’s objective is to intercept the evader, while the defender is tasked with intercepting the pursuer before it can capture the evader. Consequently, the evader and defender operate as a cooperative team, while the pursuer acts adversarially. The engagement is restricted to a planar setting, which captures characteristics of air combat scenarios at fixed altitudes.
	
	The agents are modeled as curvature-constrained vehicles with nonholonomic dynamics. The pursuer and evader follow standard unicycle-type kinematics with constant speeds $v_\mathrm{P}$ and $v_\mathrm{E}$, respectively.  The equations of motion for the pursuer and the evader agents are given by
	\begin{equation}\label{eq:basic}
		\dot{x}_i = v_i\cos\gamma_i,~\dot{y}_i = v_i\sin\gamma_i,~\dot{\gamma}_i=\dfrac{a_i}{v_i},~i\in\{\text{P, E}\},
	\end{equation}
	where $[x_i,y_i]^\top\in\mathbb{R}^2$ denotes the position of the $i$\textsuperscript{th} agent, $v_i\in\mathbb{R}_+$ is its constant translational speed, and $\gamma_i\in(-\pi,\pi]$ is its heading angle. Their accelerations, denoted by $a_i \in \mathbb{R}$,  are the steering control inputs. The defender, in contrast, is modeled with second-order speed dynamics to reflect its ability to modulate thrust or propulsion, that is,
	\begin{equation}\label{eq:basicD}
		\dot{x}_\mathrm{D} = v_\mathrm{D}\cos\gamma_\mathrm{D},~\dot{y}_\mathrm{D} = v_\mathrm{D}\sin\gamma_\mathrm{D},~\dot{\gamma}_\mathrm{D}=\dfrac{a_{\mathrm{D}_t}}{v_\mathrm{D}},~ \dot{v}_\mathrm{D} = a_{\mathrm{D}_r}.
	\end{equation}
	For the defender, $a_\mathrm{D}^2=a_{\mathrm{D}_t}^2+a_{\mathrm{D}_r}^2$, where $a_{\mathrm{D}_t}=a_\mathrm{D}\cos(\gamma_\mathrm{D}-\lambda_\mathrm{DP})$ and $a_{\mathrm{D}_r}=a_\mathrm{D}\sin(\gamma_\mathrm{D}-\lambda_\mathrm{DP})$ are the tangential and radial components of the its total acceleration, respectively.
	
	The dynamics in \eqref{eq:basic} respect nonholonomic constraints, and are more representative of real-world aerial vehicles than commonly used simplified models, such as those assuming single-integrator or omnidirectional motion via heading control. In contrast to prior works that utilize instantaneous heading angle actuation \cite{9122473, doi:10.1016/j.automatica.2011.06.010}, the present formulation retains curvature constraints inherent to aerial platforms, enabling the synthesis of implementable guidance laws that respect physical maneuverability limitations.  
	
	As illustrated in \Cref{fig:enggeo}, the engagement geometry is characterized by the pairwise relative positions of agents, denoted by scalar separation distances $r_\ell\in\mathbb{R}_+$ and associated line-of-sight (LOS) angles $\lambda_\ell\in(-\pi,\pi]$, for each agent pair indexed by $\ell\in\{\text{EP},\text{DE},\text{DP}\}$. These quantities describe the relative motion in a planar setting, whose dynamics for all pairs can be written as
	\begin{align}
		\dot{r}_{ij} &= v_j \cos(\gamma_j - \lambda_{ij}) - v_i \cos(\gamma_i - \lambda_{ij}), \label{eq:vr}\\
		r_{ij} \dot{\lambda}_{ij} &= v_j \sin(\gamma_j - \lambda_{ij}) - v_i \sin(\gamma_i - \lambda_{ij}),\label{eq:vtheta}
	\end{align}
	$\forall~(i, j) \in \mathcal{E}$, where $\mathcal{E}=\left\{ \left(\text{E},\text{P}\right),\left(\text{D},\text{E}\right),\left(\text{D},\text{P}\right) \right\}$ is the set of ordered agent pairs. Additionally, we denote $\delta_{ji}=\gamma_j-\lambda_{ij}$ to be the bearing angle of agent $j$ relative to the LOS $r_{ij}$. Similarly, $\delta_{ij}=\gamma_i-\lambda_{ij}$. These kinematic equations provide a nonlinear representation of the relative motion under the nonholonomic dynamics in \eqref{eq:basic}--\eqref{eq:basicD} and form the foundation for the derivation of cooperative guidance laws in the subsequent sections. Note that such a treatment allows the proposed solutions to remain valid for larger operating regimes.
	\begin{definition}[\cite{tahk2018impact}]
		Time-to-go for any pair of engagements is defined as the time remaining till intercept of the adversary in that engagement.
	\end{definition}
	In line with the notations introduced in this paper, the time-to-go for the pursuer-evader engagement can be denoted as $t_\mathrm{go}^\mathrm{EP}$, which is the time remaining till the evader's capture by the pursuer. Similarly, $t_\mathrm{go}^\mathrm{DP}$ is the time-to-go for the defender-pursuer engagement. Clearly, if the evader-defender team can cooperatively maneuver to ascertain $t_\mathrm{go}^\mathrm{DP}<t_\mathrm{go}^\mathrm{EP}$, the evader's safety will be guaranteed, which is to say that the defender will intercept the pursuer before it reaches the vicinity of evader. 
	
	Therefore, the primary objective of this work is to synthesize admissible nonlinear cooperative control laws $a_\mathrm{E}$ and $a_\mathrm{D}$ such that the pursuer is captured by the defender within a lesser time (or a time-margin), independent of the pursuer's control law, thereby safeguarding the evader. To this end, the time of the pursuer's capture, $t_f$, is a crucial parameter, and must be controlled by manipulating the engagement durations (or the respective time-to-go values) to ensure mission success. 
	\begin{problem*}
		Consider the aerial defense scenario in \Cref{fig:enggeo}, where agents evolve according to \eqref{eq:basic}--\eqref{eq:basicD}, with control inputs $a_i \in \mathcal{U}_i := \left\{ u \in \mathbb{R} \right\}, \quad i \in \{\mathrm{P}, \mathrm{E}, \mathrm{D}\}$. Let $\mathscr{S}(t) \in \mathcal{X} \subset \mathbb{R}^n$ denote the joint system state, comprising the agents’ positions, headings, velocities, relative ranges $r_{ij}$, and LOS angles $\lambda_{ij}$. Define the \emph{target set} as $\mathscr{T} := \left\{ \mathscr{S} \in \mathcal{X} \,\middle|\, r_{\mathrm{DP}}(t_f) = 0 \right\}$, which corresponds to successful interception of the pursuer by the defender. The evader-defender team aims to design feedback control laws $a_{\mathrm{E}}= \mu_{\mathrm{E}}(\mathscr{I}_{\mathrm{E}})$, $a_{\mathrm{D}}= \mu_{\mathrm{D}}(\mathscr{I}_{\mathrm{D}})$, where $\mathscr{I}_{\mathrm{E}}$ and $\mathscr{I}_{\mathrm{D}}$ represent the sets of measurable relative information such as $\{r_{ij}, \lambda_{ij}, \delta_{ij}\}$, and where the admissibility conditions must hold. The control objective is to ensure the reachability of the target set, that is, $\exists~ t_f < \infty$ such that $\mathscr{S}(t_f) \in \mathscr{T}, ~ \forall\, \mathscr{S}(0) \in \mathcal{X}_0$ for a feasible initial condition set $\mathcal{X}_0 \subset \mathcal{X}$, under admissible controls.
	\end{problem*}
	\begin{lemma}\label{lem:LOSratedotDP}
		The LOS rate dynamics of the defender-pursuer engagement has a relative degree of one with respect to the acceleration inputs of the agents involved.
	\end{lemma}
	\begin{proof}
		Differentiating \eqref{eq:vtheta} with respect to time for $(i,j)=(\text{D,P})$ yields $\dot{r}_\mathrm{DP} \dot{\lambda}_\mathrm{DP} + r_\mathrm{DP} \ddot{\lambda}_\mathrm{DP} 
		= v_\mathrm{P} \cos \delta_\mathrm{PD} \dot{\delta}_\mathrm{PD} - v_\mathrm{D} \cos \delta_\mathrm{DP} \dot{\delta}_\mathrm{DP}-\dot{v}_\mathrm{D}\sin\delta_\mathrm{DP}$, which can be simplified further using \eqref{eq:basic}
		
		\begin{align} \label{eq: LOS rate Dynamnics DP_1}
			\dot{r}_\mathrm{DP}\dot{\lambda}_\mathrm{DP} + r_\mathrm{DP}\ddot{\lambda}_\mathrm{DP} =&~ v_\mathrm{P}\cos \delta_\mathrm{PD}\left(\dfrac{a_\mathrm{P}}{v_\mathrm{P}}- \dot{\lambda}_\mathrm{DP} \right) -  a_\mathrm{D_\mathrm{r}}\sin \delta_\mathrm{DP} \nonumber \\
			&- v_\mathrm{D}\cos \delta_\mathrm{DP}\left(\dfrac{a_\mathrm{D_\mathrm{t}}}{v_\mathrm{D}} - \dot{\lambda}_\mathrm{DP}\right) .
		\end{align}
		Substituting $a_\mathrm{D_\mathrm{r}}=a_\mathrm{D}\sin\delta_\mathrm{DP}$ and $a_\mathrm{D_\mathrm{t}} = a_\mathrm{D}\cos \delta_\mathrm{DP}$ into \eqref{eq: LOS rate Dynamnics DP_1} and simplifying yields
		\begin{align}\label{eq: LOS rate Dynamnics DP_2}
			\dot{r}_\mathrm{DP}\dot{\lambda}_\mathrm{DP} + r_\mathrm{DP}\ddot{\lambda}_\mathrm{DP}     
			=&~ a_\mathrm{P}\cos \delta_\mathrm{PD} - v_\mathrm{P}\cos \delta_\mathrm{PD} \dot{\lambda}_\mathrm{DP} -a_\mathrm{D} \nonumber \\
			& + v_\mathrm{D}\cos \delta_\mathrm{DP} \dot{\lambda}_\mathrm{DP}
		\end{align}
		
		Using \eqref{eq:vr} in the above expression and after rearranging the terms, one may obtain
		\begin{align}\label{eq: los ddot DP}
			\ddot{\lambda}_\mathrm{DP} =&~ \dfrac{-2 \dot{r}_\mathrm{DP}\dot{\lambda}_\mathrm{DP} - a_\mathrm{D} + a_\mathrm{P}\cos \delta_\mathrm{PD}}{r_\mathrm{DP}}.
		\end{align}
		This concludes the proof.
		
	\end{proof}
	\begin{lemma}\label{lem:LOSratedotEP}
		The LOS rate dynamics of the pursuer-evader engagement has a relative degree of one with respect to the acceleration inputs of the agents involved.
	\end{lemma}
	\begin{proof}
		Similar to the proof of \Cref{lem:LOSratedotDP}, excluding speed dynamics.
		\begin{equation}\label{eq: LOS dynamics of evader-pursuer}
			\ddot{\lambda}_\mathrm{EP} = -\dfrac{2\dot{r}_\mathrm{EP}\dot{\lambda}_\mathrm{EP}}{r_\mathrm{EP}} - \dfrac{\cos\delta_\mathrm{EP}}{r_\mathrm{EP}}a_\mathrm{E} + \dfrac{\cos \delta_\mathrm{PE}}{r_\mathrm{EP}}a_\mathrm{P}. 
		\end{equation}
		
		
	\end{proof}
	\begin{lemma}[\cite{polyakov2011nonlinear}]\label{lemma: fixed-time lemma}
		Suppose there exists a continuous, radially unbounded function $V: \mathbb{R}^\mathrm{n} \to \mathbb{R}_{\geq 0}$ such that $V(x)=0\;\Rightarrow\; x \in D$ where $D := \{x \in \mathbb{R}^n \mid V(x) = 0\}$ denotes the equilibrium set. Any solution $x(t)$ of $\dot{x} = g(t,x)$ with $x(0) = x_0$ that satisfies the inequality $\dot{V}(x(t)) \leq -(\zeta V^{\alpha}(x(t)) + \xi V^{\beta}(x(t)))^\kappa$ for some $\alpha,\beta, \kappa,\zeta,\xi >0$ and $\alpha \kappa<1, \beta \kappa>1$ then $D \subset \mathbb{R}^n$ is globally fixed-time attractive set for the system and $T(x_0) \leq \frac{1}{\zeta^{\kappa}(1 - \alpha \kappa)} + \frac{1}{\xi^\kappa(\beta \kappa -1)}, \forall x_0  \in \mathbb{R}^n$ denotes the settling time of the trajectory with initial condition $x(0)=x_0$.
	\end{lemma}
	
	\section{Design of the Cooperative Guidance Law}\label{sec:main}
	From a time-critical standpoint, the interception time is equally crucial to metrics such as miss distance or relative distance, particularly from the evader’s perspective. Specifically, if the evader possesses knowledge of the estimated time of its interception by the pursuer, it can communicate this information to the defender, thereby allowing the defender to shape its control action to ensure interception of the pursuer strictly before this time. This allows us to transform the cooperative guidance task into designing a time-constrained guidance strategy for the defender, where the evader is executing certain maneuvers to aid the former.
	
	Since the pursuer may execute arbitrary and possibly adversarial maneuvers, accurately predicting the time of the evader's capture may generally be infeasible. However, under the special case where the evader and the pursuer are on a collision course, the expected time of capture can be computed precisely, and is given by
	\begin{equation}\label{eq:tgoEP}
		t_\mathrm{go}^\mathrm{EP} =  \dfrac{r_\mathrm{EP}}{v_{\mathrm{E}} \cos \delta_{\mathrm{EP}}-v_{\mathrm{P}} \cos \delta_{\mathrm{PE}}},
	\end{equation}
	provided that the denominator of \eqref{eq:tgoEP} is positive (the relative closing velocity of the pursuer with respect to the evader is strictly positive). In a general case, \eqref{eq:tgoEP} may serve as a conservative estimate of the evader's survival horizon. To induce such predictable behavior, the evader may adopt a deception-based maneuvering strategy. Specifically, the evader may deliberately nullify its LOS rate with respect to the pursuer, independent of their initial engagement configuration. This deceptive behavior may create an illusion of vulnerability, which may entice the pursuer into ceasing its own evasive maneuvers, interpreting the current trajectory as favorable for interception. From the pursuer's perspective, such a trajectory appears advantageous, and hence it is likely to commit to a straight-line pursuit. While this induced non-maneuvering behavior of the pursuer is not strictly required for ensuring the evader’s safety, it significantly reduces the maneuverability burden on the defender, as will be established in later sections. To this end, we consider the manifold $s_1 = \dot{\lambda}_{\mathrm{EP}}$, driving which to zero ensures that the evader and pursuer arrive on a collision course. This forms the basis for a deception strategy that shapes the engagement into a predictable regime, enabling reliable estimation of interception time and reduced maneuverability burden on the defender.
	\begin{theorem}
		Consider the aerial defense scenario shown in \Cref{fig:enggeo}, governed by \eqref{eq:vr}-\eqref{eq:vtheta}. The evader's strategy,
		\begin{align}
			a_\mathrm{E}=&-\dfrac{2\dot{r}_\mathrm{EP}\dot{\lambda}_\mathrm{EP}}{\cos\delta_\mathrm{EP}}+ \dfrac{r_\mathrm{EP}}{\cos\delta_\mathrm{EP}}\left[\left(\zeta_1 |s_1|^{\alpha_1} + \xi_1 |s_1|^{\beta_1}\right)^{\kappa_1}+ \right.  \nonumber\\ &\left. \sec\delta_\mathrm{EP}\,\epsilon_1 
			\right]\sign(s_1),\label{eq:aE}
		\end{align}
		where the design parameters satisfy $\zeta_1, \xi_1, \alpha_1, \beta_1, \kappa_1 > 0,~ \alpha_1 \kappa_1 < 1, \beta_1 \kappa_1 > 1$ and $\epsilon_1>\sup_{t\geq 0}\frac{{a}_\mathrm{P}^\mathrm{max}}{r_\mathrm{EP}}$ guarantees that $s_1$ converges to zero within a fixed time independent of the initial configuration of the pursuer-evader engagement and the pursuer's strategy.
	\end{theorem}
	\begin{proof}
		Consider the Lyapunov function candidate $V_1=|s_1|$. Differentiating $V_1$ with respect to time and using the results in \Cref{lem:LOSratedotEP} yields   
		\begin{align}\label{eq:V1dotstep1}
			\dot{V}_1 
			=&~\sign(s_1)\left[-\dfrac{2\dot{r}_\mathrm{EP}\dot{\lambda}_\mathrm{EP}}{{r}_\mathrm{EP}}-\dfrac{\cos\delta_\mathrm{EP}}{{r}_\mathrm{EP}}a_\mathrm{E}+\dfrac{\cos\delta_\mathrm{PE}}{{r}_\mathrm{EP}}a_\mathrm{P}\right].
		\end{align}
		If the evader's strategy is chosen as \eqref{eq:aE}, then \eqref{eq:V1dotstep1} becomes 
		\begin{align} \label{eq: evader derivative final}
			\dot{V}_1 =&~ \sign(s_1)\left[\dfrac{\cos \delta_\mathrm{PE}}{r_\mathrm{EP}}a_\mathrm{P} - \left( \left(\zeta_1 |s_1|^{\alpha_1} + \xi_1 |s_1|^{\beta_1}\right)^{\kappa_1}\right.\right.\nonumber \\ &\left.\left.+\sec\delta_\mathrm{EP}\epsilon_1\right)\sign(s_1) \right] \nonumber \\
			=& -\left(\zeta_1 |s_1|^{\alpha_1} + \xi_1 |s_1|^{\beta_1}\right)^{\kappa_1}  \nonumber \\
			& -\left(\sec \delta_\mathrm{EP}\epsilon_1 - \dfrac{\cos \delta_\mathrm{PE}}{r_\mathrm{EP}}a_\mathrm{P} \sign(s_1)\right) \nonumber \\
			\leq& -\left(\zeta_1 |s_1|^{\alpha_1} + \xi_1 |s_1|^{\beta_1}\right)^{\kappa_1} -\left(\epsilon_1 - \dfrac{a_\mathrm{P}^\mathrm{max}}{r_\mathrm{EP}}\right) \nonumber \\
			\leq& -\left(\zeta_1 |s_1|^{\alpha_1} + \xi_1 |s_1|^{\beta_1}\right)^{\kappa_1} \quad \forall ~ s_1 \neq 0
		\end{align}
		when $\epsilon_1>\sup_{t\geq 0}\frac{{a}_\mathrm{P}^\mathrm{max}}{r_\mathrm{EP}}$. It follows from \eqref{eq: evader derivative final} and the results in \Cref{lemma: fixed-time lemma} that sliding mode is enforced on $s_1$ within a fixed time whose upper bound is $t_1\leq \frac{1}{\zeta_1^{\kappa_1} (1 - \alpha_1 \kappa_1)} + \frac{1}{\xi_1^{\kappa_1} (\beta_1 \kappa_1 - 1)}$ irrespective of initial value of $\dot{\lambda}_\mathrm{EP}$.
	\end{proof}
	\begin{remark}
		The evader’s control input, as given in \eqref{eq:aE}, possesses meaningful structure from a guidance perspective, and can be viewed as a nonlinear generalization of the classic proportional navigation guidance law, with a nonlinear navigation term of $-\frac{2\dot{r}_\mathrm{EP}}{\cos\delta_\mathrm{EP}}$, which is the range rate normalized over the projection of the relative velocity. The second term in \eqref{eq:aE} enforces convergence of the sliding manifold $s_1$ and brings course correction to the evader's maneuver. Overall, the control law is structured such that $a_\mathrm{E}\propto \dot{\lambda}_{\mathrm{EP}}$ with vanishing effect once sliding mode is enforced on $s_1$.  In this scenario, the defender benefits from reduced control effort for successful interception, since neutralizing a non-maneuvering pursuer demands significantly less agility than intercepting an adversarially maneuvering one.
	\end{remark}
	\begin{remark}
		The evader's guidance law is nonsingular by design despite the presence of $\delta_\mathrm{EP}$ in the denominator. Analyzing its dynamics for the equilibrium point leads to the quadratic relation in $\delta_\mathrm{EP}$ as $\cos^2\delta_{\mathrm{EP}}- \frac{r_{\mathrm{EP}} \, \sign(s_1)\,\epsilon_1}{v_\mathrm{E}^2 \dot{\lambda}_{\mathrm{EP}}}+ \frac{\left(2 v_\mathrm{E} \dot{r}_{\mathrm{EP}} \dot{\lambda}_{\mathrm{EP}}
			- r_{\mathrm{EP}} \, \sign(s_1)\,(\zeta_1|s_1|^{\alpha_1} + \xi_1|s_1|^{\beta_1})^{\kappa_1}\right)\cos\delta_{\mathrm{EP}}}{v_\mathrm{E}^2 \dot{\lambda}_{\mathrm{EP}}}
		=0$,
		which is non-zero on the LHS when $\delta_\mathrm{EP}=\pm \pi/2$, confirming that the system trajectories never settle on those isolated points.
	\end{remark}
	\begin{remark}
		It is important to emphasize that the defender's ability to successfully intercept the pursuer is not strictly contingent upon the pursuer becoming non-maneuvering. That is, even if the evader is unable to maintain $\dot{\lambda}_{\mathrm{EP}} = 0$ due to persistent maneuvering by the pursuer, the proposed cooperative guidance strategy still guarantees interception by the defender, provided sufficient control authority is available. The evader’s strategy to regulate the LOS rate, then, serves a dual purpose in the cooperative defense framework.
	\end{remark}
	Note that most existing impact time-constrained guidance strategies (e.g., \cite{cho2016modified,dong2023varying,erer2024computational,kim2018backstepping,tahk2018impact,kim2018sliding}) rely on the pure proportional navigation principle, which have been shown to be effective against stationary adversaries. However, their performance often degrades against maneuvering adversaries due to limited control authority and a lack of anticipatory motion information. This work approaches the time-constrained guidance design from the perspective of true proportional navigation, previously explored only for unconstrained interception \cite{shneydor1998missile}, except in our own recent works \cite{kumar2022true,sinha2024time}. Unlike pure proportional navigation, where acceleration is applied perpendicular to the velocity vector (lateral only), true proportional navigation applies acceleration perpendicular to the LOS, enabling both radial and tangential control. As a result, the defender employing such a principle becomes more agile, making it well-suited for an autonomous vehicle engaging a dynamic, maneuvering adversary under time constraints. Under such a principle, the time-to-go for the defender-pursuer engagement is given by 
	\begin{align}
		t_\mathrm{go}^\mathrm{DP}=-\dfrac{r_\mathrm{DP}(v_\mathrm{P} \cos\delta_\mathrm{PD}-v_\mathrm{D}\cos\delta_\mathrm{DP}+2c)}{v_\mathrm{D}^2 + v_\mathrm{P}^2-2v_\mathrm{P}v_\mathrm{D}\cos\left(\delta_\mathrm{PD}-\delta_\mathrm{DP}\right)+2c \dot{r}_\mathrm{DP}},\label{tgo_DP}
	\end{align}
	where $c$ is a parameter that can be chosen as a constant or designed as a function of engagement variables such as the defender's speed or the closing velocity.
	
	The defender's objective is to capture the pursuer in a lesser time. To that end, consider a manifold $s_2=t_{\mathrm{go}}^\mathrm{DP} - (t_\mathrm{go}^\mathrm{EP} - \tau)$, which is the error between the time-to-go of the defender–pursuer engagement and that of the pursuer–evader engagement, offset by a desired time-margin $\tau\in\mathbb{R}_+$. This margin is the desired time lead with which the defender intercepts the pursuer before it can reach the evader. 
	\begin{theorem}\label{theorem: Defender's law}
		Consider the aerial defense scenario shown in \Cref{fig:enggeo}, governed by \eqref{eq:vr}-\eqref{eq:vtheta}. The defender's strategy, 
		\begin{align}
			a_\mathrm{D}=& c\dot{\lambda}_\mathrm{DP} -\dfrac{r^2_\mathrm{EP}(\dot{r}_\mathrm{DP}^2 + r^2_\mathrm{DP}\dot{\lambda}_\mathrm{DP}^2 + 2c\dot{r}_\mathrm{DP})^2 }{2\dot{r}_\mathrm{EP}^2 r^2_\mathrm{DP}\dot{\lambda}_\mathrm{DP}(\dot{r}_\mathrm{DP} + 2c)}\dot{\lambda}^2_\mathrm{EP} \nonumber \\
			& - \dfrac{r_\mathrm{EP} \sin{\delta_\mathrm{EP}}(\dot{r}_\mathrm{DP}^2 + r^2_\mathrm{PD}\dot{\lambda}_\mathrm{DP}^2 + 2c\dot{r}_\mathrm{DP})^2}{2\dot{r}_\mathrm{EP}^2 r^2_\mathrm{DP}\dot{\lambda}_\mathrm{DP}(\dot{r}_\mathrm{DP} + 2c)}a_\mathrm{E} \nonumber \\
			& + \dfrac{(\dot{r}_\mathrm{DP}^2 + r^2_\mathrm{DP}\dot{\lambda}_\mathrm{DP}^2 + 2c\dot{r}_\mathrm{DP})^2}{2 r_\mathrm{DP}^2 \dot{\lambda}_\mathrm{DP}(\dot{r}_\mathrm{DP} + 2c)} \nonumber \times \\
			&\left[(\zeta_2 |s_2|^{\alpha_2} + \xi_2|s_2|^{\beta_2})^{\kappa_2} + \sec{\delta_\mathrm{DP}\epsilon_2}  \right] \sign(s_2),  \label{eq:aD_with_aE}
		\end{align}
		where the design parameters satisfy $\zeta_2, \xi_2, \alpha_2, \beta_2, \kappa_2, c > 0,~ \alpha_2 \kappa_2 < 1, \beta_2 \kappa_2 > 1$ and 
		$\epsilon_2 >\sup_{t\ge 0} \left( \frac{(v_\mathrm{P}+ v_\mathrm{D})^2 + 4r_\mathrm{DP}c^2 + r_\mathrm{DP}v_\mathrm{D} (4c +v_\mathrm{D} + v_\mathrm{P})}{(2(v_\mathrm{P} + v_\mathrm{D})^2+ 2c(v_\mathrm{P} + v_\mathrm{D}))^2} + \frac{r_\mathrm{EP}}{(v_\mathrm{P} \\ + v_\mathrm{E})^2}\right)a^\mathrm{max}_\mathrm{P}$,  guarantees that $s_2$ converges to zero within a fixed time independent of the initial configuration of the defender-pursuer engagement and the pursuer's strategy. Consequently, the defender intercepts the pursuer with the prescribed time margin $\tau$.
	\end{theorem}

	\begin{proof}
		Consider a Lyapunov function candidate $V_2=|s_2|$, whose time differentiation yields $\dot{V}_2=\sign(s_2)\dot{s}_2 = \sign(s_2)\left(\dot{t}_{\mathrm{go}}^\mathrm{DP} -\dot{t}_{\mathrm{go}}^\mathrm{EP} \right)$ since $\tau$ is treated constant here. On further simplifications, one may obtain $\dot{V}_2=\sign(s_2)\left(\dot{t}_{\mathrm{go}}^\mathrm{DP} -\dot{t}_{\mathrm{go}}^\mathrm{EP} \right)$.
		On differentiating \eqref{tgo_DP} with respect to time, one may obtain
		
		\begin{align}\label{eq: time_to_go_dot_DP}
			\dot{t}_\mathrm{go}^\mathrm{DP} &= -1+ \dfrac{2cr^2_\mathrm{DP}(\dot{r}_\mathrm{DP} + 2c)}{(\dot{r}^2_\mathrm{DP} + r^2_\mathrm{DP}\dot{\lambda}^2_\mathrm{DP} + 2c\dot{r}_\mathrm{DP})^2}\dot{\lambda}^2_\mathrm{DP} \nonumber \\
			&\quad - \dfrac{2r^2_\mathrm{DP}\dot{\lambda}_\mathrm{DP}(\dot{r}_\mathrm{DP} + 2c)}{(\dot{r}^2_\mathrm{DP} + r^2_\mathrm{DP}\dot{\lambda}^2_\mathrm{DP} + 2c\dot{r}_\mathrm{DP})^2}  a_\mathrm{D} \nonumber \\
			&\quad + \left( \dfrac{r_\mathrm{DP}\dot{\lambda}_\mathrm{DP}(v_\mathrm{P} - v_\mathrm{D}\cos(\gamma_\mathrm{P} - \gamma_\mathrm{D})) - 4c^2\sin \delta_\mathrm{PD}}{( r_\mathrm{DP}^2 + r^2_\mathrm{DP}\dot{\lambda}^2_\mathrm{DP} + 2c\dot{r}_\mathrm{DP})^2} \right. \nonumber \\
			& \left. \quad + \dfrac{(4c + \dot{r}_\mathrm{DP})v_\mathrm{D}\sin(\gamma_\mathrm{P} - \gamma_\mathrm{D})}{( r_\mathrm{DP}^2  +r^2_\mathrm{DP}\dot{\lambda}^2_\mathrm{DP} + 2c\dot{r}_\mathrm{DP})^2} \right) r_\mathrm{DP} a_\mathrm{P}.
		\end{align}

		Similarly, differentiating \eqref{eq:tgoEP} with respect to time yields
		\begin{equation}\label{eq:tgoEPdot}
			\dot{t}_\mathrm{go}^\mathrm{EP} = -1 + \dfrac{r_\mathrm{EP}^2\dot{\lambda}_\mathrm{EP}^2}{\dot{r}_\mathrm{EP}^2} + \dfrac{{r}_\mathrm{EP}\sin\delta_\mathrm{EP}}{\dot{r}_\mathrm{EP}^2}a_\mathrm{E} - \dfrac{{r}_\mathrm{EP}\sin\delta_\mathrm{PE}}{\dot{r}_\mathrm{EP}^2}a_\mathrm{P}.
		\end{equation}
		
		On substituting \eqref{eq: time_to_go_dot_DP} and \eqref{eq:tgoEPdot} in $\dot{V}_2$, and carrying out further simplifications using \eqref{eq:aD_with_aE}, one may obtain 
		
		
		\begin{align} \label{eq: V_2_dot_final}
			\dot{V}_2 =& -(\zeta_2|s_2|^{\alpha_2} + \xi_2|s_2|^{\beta_2})^{\kappa_2} - \left(\sec \delta_\mathrm{DP}\epsilon_2 - \right. \nonumber \\
			& \left.  \left(  
			\dfrac{r^2_\mathrm{DP}\dot{\lambda}_\mathrm{DP}(v_\mathrm{P} - v_\mathrm{D}\cos(\gamma_\mathrm{P} - \gamma_\mathrm{D})) -4r_\mathrm{DP}c^2\sin \delta_\mathrm{PD}}{( \dot{r}_\mathrm{DP}^2 + r^2_\mathrm{DP}\dot{\lambda}^2_\mathrm{DP} + 2c\dot{r}_\mathrm{DP})^2} + \right. \right. \nonumber \\ 
			& \left. \left.  \dfrac{r_\mathrm{DP}v_\mathrm{D}(4c + \dot{r}_\mathrm{DP}) \sin (\gamma_\mathrm{P}- \gamma_\mathrm{D})}{( \dot{r}_\mathrm{DP}^2 + r^2_\mathrm{DP}\dot{\lambda}^2_\mathrm{DP} + 2c\dot{r}_\mathrm{DP})^2} + \dfrac{r_\mathrm{EP}\sin \delta_\mathrm{PE}}{\dot{r}_\mathrm{EP}^2} \right)\right. \nonumber \\
			& \left. \times \sign(s_2)a_\mathrm{P} \right) \nonumber \\
			\leq& -(\zeta_2|s_2|^{\alpha_2} + \xi_2|s_2|^{\beta_2})^{\kappa_2} - \left( \epsilon_2 - \left( \dfrac{r_\mathrm{EP}}{(v_\mathrm{P} + v_\mathrm{E})^2} \right. \right. \nonumber \\
			&\left. \left. + \dfrac{(v_\mathrm{P} + v_\mathrm{D})^2 + 4r_\mathrm{DP}c^2 + r_\mathrm{DP}v_\mathrm{D}(4c + v_\mathrm{P} + v_\mathrm{D})}{(2(v_\mathrm{P} + v_\mathrm{D})^2 + 2c(v_\mathrm{P} + v_\mathrm{D}))^2}
			\right)a_\mathrm{P}^{\mathrm{max}}
			\right) \nonumber \\
			\leq& -(\zeta_2|s_2|^{\alpha_2} + \xi_2|s_2|^{\beta_2})^{\kappa_2} <0, ~ \forall ~s_2 \neq 0,
		\end{align}
		and when the condition on $\epsilon_2$ given in \Cref{theorem: Defender's law} holds. Consequently, $s_2$ converges within a fixed time $t_2 \leq \frac{1}{\zeta^{\kappa_2}_2(1- \alpha_2 \kappa_2)}+ \frac{1}{\xi_2^{\kappa_2}(\beta_2 \kappa_2 - 1)}$ , regardless of its initial value, leading to an interception of the pursuer within margin $\tau$. 
		
		
	\end{proof}

	\begin{remark}
		In \eqref{eq:aD_with_aE}, each term in the denominator is bounded and non-vanishing over the engagement prior to interception ($r_\mathrm{DP} \to 0$). Moreover, as $s_1 \to 0$, $a_\mathrm{E} \to 0$, and $\dot{\lambda}_\mathrm{EP} \to 0$, so the associated terms decay. Hence, the control input, $a_\mathrm{D}$, remains bounded.
	\end{remark}
	The defender's proposed time-constrained cooperative guidance strategy leverages the knowledge of the evader's maneuver in shaping its own interception trajectory to protect the latter. This may necessitate higher communication between the evader-defender team. However, the proposed design remains robust to partial information from the evader. In scenarios where the defender does not have direct access to the evader's maneuver, the defender can still execute an interception strategy to neutralize the pursuer by relying on partial engagement information. This is discussed in the next corollary.
	
	\begin{corollary}
		In scenarios where the defender does not have access to the evader's maneuver information, the proposed guidance strategy, \eqref{eq:aD_with_aE}, still remains effective, ensuring the interception of the pursuer by the defender before the former can capture the evader. The resulting guidance strategy, in the absence of knowledge of the evader's acceleration, is  
		\begin{align}\label{eq:aD_without_aE}
			a_\mathrm{D}=& c\dot{\lambda}_\mathrm{DP} -\dfrac{r^2_\mathrm{EP}(\dot{r}_\mathrm{DP}^2 + r^2_\mathrm{DP}\dot{\lambda}_\mathrm{DP}^2 + 2c\dot{r}_\mathrm{DP})^2 }{2 r^2_\mathrm{DP}\dot{r}_\mathrm{EP}^2\dot{\lambda}_\mathrm{DP}(\dot{r}_\mathrm{DP} + 2c)}\dot{\lambda}^2_\mathrm{EP} \nonumber \\
			& + \dfrac{(\dot{r}_\mathrm{DP}^2 + r^2_\mathrm{DP}\dot{\lambda}_\mathrm{DP}^2 + 2c\dot{r}_\mathrm{DP})^2}{2 r_\mathrm{DP}^2 \dot{\lambda}_\mathrm{DP}(\dot{r}_\mathrm{DP} + 2c)} \notag  \\
			&\times\left[(\zeta_2 |s_2|^{\alpha_2} + \xi_2|s_2|^{\beta_2})^{\kappa_2} + \sec{\delta_\mathrm{DP}\epsilon_3}  \right] \sign(s_2),  
		\end{align}
		where the design parameters satisfy $\zeta_2, \xi_2, \alpha_2, \beta_2, \kappa_2, c > 0,~ \alpha_2 \kappa_2 < 1, \beta_2 \kappa_2 > 1$ and $\epsilon_3 >\sup_{t\ge 0} \left[ \frac{r_\mathrm{DP}(v_\mathrm{P}+v_\mathrm{D})^2 + 4c^2r_\mathrm{DP}  + (4c + v_\mathrm{P} + v_\mathrm{D})r_\mathrm{DP}}{((1+r_\mathrm{DP})(v_\mathrm{P} + v_\mathrm{D})^2 +2c(v_\mathrm{P}+v_\mathrm{D}))^2}+ \frac{r_\mathrm{EP}}{(v_\mathrm{E}+v_\mathrm{P})^2} \right] a_\mathrm{P}^{\max} + \frac{r_\mathrm{EP}}{(v_\mathrm{P} + v_\mathrm{E})^2}a_\mathrm{E}^\mathrm{max}$ guarantees that $s_2$ converges to zero within a fixed time independent of the initial configuration of the defender-pursuer engagement and the pursuer's strategy.
	\end{corollary}
	\begin{proof}
		The proof is similar to that of \Cref{theorem: Defender's law} with the exception of a different sufficient condition reflected in $\epsilon_3$. The proof is thus omitted due to space constraints.

	\end{proof}
	\begin{remark}
		The time margin protects the evader from last-moment hits, especially if the pursuer has some unpredictable behavior in the terminal interception phase. However, the selection of time margin is critical during practice and obtaining explicit bounds of $\tau$ may not be tractable for a general nonlinear engagement. However, a conservative bound on $\tau$ can still be given as $\tau<\min\left\{\frac{r_\mathrm{DP}(0)}{v_\mathrm{D}^{\max}+v_\mathrm{P}},t_\mathrm{go}^\mathrm{EP}(0) - t_2\right\}$ assuming point capture.
	\end{remark}
	
	\section{Simulations}\label{sec:results}
	
	The effectiveness of the proposed time-constrained guidance strategies for the evader-defender team, aimed at protecting the evader from interception by a pursuer through the use of the defender, is demonstrated in this section via numerical simulations. In the first scenario, the defender has access to the evader's maneuver and uses
	the strategy given by \eqref{eq:aD_with_aE}, whereas in the second scenario, such information is unavailable to the defender, and it uses the guidance strategy \eqref{eq:aD_without_aE}. For both scenarios, three different cases are considered, where the pursuer executes various variants of the PN principle (pure PN, augmented PN, and realistic true PN) to effect an optimal class of maneuvers. The controller parameters are chosen as $\alpha_1 = \alpha_2 = 0.3, \beta_1=2, \kappa_1=\kappa_2=1, \zeta_1= \zeta_2 = 0.05, \xi_1= 0.005$. The time margin $\tau$ is set to $5$\,s, unless noted otherwise. The speeds of the evader and the pursuer are set as $100$ m/s and $375$ m/s, respectively. The maximum lateral acceleration bound on the evader is $|5|$ g, while for the pursuer and the defender, it is set at $|40|$ g. The initial positions of the agents are indicated by diamonds $(\diamond)$, and the interception position is depicted by a cross $(\times)$. The initial range and LOS angle between the evader and the pursuer are $r_\mathrm{EP}=15$ km and $\lambda_\mathrm{EP}=-45^\circ$, respectively, while the initial heading, $\gamma_\mathrm{P}$, of the pursuer is set to $165^\circ$.

	\subsection{Defender can access the Evader's Strategy}
\begin{figure}[t]
		\centering
		\captionsetup[sub]{justification=centering}
		\begin{subfigure}[t]{.49\linewidth}
			\centering
			\includegraphics[width=1.1\linewidth]{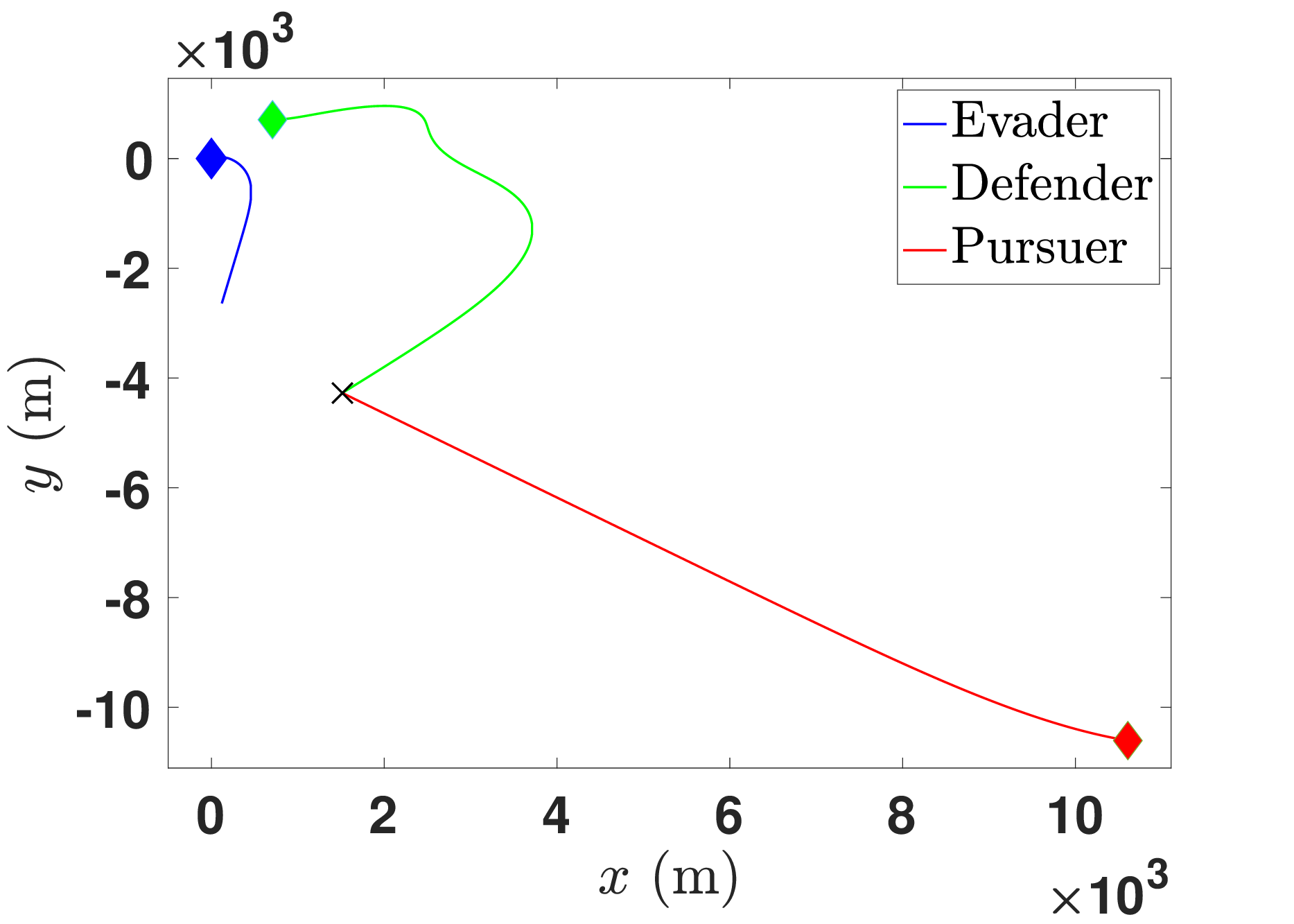}
			\caption{Trajectories.}
			\label{fig:traj_PPN_with_aE_access}
		\end{subfigure}%
		\hfill
		\begin{subfigure}[t]{.49\linewidth}
			\centering
			\includegraphics[width= 1.1\linewidth]{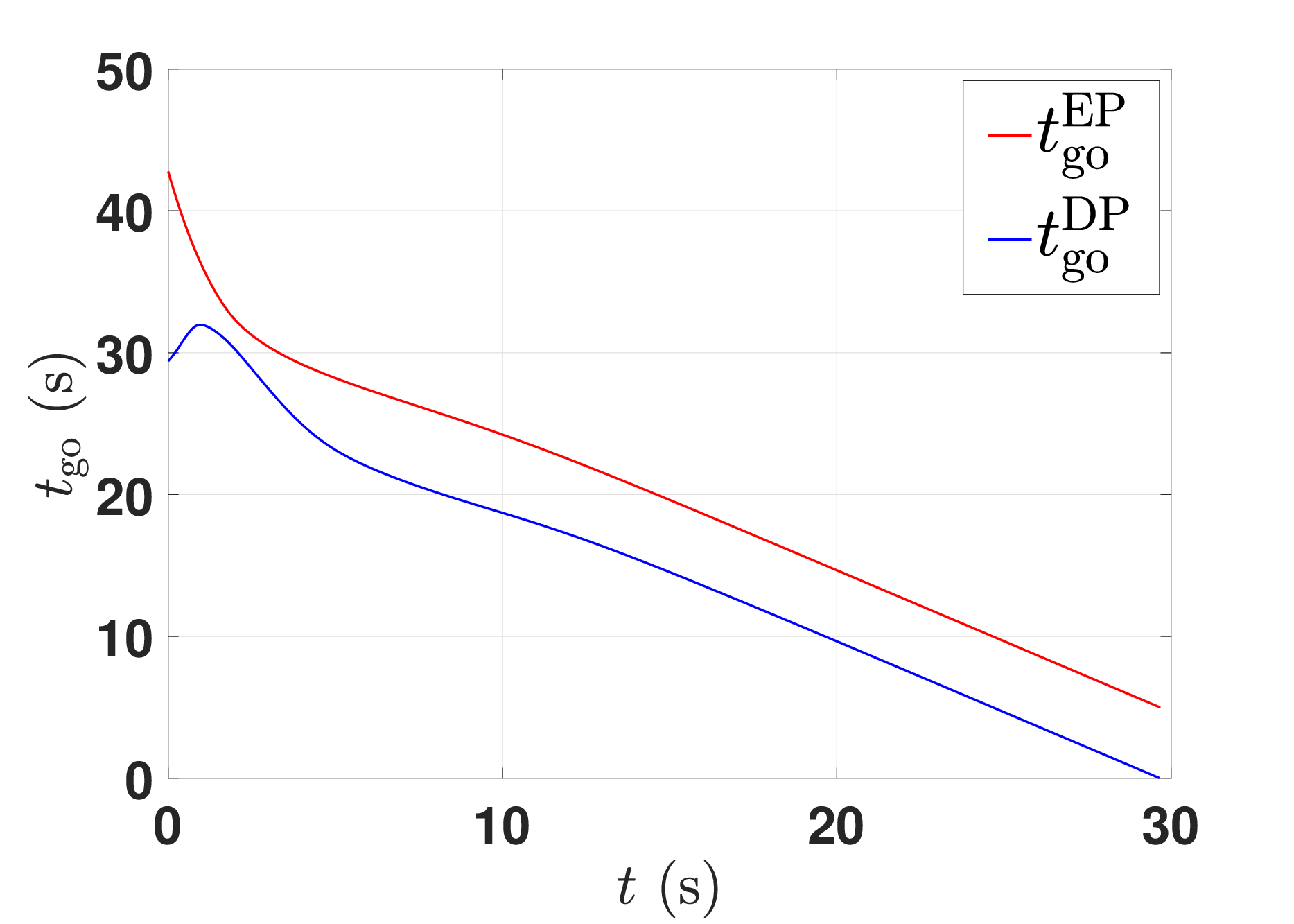}
			\caption{Time-to-go.}
			\label{fig:time_to_go_PPN_with_aE_access}
		\end{subfigure}
		\vspace{0.6em}
		\begin{subfigure}[t]{.49\linewidth}
			\centering
			\includegraphics[width=1.1\linewidth]{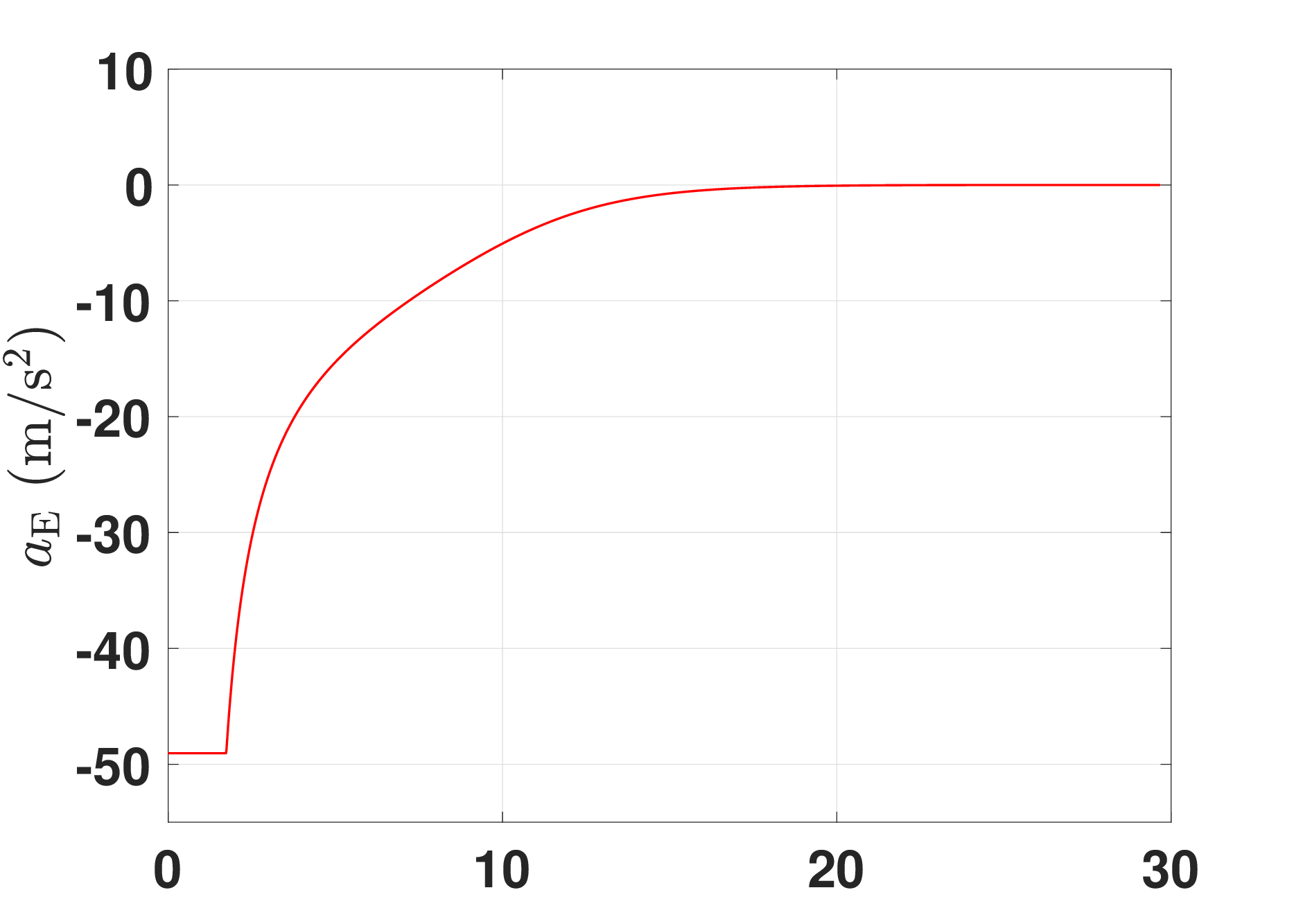}\vspace{0.3em}
			\includegraphics[width=1.1\linewidth]{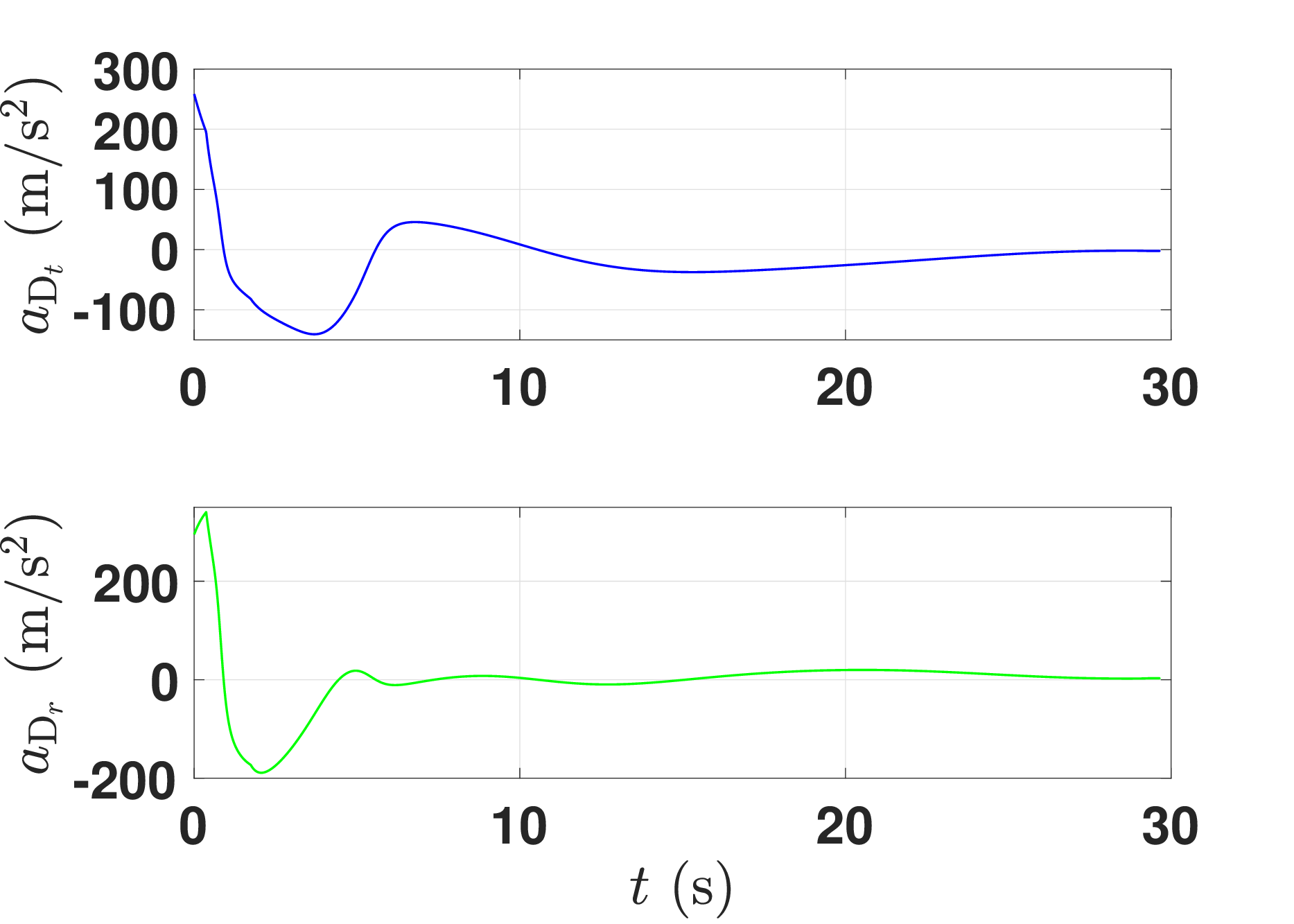}
			\caption{Lateral accelerations (steering controls).}
			\label{fig:acceleration_PPN_with_aE_access}
		\end{subfigure}
		\hfill
		\begin{subfigure}[t]{.49\linewidth}
			\centering
			\includegraphics[width=1.1\linewidth]{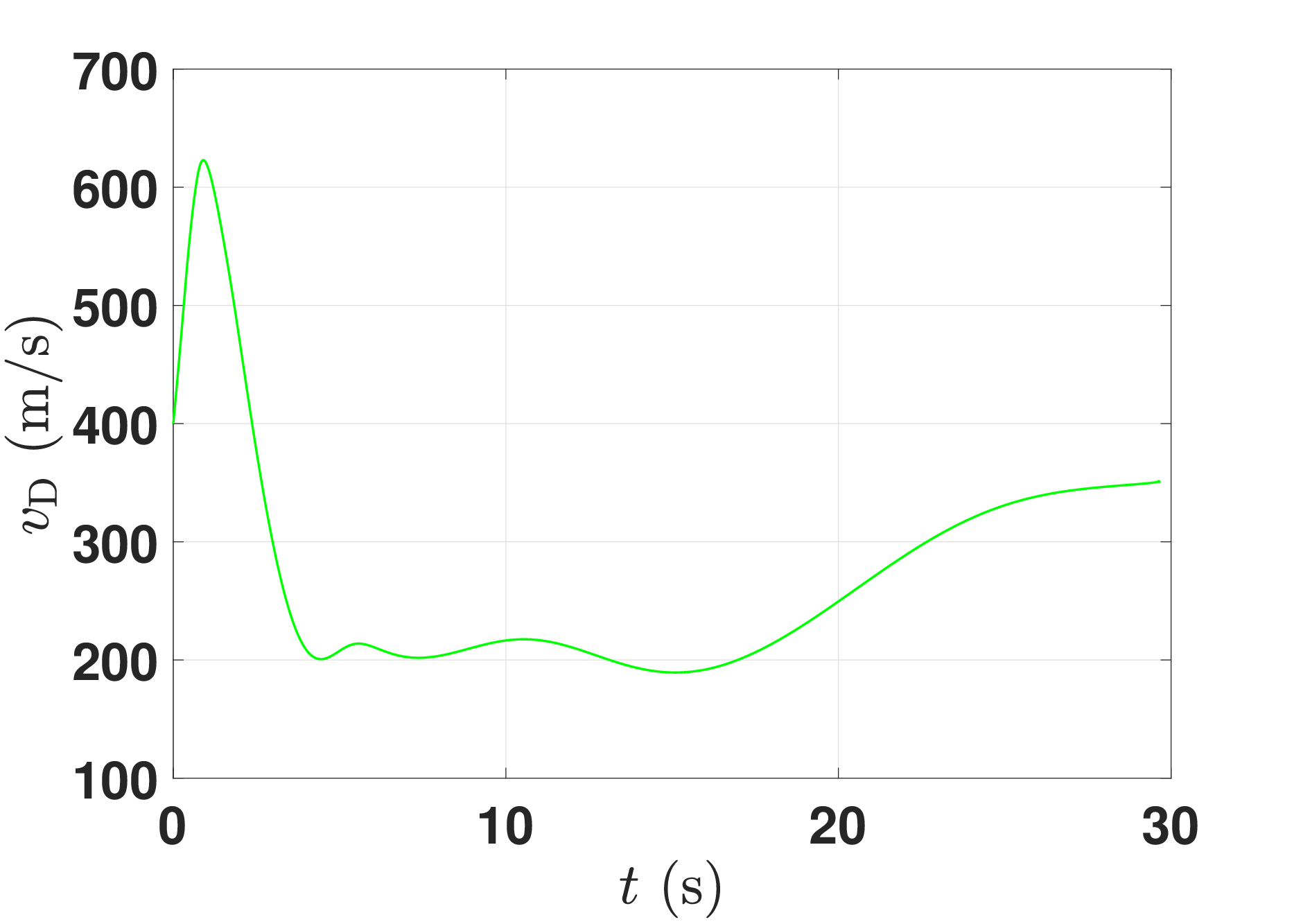}\vspace{0.3em}
			\includegraphics[width=1.1\linewidth]{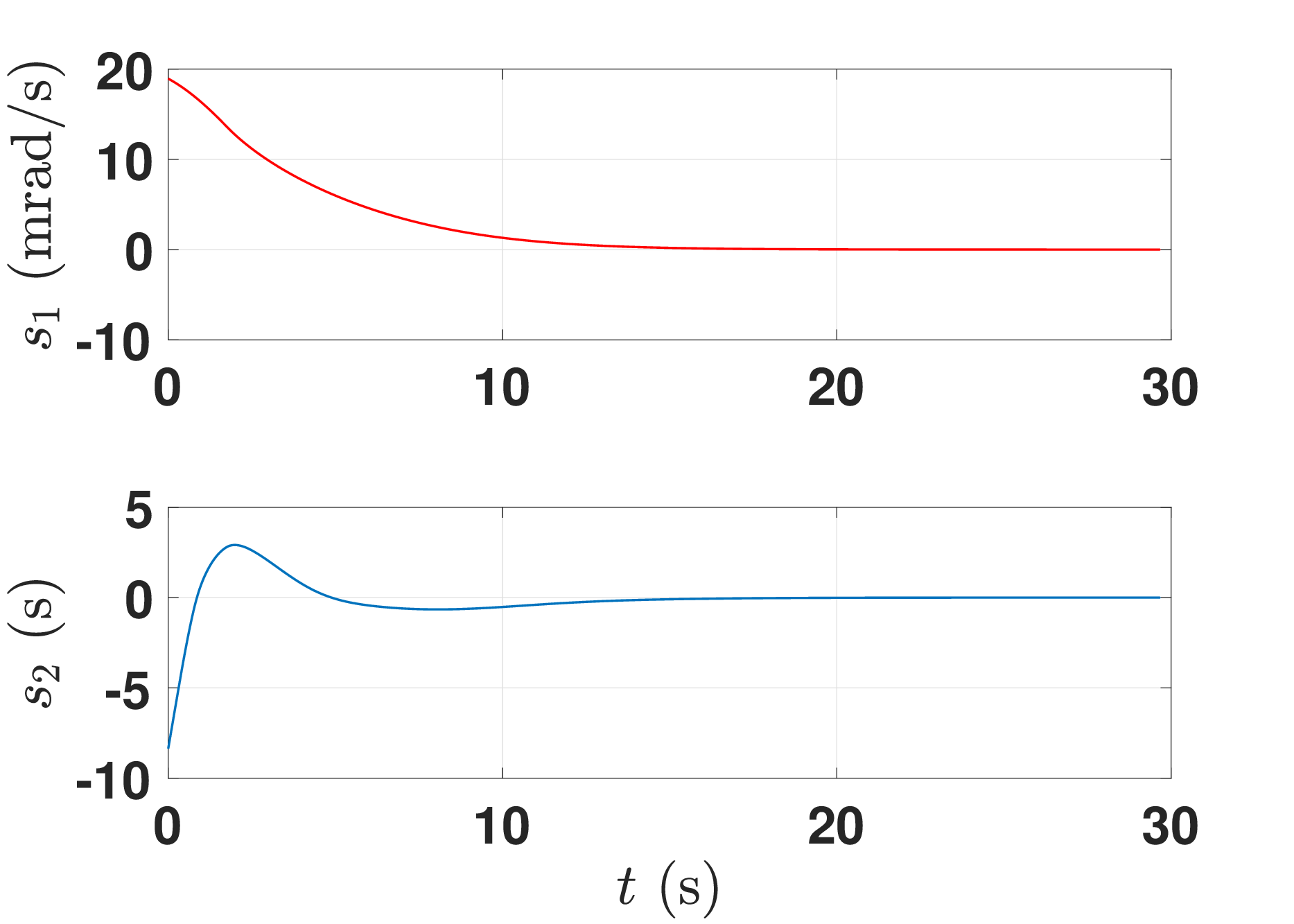}
			\caption{Defender's velocity and sliding manifolds (error profiles).}
			\label{fig:errors_PPN_with_aE_access}
		\end{subfigure}
		\caption{Performance evaluation under \(a_{\mathrm{P}}=N v_{\mathrm{P}}\dot{\lambda}_{\mathrm{EP}}\).}
		\label{fig:Results_PPN_aE_access}
	\end{figure}

	Consider the first case, where the pursuer is assumed to execute a pure PN guidance strategy, given as $a_\mathrm{P}= Nv_\mathrm{P}\dot{\lambda}_\mathrm{EP}$, with the navigation constant $N$ chosen as $5$. The design parameters are selected as $\alpha_2=0.3$, $\zeta_2=0.05$, $\xi_2=0.05$ and $\beta_2 = 0.8$. The initial speed of the defender is set as $400$ m/s. The results for this case are illustrated in \Cref{fig:Results_PPN_aE_access}. The initial range and the LOS angle between the evader and the defender are $r_\mathrm{DE}=1$ km and $\lambda_\mathrm{DE}=45^\circ$, while the initial headings are $\gamma_\mathrm{E} =30^\circ$, and $\gamma_\mathrm{D} =0^\circ$. The trajectories of the agents are depicted in \Cref{fig:traj_PPN_with_aE_access}. It is apparent from \Cref{fig:traj_PPN_with_aE_access} that the evader adapts its motion according to \eqref{eq:aE}, such that it nullifies the rate of relative LOS angle between the pursuer and itself, $\dot{\lambda}_\mathrm{EP}$, while the defender employs the guidance strategy \eqref{eq:aD_with_aE}, and successfully protects the evader by intercepting the pursuer before the latter can capture the evader. \Cref{fig:time_to_go_PPN_with_aE_access} illustrates that the defender follows the required trajectory to intercept the pursuer within the prespecified time margin, $\tau$. The time-to-go profile exhibits an initial increase followed by a decrease, which arises from the defender’s trajectory initially deviating and subsequently converging toward the pursuer’s trajectory (as depicted in \Cref{fig:traj_PPN_with_aE_access}). The behavior of the lateral accelerations is portrayed in \Cref{fig:acceleration_PPN_with_aE_access}. It can be observed that the evader’s lateral acceleration is initially high to nullify the relative LOS angle rate, $\dot{\lambda}_\mathrm{EP}$, and subsequently converges to zero, while during the transient phase, the defender’s lateral acceleration is high to achieve course correction, and subsequently converges close to zero in the steady state. The defender's speed and the sliding manifold profiles are shown in \Cref{fig:errors_PPN_with_aE_access}. This shows that with the proposed guidance strategy of the evader, the LOS rate, $\dot{\lambda}_\mathrm{EP}$, becomes zero in approximately $15$ s, inducing the pursuer to become non-maneuvering by executing a deceptive maneuver. The sliding manifold for the time-to-go error exhibits an initial overshoot due to the deviation in the defender’s trajectory, before converging to zero in about $18$ s.

	
	\begin{figure}[t]
		\centering
		\captionsetup[sub]{justification=centering}
		\begin{subfigure}[t]{.49\linewidth}
			\centering
			\includegraphics[width=1.1\linewidth]{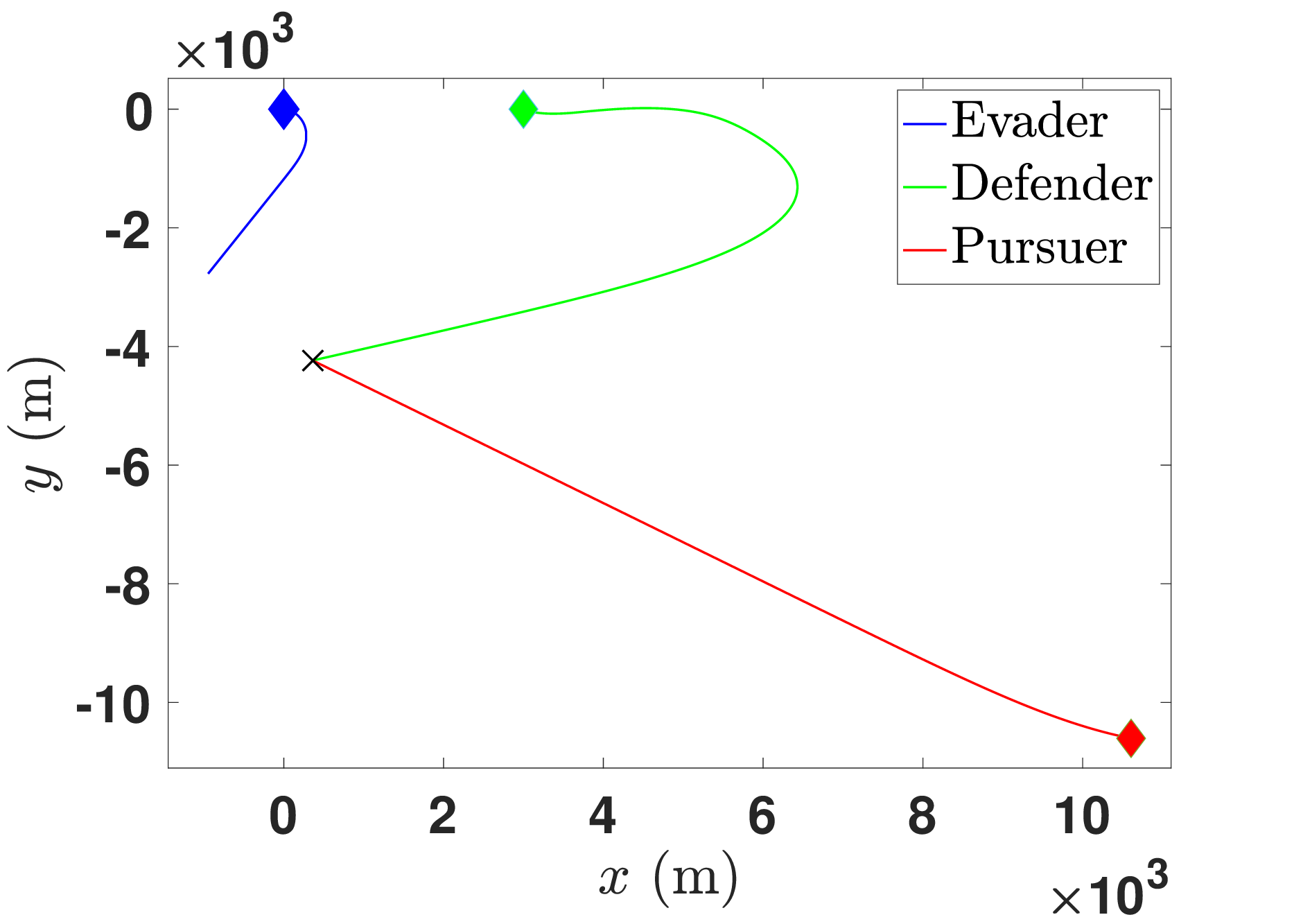}
			\caption{Trajectories.}
			\label{fig:traj_RTPN_with_aE_access}
		\end{subfigure}%
		\hfill
		\begin{subfigure}[t]{.49\linewidth}
			\centering
			\includegraphics[width= 1.1\linewidth]{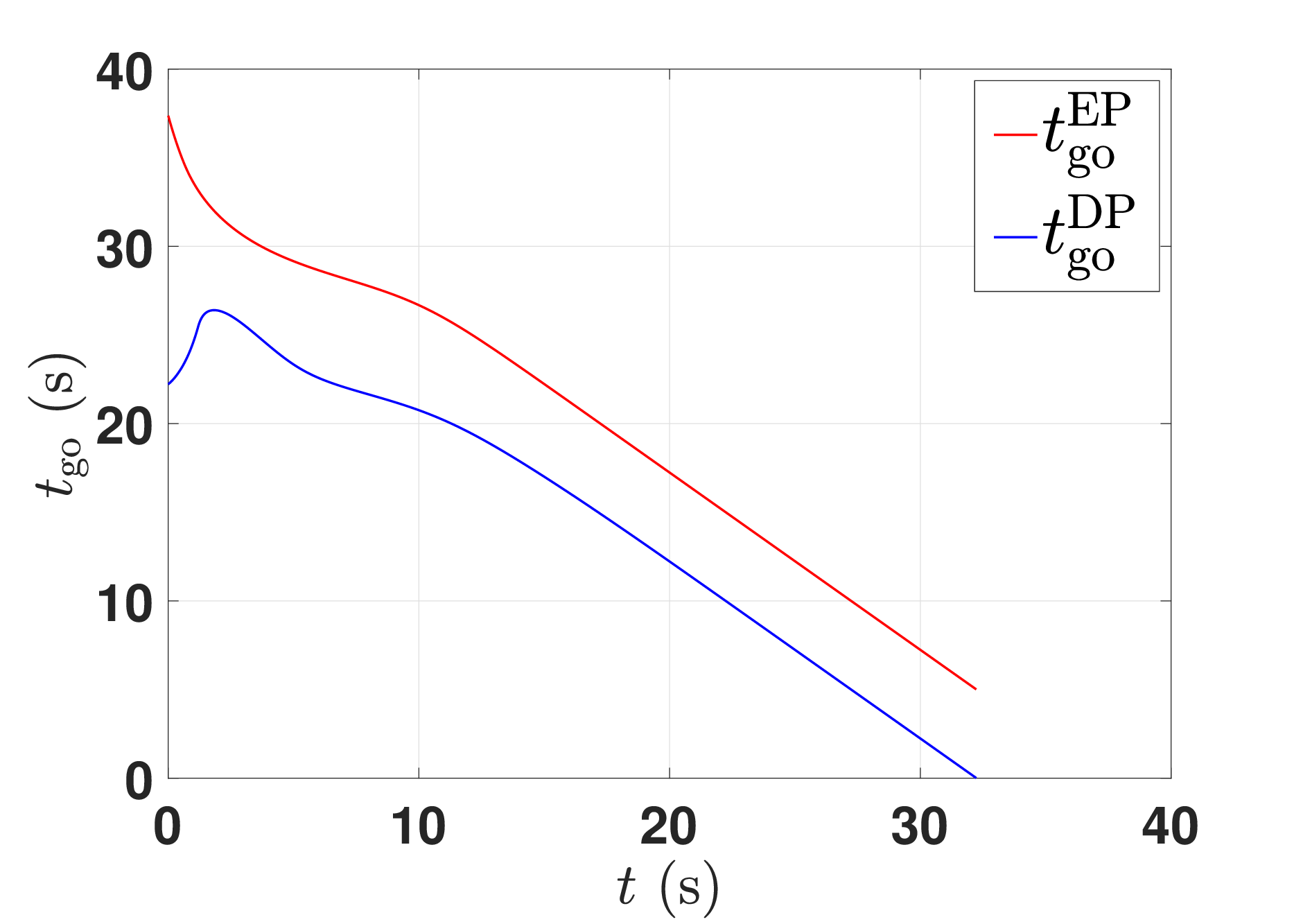}
			\caption{Time-to-go.}
			\label{fig:time_to_go_RTPN_with_aE_access}
		\end{subfigure}
		\vspace{0.6em}
		\begin{subfigure}[t]{.49\linewidth}
			\centering
			\includegraphics[width=1.1\linewidth]{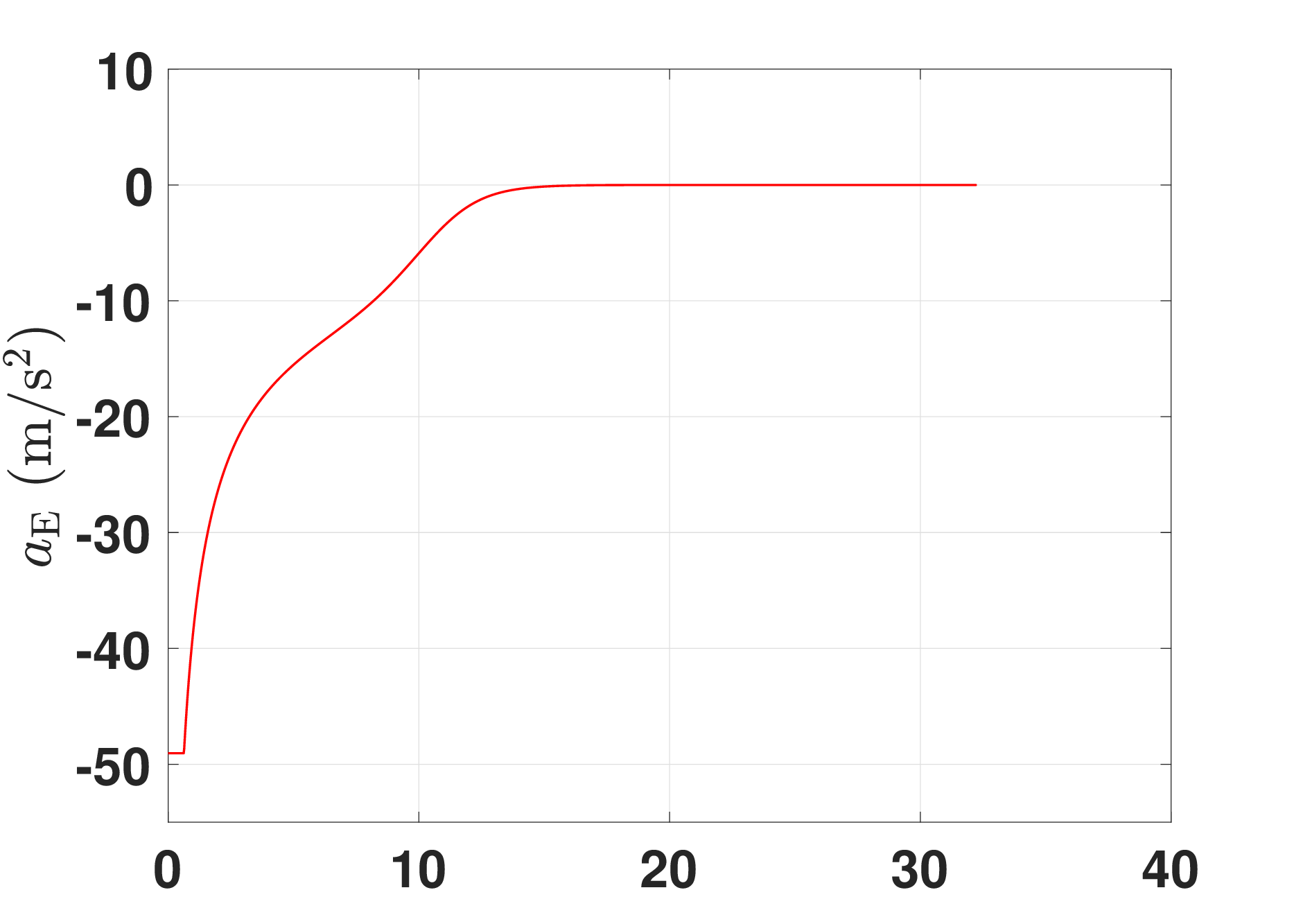}\vspace{0.3em}
			\includegraphics[width=1.1\linewidth]{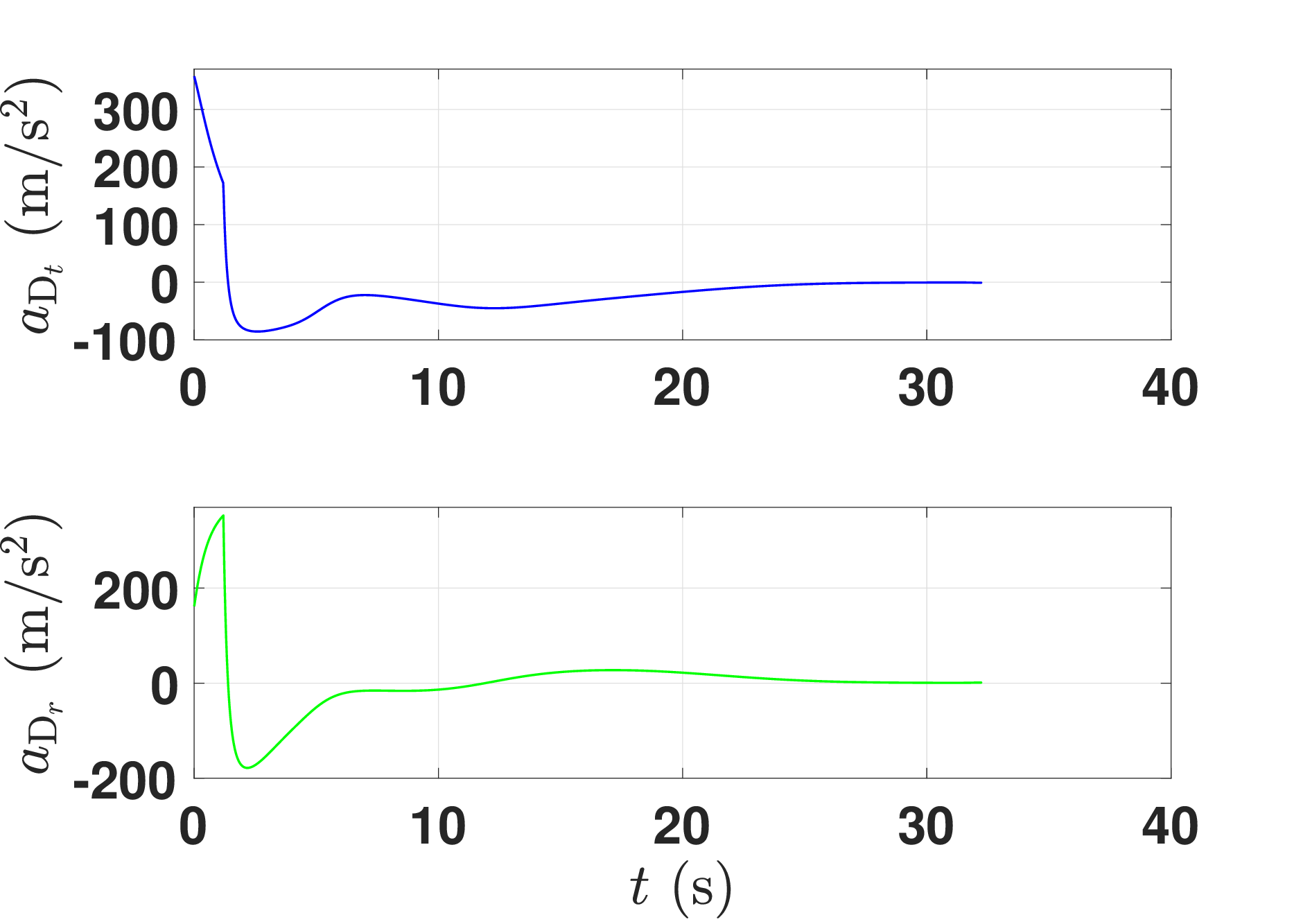}
			\caption{Lateral accelerations (steering controls).}
			\label{fig:acceleration_RTPN_with_aE_access}
		\end{subfigure}
		\hfill
		\begin{subfigure}[t]{.49\linewidth}
			\centering
			\includegraphics[width=1.1\linewidth]{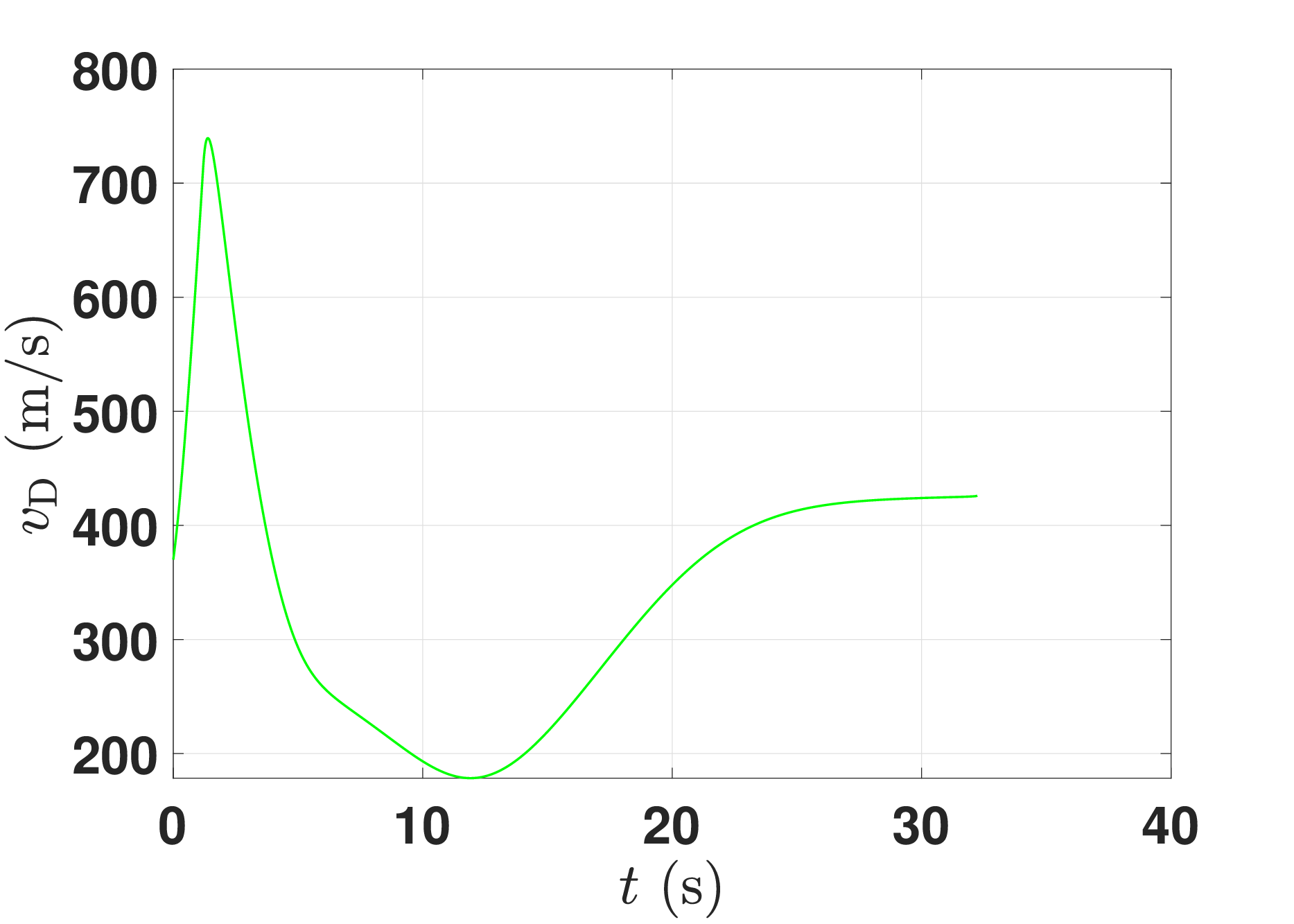}\vspace{0.3em}
			\includegraphics[width=1.1\linewidth]{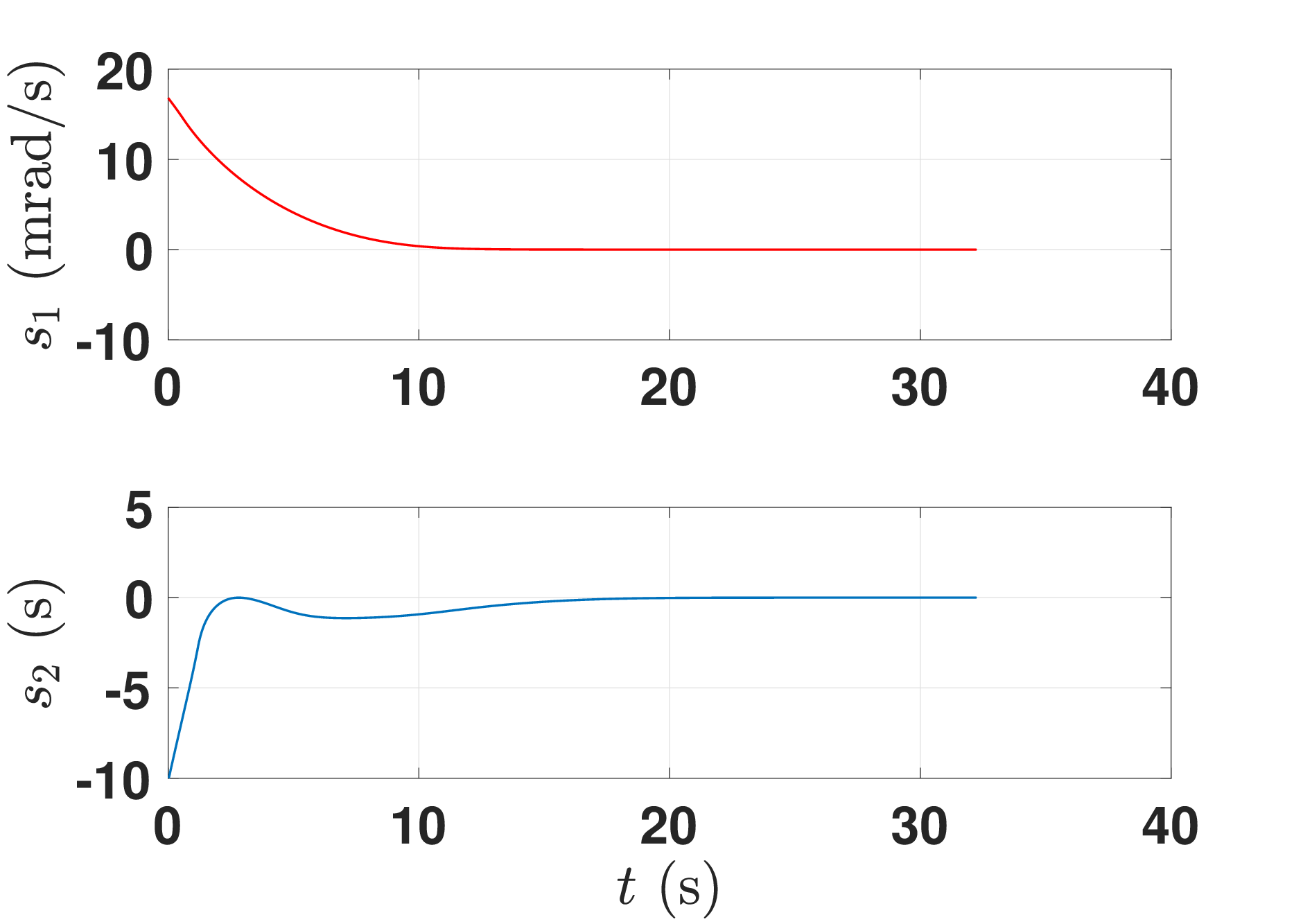}
			\caption{Defender's velocity and sliding manifolds.}
			\label{fig:errors_RTPN_with_aE_access}
		\end{subfigure}
		\caption{Performance evaluation under \(a_{\mathrm{P}}=-N \dot{r}_\mathrm{EP}\dot{\lambda}_{\mathrm{EP}}\).}
		\label{fig:Results_RTPN_aE_access}
	\end{figure}

	
	\begin{figure}[t]
		\centering
		\captionsetup[sub]{justification=centering}
		\begin{subfigure}[t]{.49\linewidth}
			\centering
			\includegraphics[width=1.1\linewidth]{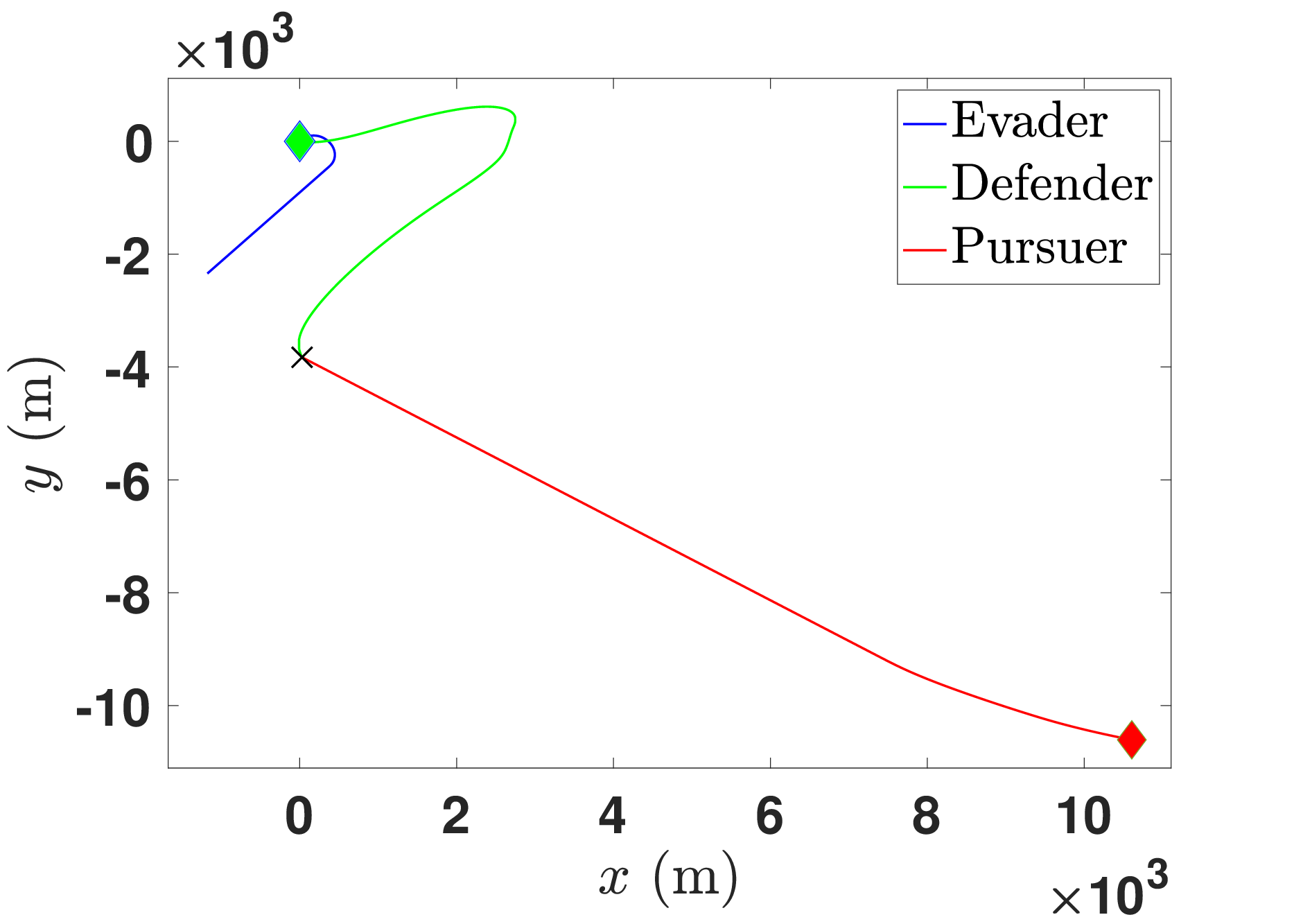}
			\caption{Trajectories.}
			\label{fig:traj_APN_with_aE_access}
		\end{subfigure}%
		\hfill
		\begin{subfigure}[t]{.49\linewidth}
			\centering
			\includegraphics[width= 1.1\linewidth]{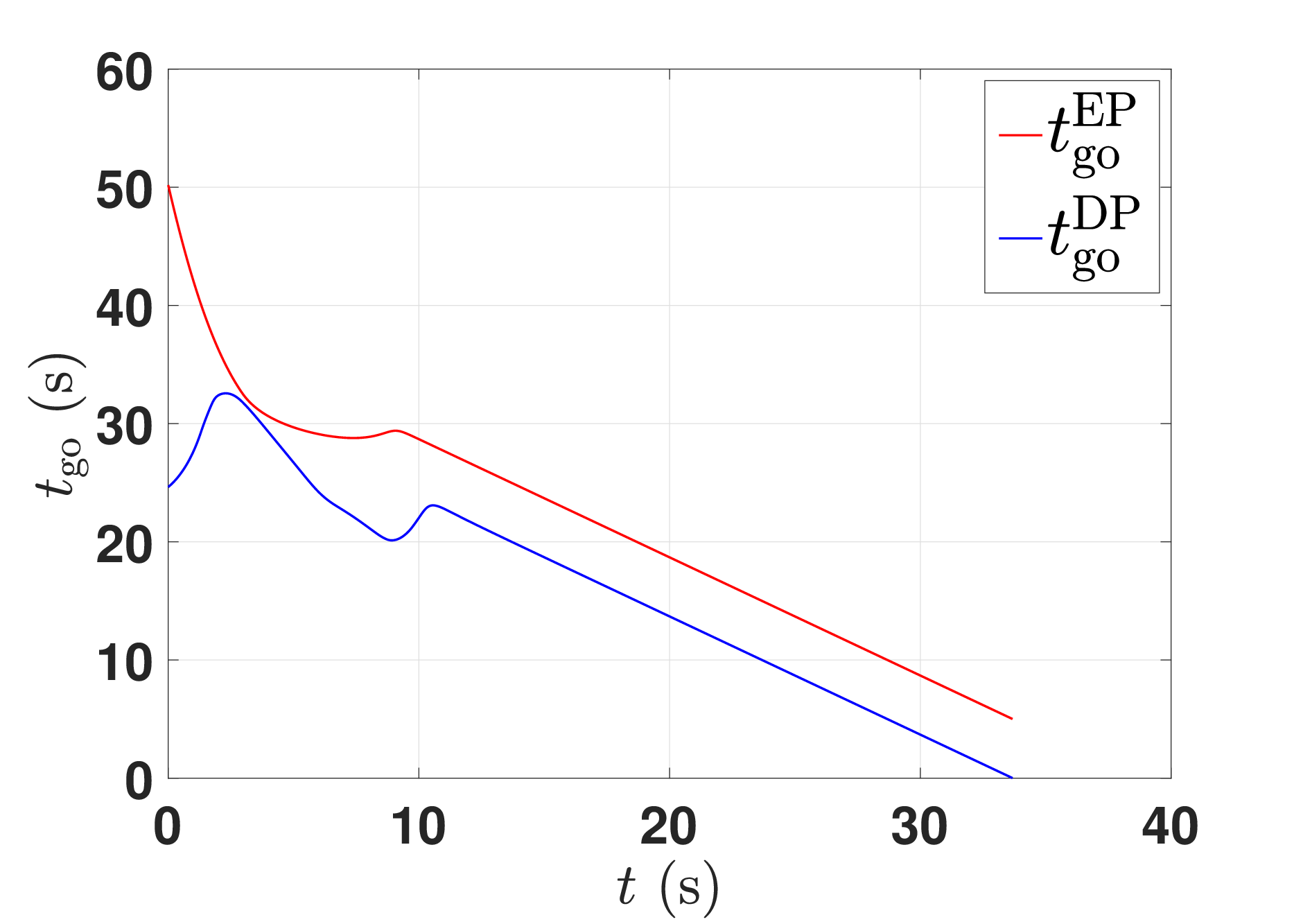}
			\caption{Time-to-go.}
			\label{fig:time_to_go_APN_with_aE_access}
		\end{subfigure}
		\vspace{0.6em}
		\begin{subfigure}[t]{.49\linewidth}
			\centering
			\includegraphics[width=1.1\linewidth]{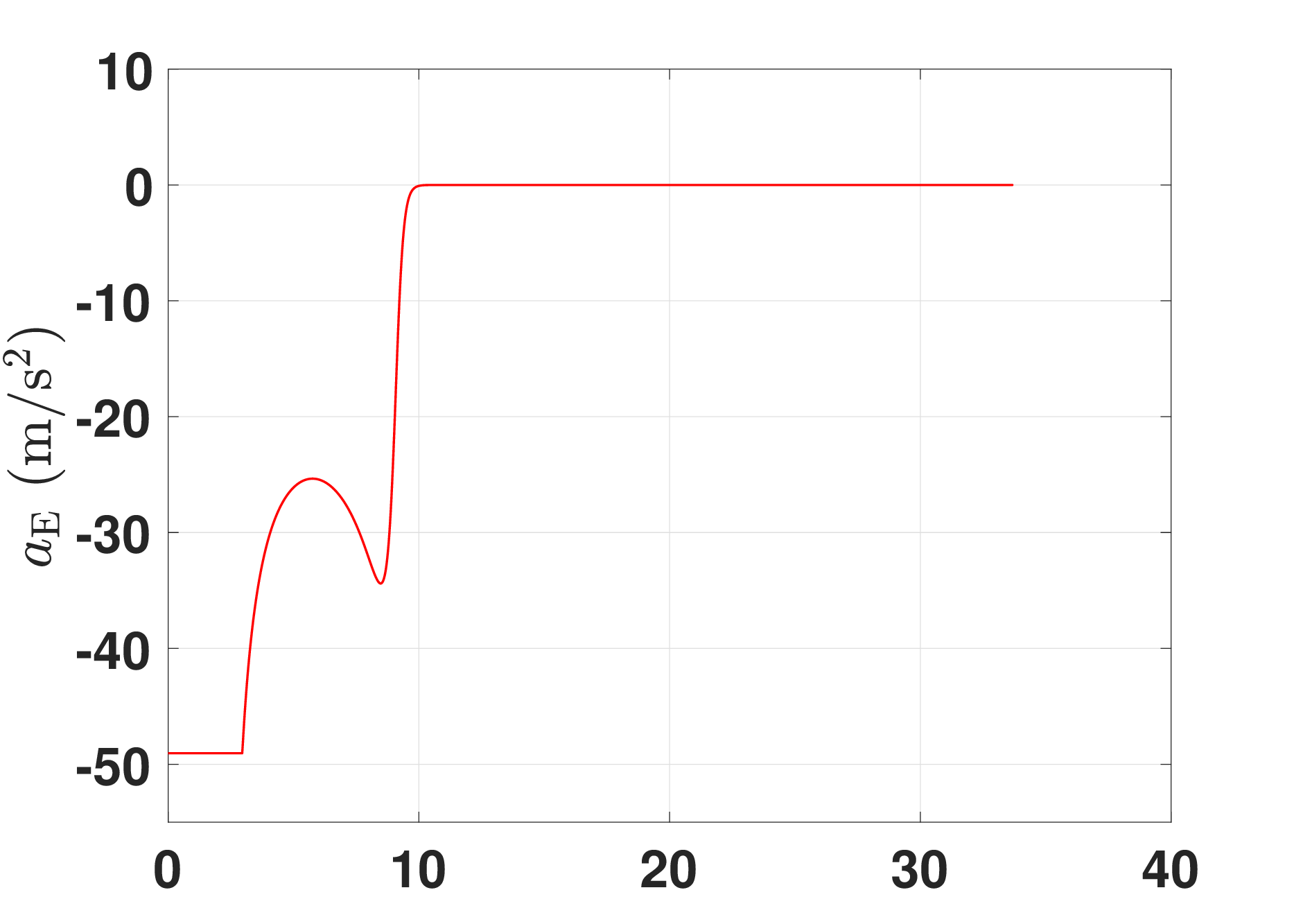}\vspace{0.3em}
			\includegraphics[width=1.1\linewidth]{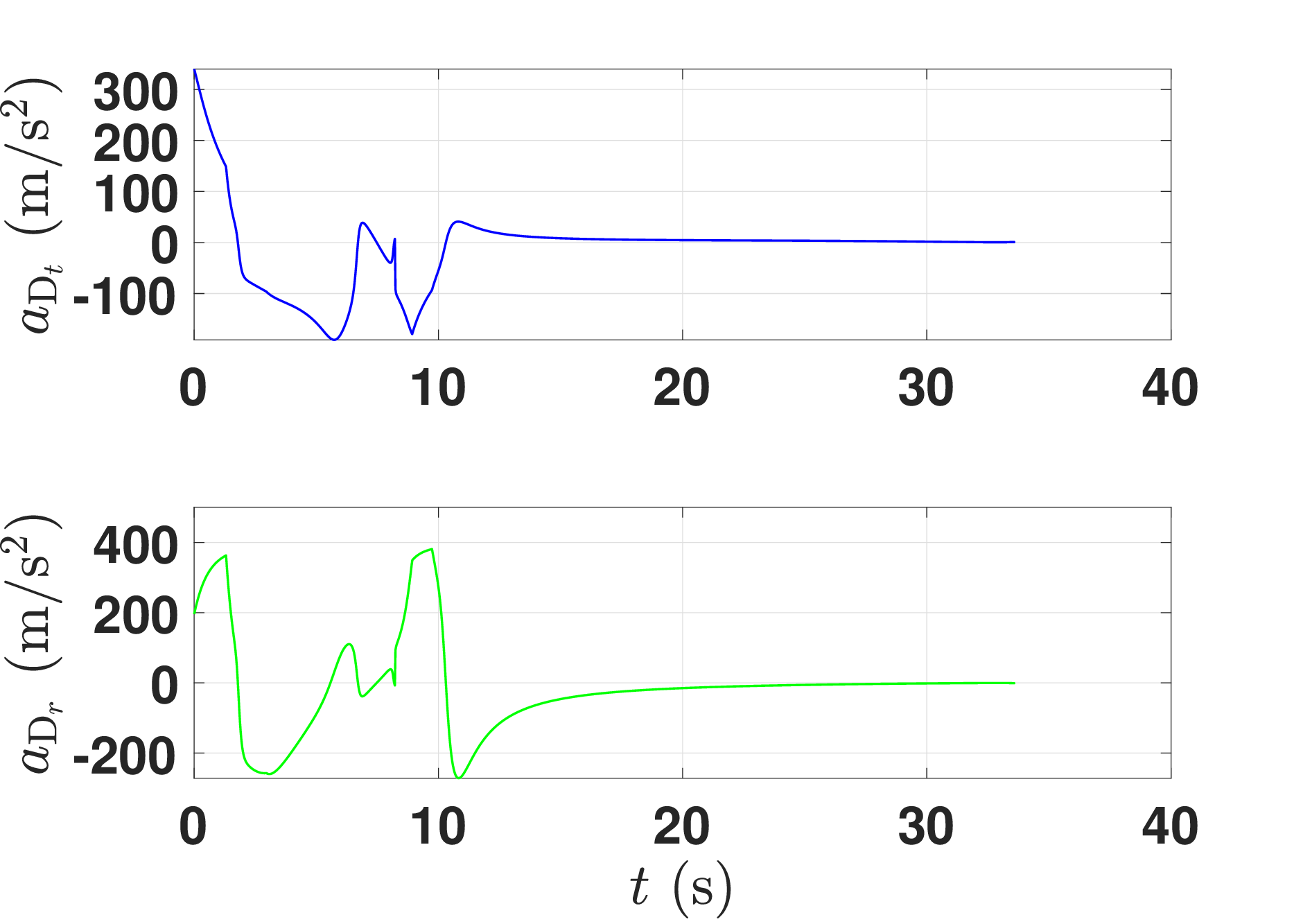}
			\caption{Lateral accelerations (steering controls).}
			\label{fig:acceleration_APN_with_aE_access}
		\end{subfigure}
		\hfill
		\begin{subfigure}[t]{.49\linewidth}
			\centering
			\includegraphics[width=1.1\linewidth]{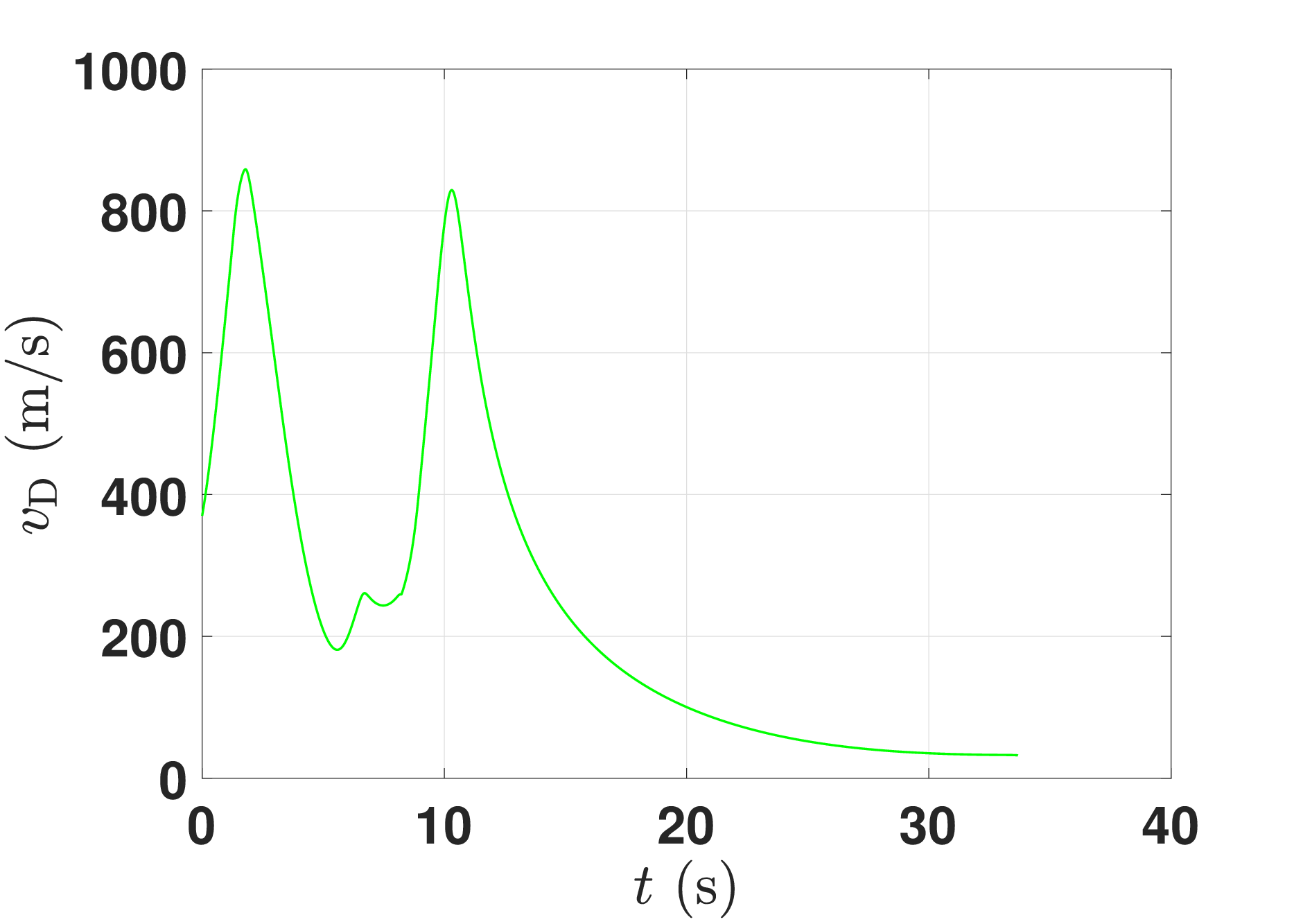}\vspace{0.3em}
			\includegraphics[width=1.1\linewidth]{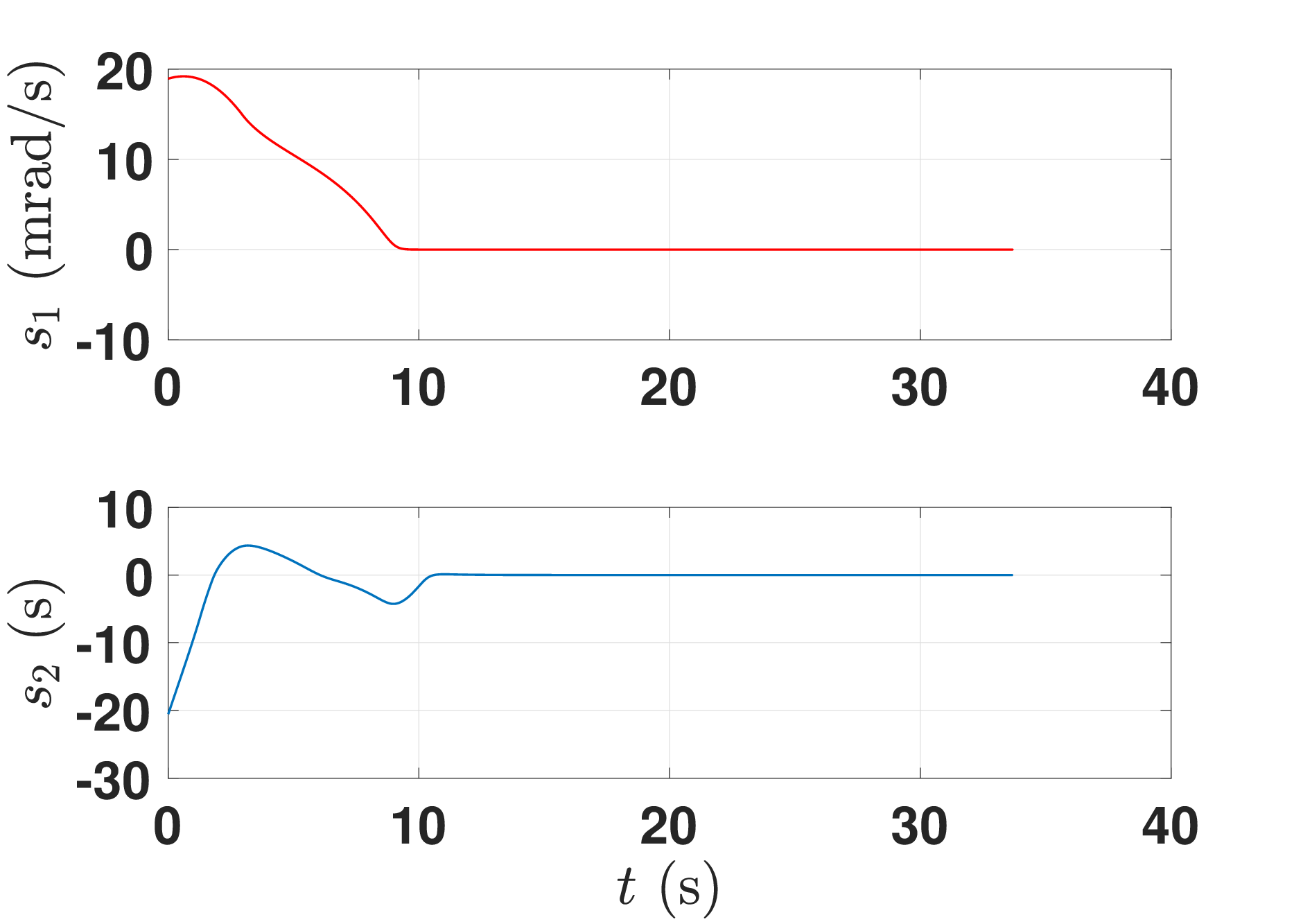}
			\caption{Defender's velocity and sliding manifolds.}
			\label{fig:errors_APN_with_aE_access}
		\end{subfigure}
		\caption{Performance evaluation under \(a_{\mathrm{P}}= -Nv_{\mathrm{P}}\dot{\lambda}_{\mathrm{EP}} +k_\mathrm{P}a_\mathrm{E}\).}
		\label{fig:Results_APN_aE_access}
	\end{figure}
	In the second case, the pursuer employs a realistic true PN guidance strategy, $a_\mathrm{P}=-N\dot{r}_\mathrm{EP}\dot{\lambda}_\mathrm{EP}$, with the navigation constant kept the same as in the previous case. The design parameters are chosen as $\alpha_2=0.3$, $\zeta_2=0.05$, $\beta_2 = 2$ and $\xi_2 =1.2$. The initial speed of the defender is set as $370$ m/s. The results are presented in \Cref{fig:Results_RTPN_aE_access}. The initial range and LOS angle between the evader and the defender are $3$ km and $0^\circ$, respectively, while the initial headings are $\gamma_\mathrm{E}= -5^\circ$, and $\gamma_\mathrm{D}= -30^\circ$. Despite the pursuer adopting a different strategy, the defender intercepts it and protects the evader. The defender’s trajectory, however, exhibits a smoother profile compared to the earlier scenario. Similar to the case when the pursuer employs the pure PN strategy, the time-to-go profiles indicate a similar trend in \Cref{fig:time_to_go_RTPN_with_aE_access}. The lateral acceleration profiles are shown in \Cref{fig:acceleration_RTPN_with_aE_access}, which indicate that the evader’s control converges to zero within $20$ s, while a higher control effort from the defender is required during the transient phase to correct the trajectory. The defender's velocity and the error profiles are depicted in \Cref{fig:errors_RTPN_with_aE_access}. From \Cref{fig:errors_RTPN_with_aE_access}, it is apparent that the defender's velocity is smoother than in the prior case. The LOS rate error vanishes in around $15$ s, while the time-to-go error reduces to zero in approximately $20$ s. 

	In the third case, the pursuer employs augmented PN guidance, given as $a_\mathrm{P}= Nv_\mathrm{P}\dot{\lambda}_\mathrm{EP} + k_\mathrm{P}a_\mathrm{E}$, where $k_\mathrm{P}$ is a design parameter whose value is set as $1$.  As evident from the expression, it consists of two terms-- the first corresponds to pure PN, and the second accounts for the evader’s maneuver. The pursuer exploits the evader's maneuvering information to adjust the trajectory accordingly. The parameters are selected as $\alpha_2 = 0.99$, $\zeta_2= 0.01$, $\xi_2=0.07$. The initial speed of the defender is set as $370$ m/s. The results are demonstrated in \Cref{fig:Results_APN_aE_access}. The initial range for the evader-defender engagement is $r_\mathrm{DE}=0$ km, while the initial heading of the agents is $\gamma_\mathrm{E}=60^\circ$, and $\gamma_\mathrm{D}=-15^\circ$.  Despite the pursuer having access to the evader's strategy, the defender successfully intercepts it before the pursuer approaches the vicinity of the evader. It is now evident that if the evader becomes non-maneuvering, then the pursuer's strategy becomes a function of their LOS rate, which will be driven to zero to render the pursuer on a straight line pursuit. This showcases the advantage of introducing a deceptive maneuver. Thereafter, we notice that the prespecified time margin is maintained by the defender to intercept the pursuer regardless of the strategy employed by the pursuer.
	
	\subsection{Defender can't access the Evader's Strategy}
	
\begin{figure}[t]
		\centering
		\captionsetup[sub]{justification=centering}
		\begin{subfigure}[t]{.49\linewidth}
			\centering
			\includegraphics[width=1.1\linewidth]{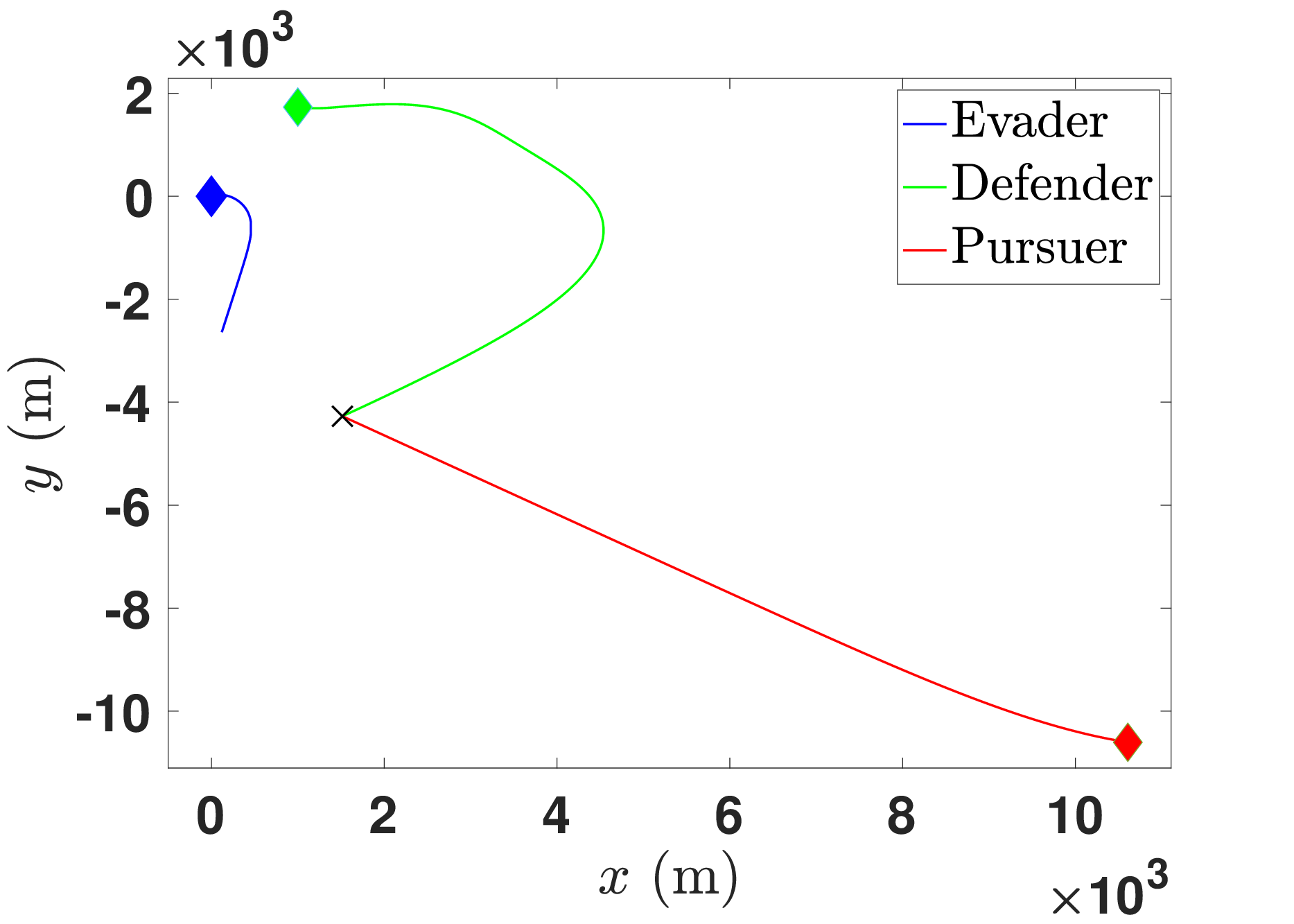}
			\caption{Trajectories.}
			\label{fig:traj_PPN_without_aE_access}
		\end{subfigure}%
		\hfill
		\begin{subfigure}[t]{.49\linewidth}
			\centering
			\includegraphics[width= 1.1\linewidth]{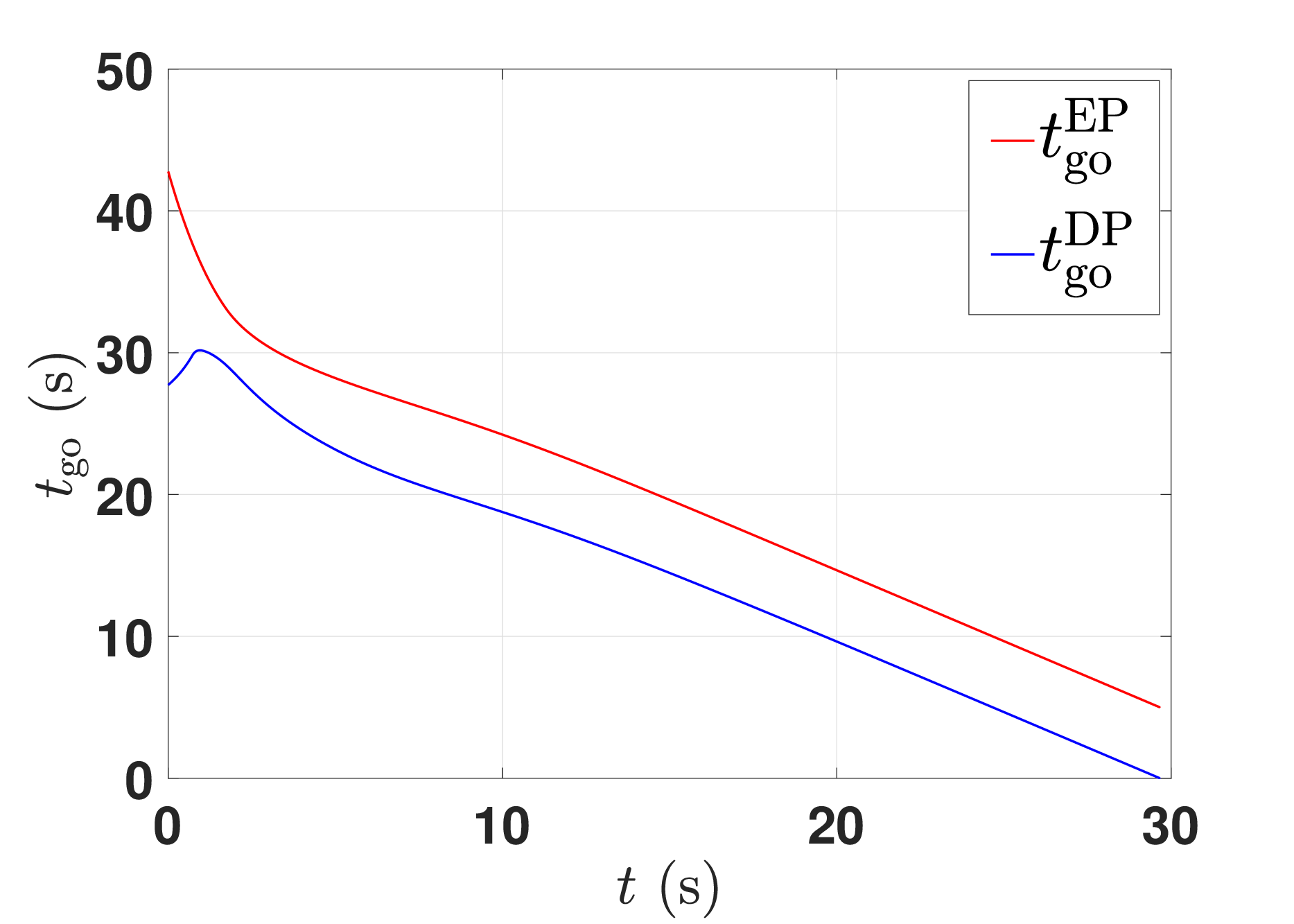}
			\caption{Time-to-go.}
			\label{fig:time_to_go_PPN_without_aE_access}
		\end{subfigure}
		\vspace{0.6em}
		\begin{subfigure}[t]{.49\linewidth}
			\centering
			\includegraphics[width=1.1\linewidth]{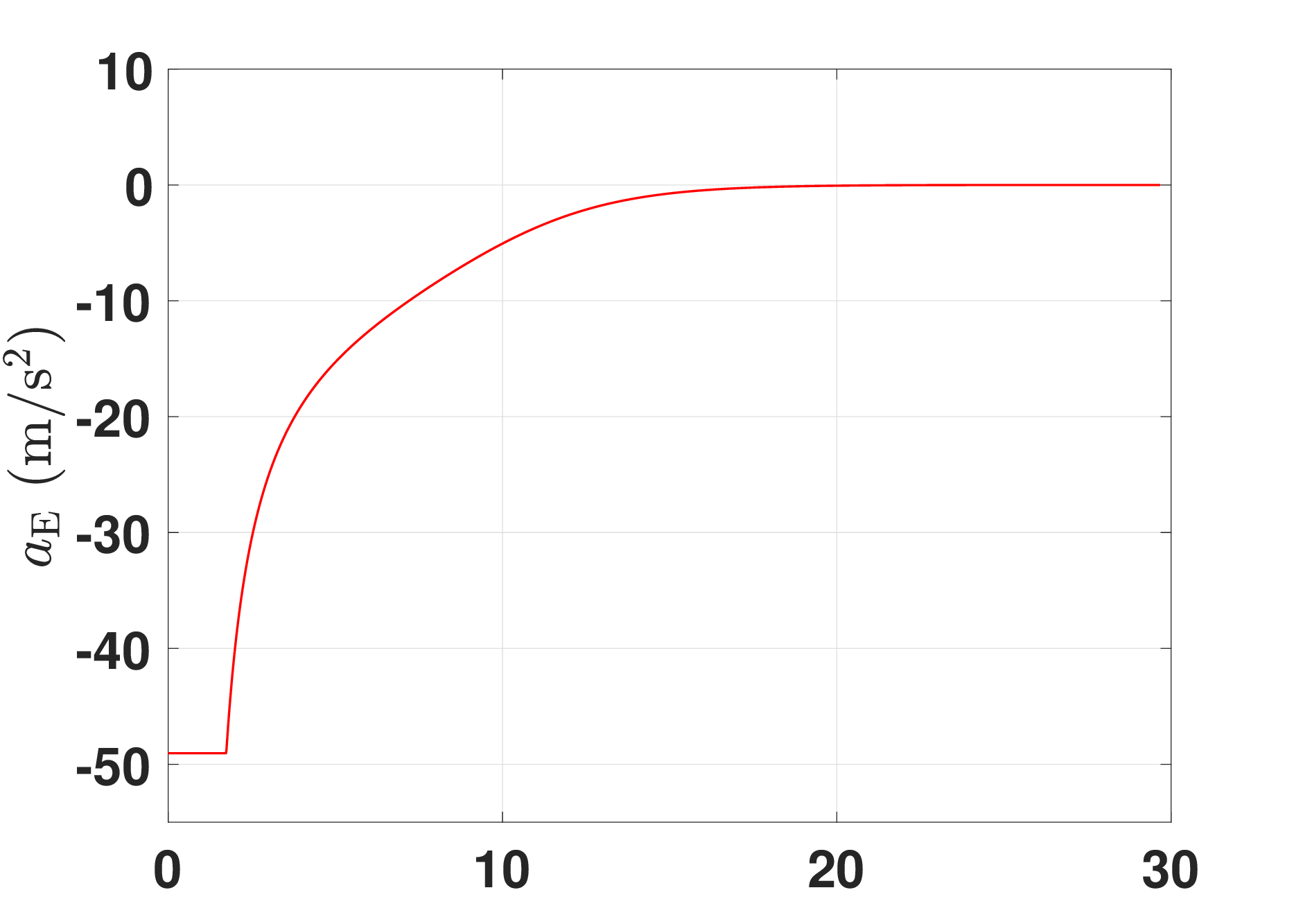}\vspace{0.3em}
			\includegraphics[width=1.1\linewidth]{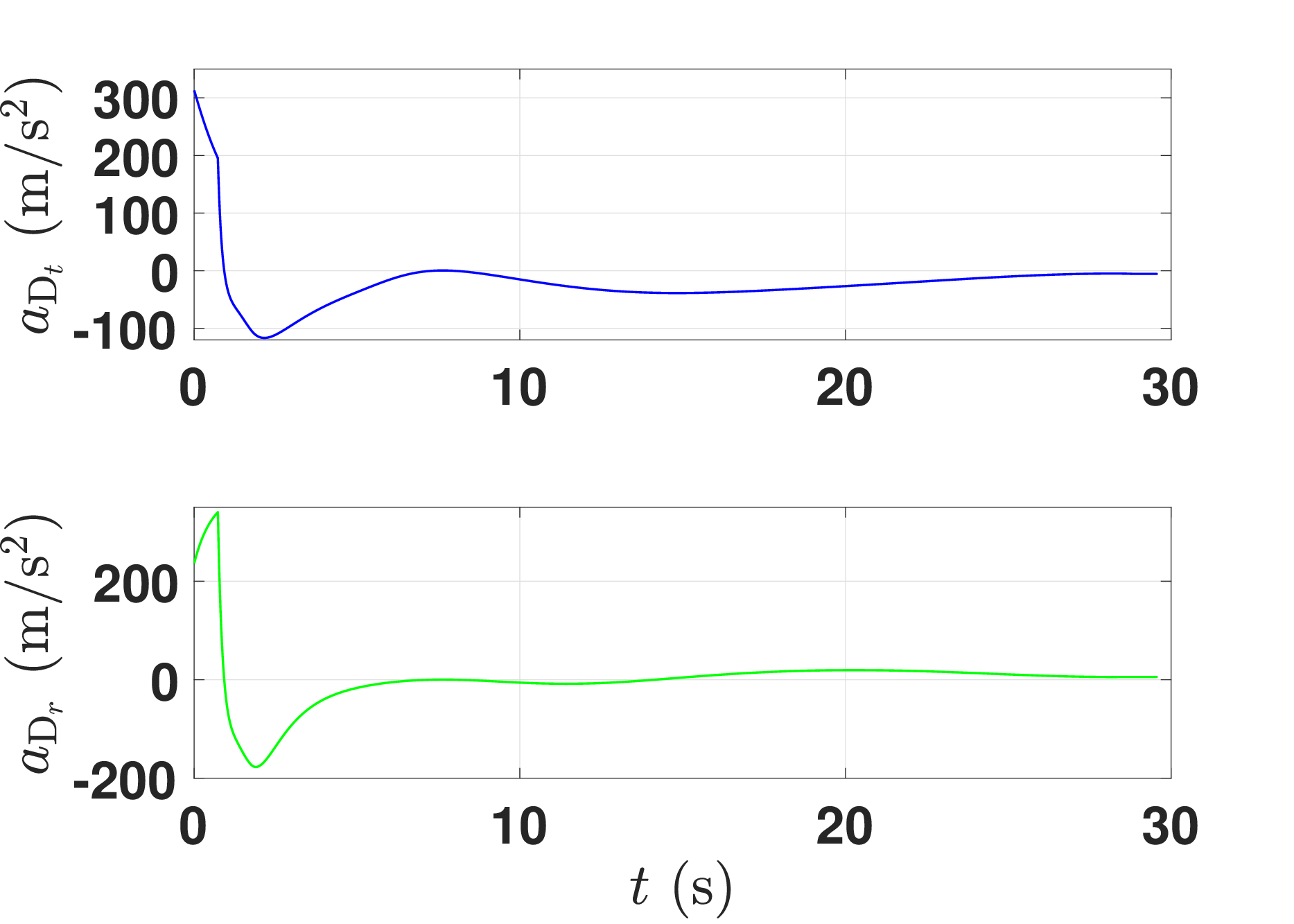}
			\caption{Lateral accelerations (steering controls).}
			\label{fig:acceleration_PPN_without_aE_access}
		\end{subfigure}
		\hfill
		\begin{subfigure}[t]{.49\linewidth}
			\centering
			\includegraphics[width=1.1\linewidth]{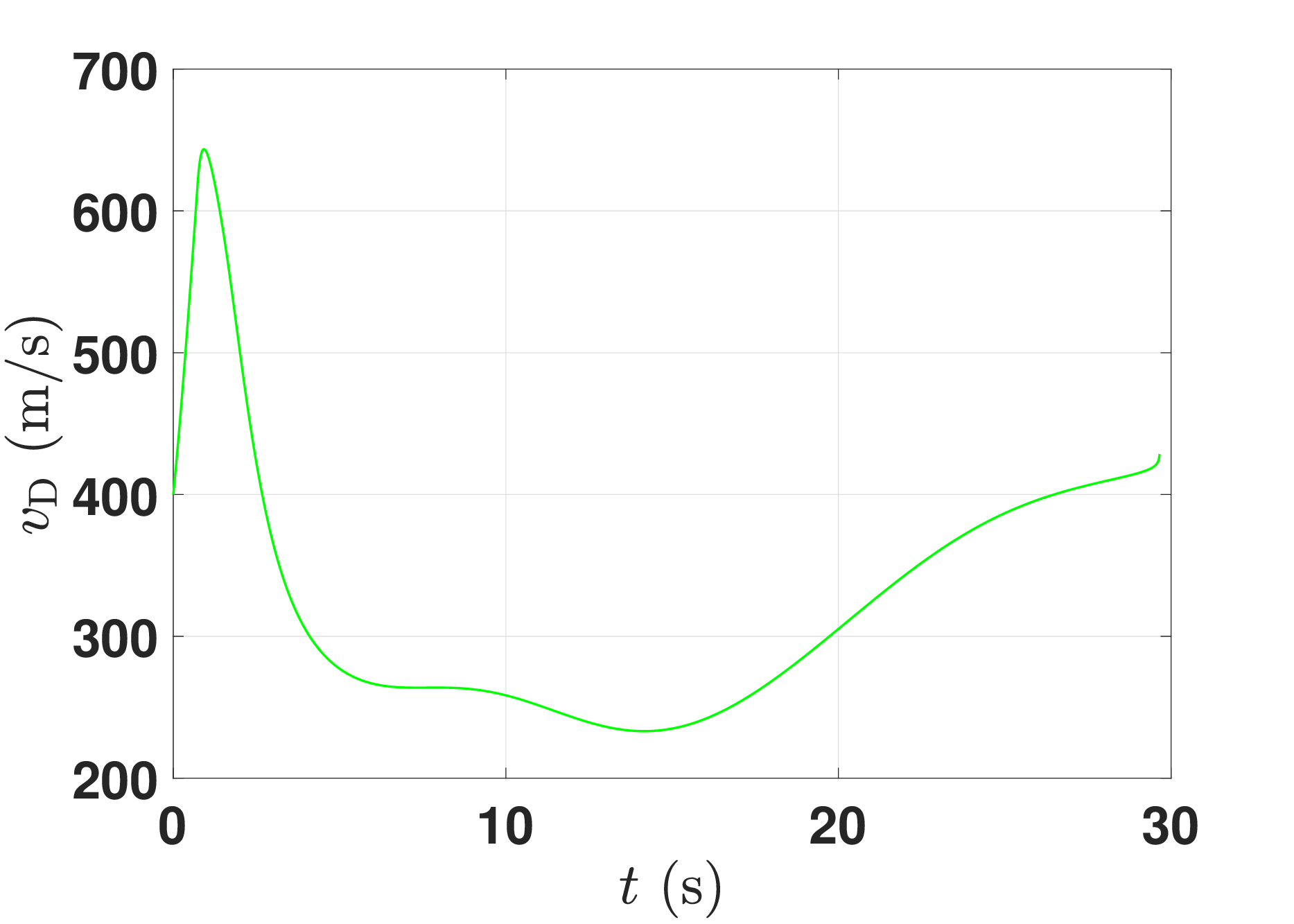}\vspace{0.3em}
			\includegraphics[width=1.1\linewidth]{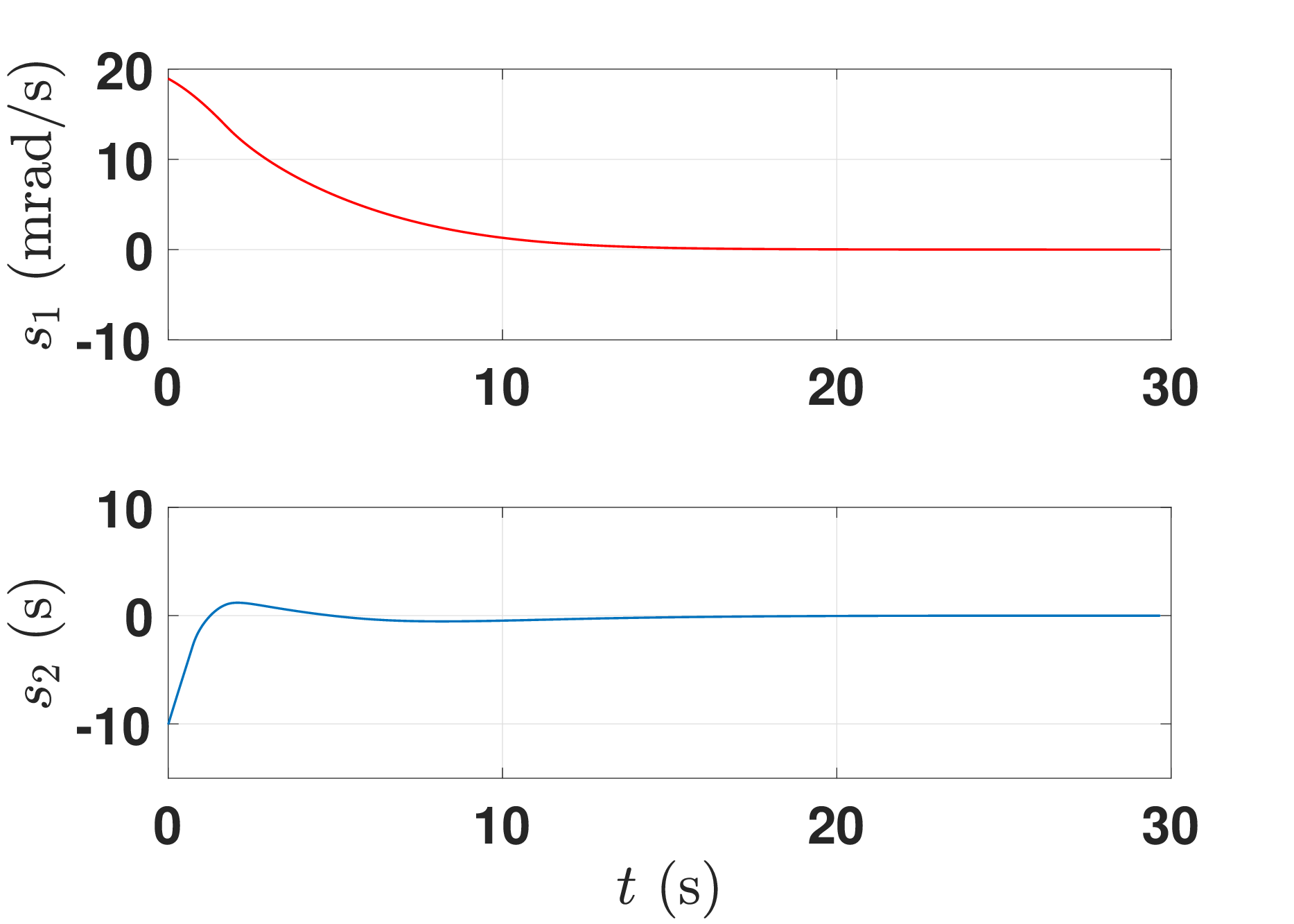}
			\caption{Defender's velocity and sliding manifolds.}
			\label{fig:errors_PPN_without_aE_access}
		\end{subfigure}
		\caption{Performance evaluation under \(a_{\mathrm{P}}=N v_{\mathrm{P}}\dot{\lambda}_{\mathrm{EP}}\) and $a_\mathrm{E}$ is unavailable.}
		\label{fig:Results_PPN_aE_without_access}
	\end{figure}

\begin{figure}[t]
		\centering
		\captionsetup[sub]{justification=centering}
		\begin{subfigure}[t]{.49\linewidth}
			\centering
			\includegraphics[width=1.1\linewidth]{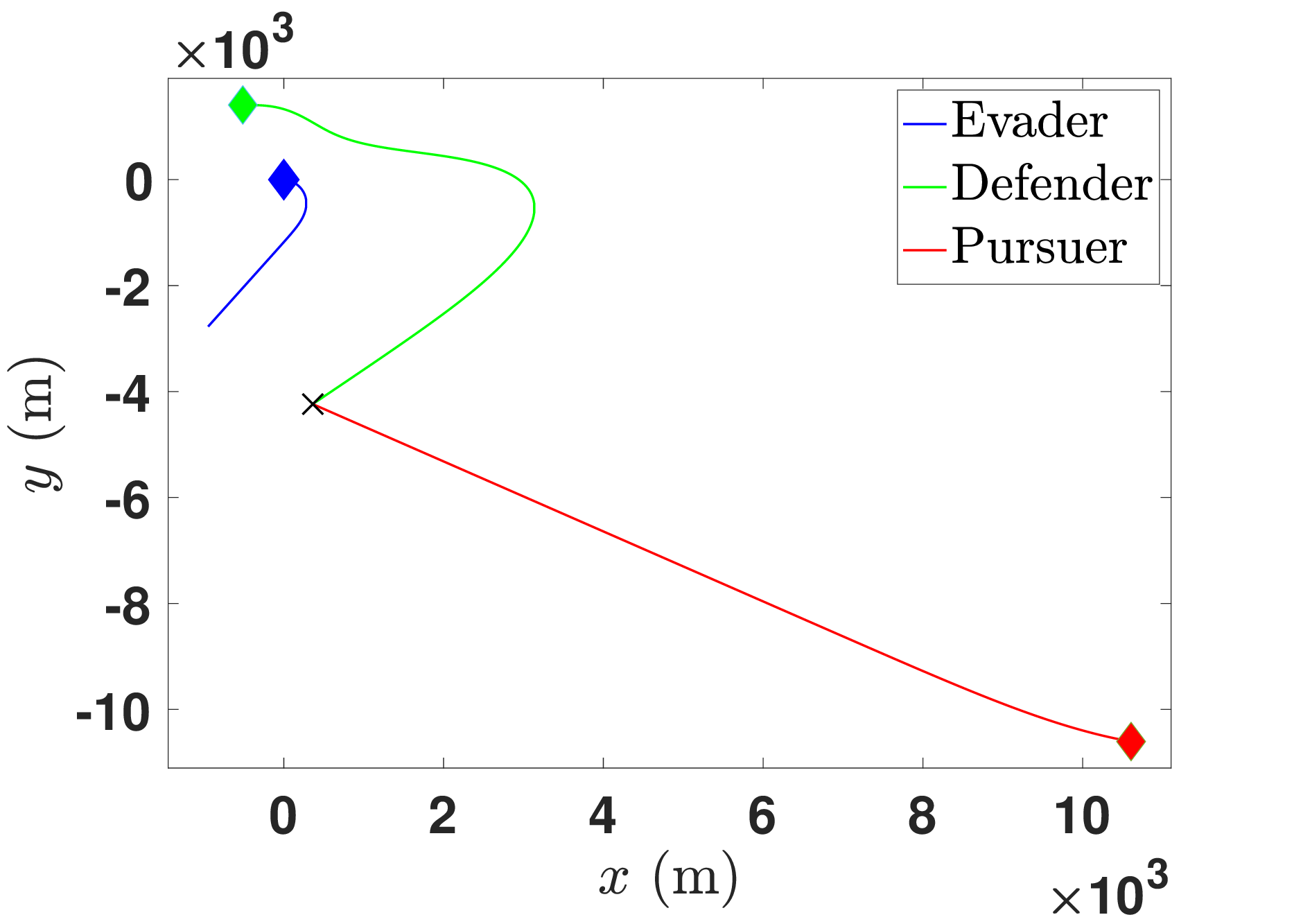}
			\caption{Trajectories.}
			\label{fig:traj_RTPN_without_aE_access}
		\end{subfigure}%
		\hfill
		\begin{subfigure}[t]{.49\linewidth}
			\centering
			\includegraphics[width= 1.1\linewidth]{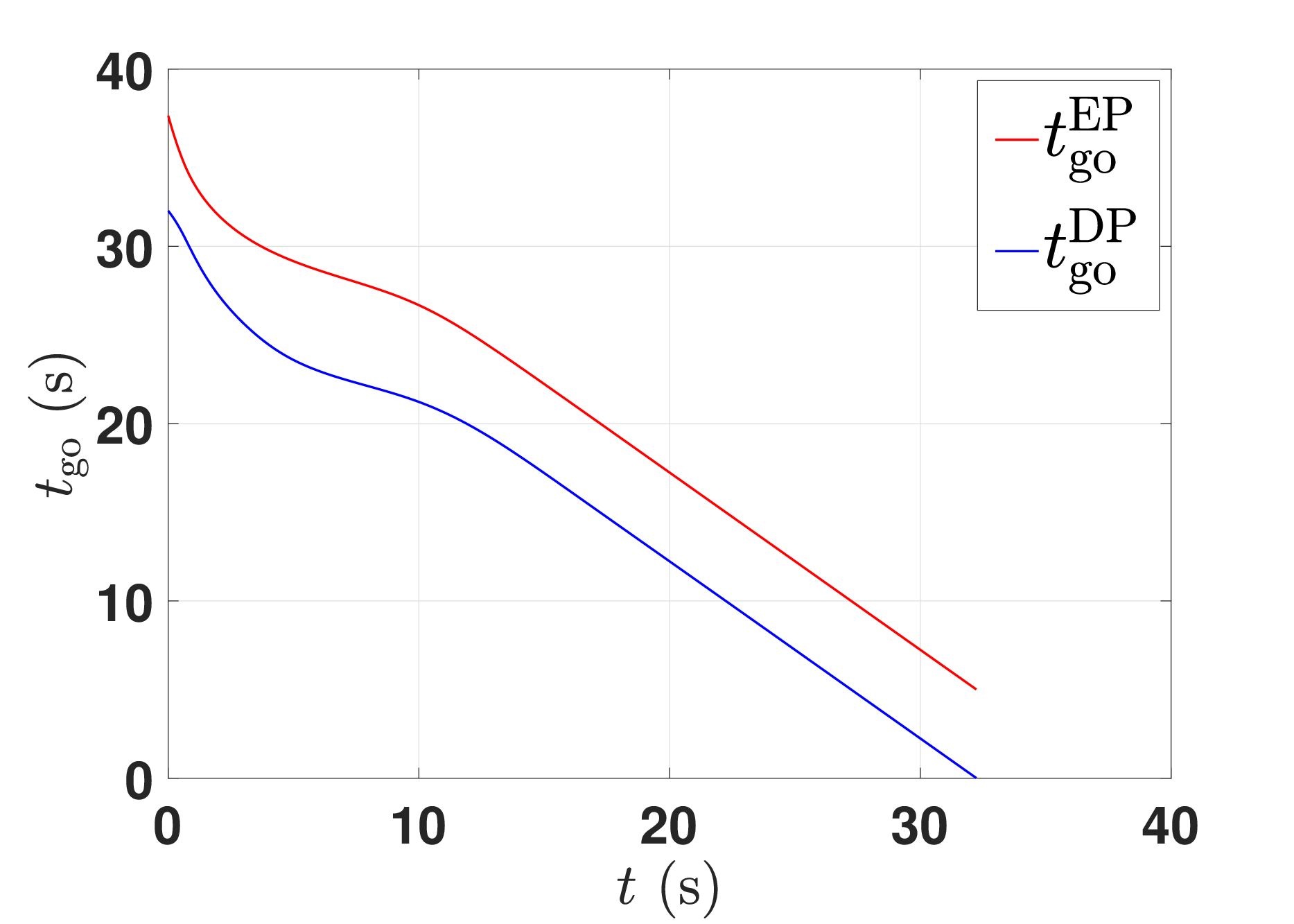}
			\caption{Time-to-go.}
			\label{fig:time_to_go_RTPN_without_aE_access}
		\end{subfigure}
		\vspace{0.6em}
		\begin{subfigure}[t]{.49\linewidth}
			\centering
			\includegraphics[width=1.1\linewidth]{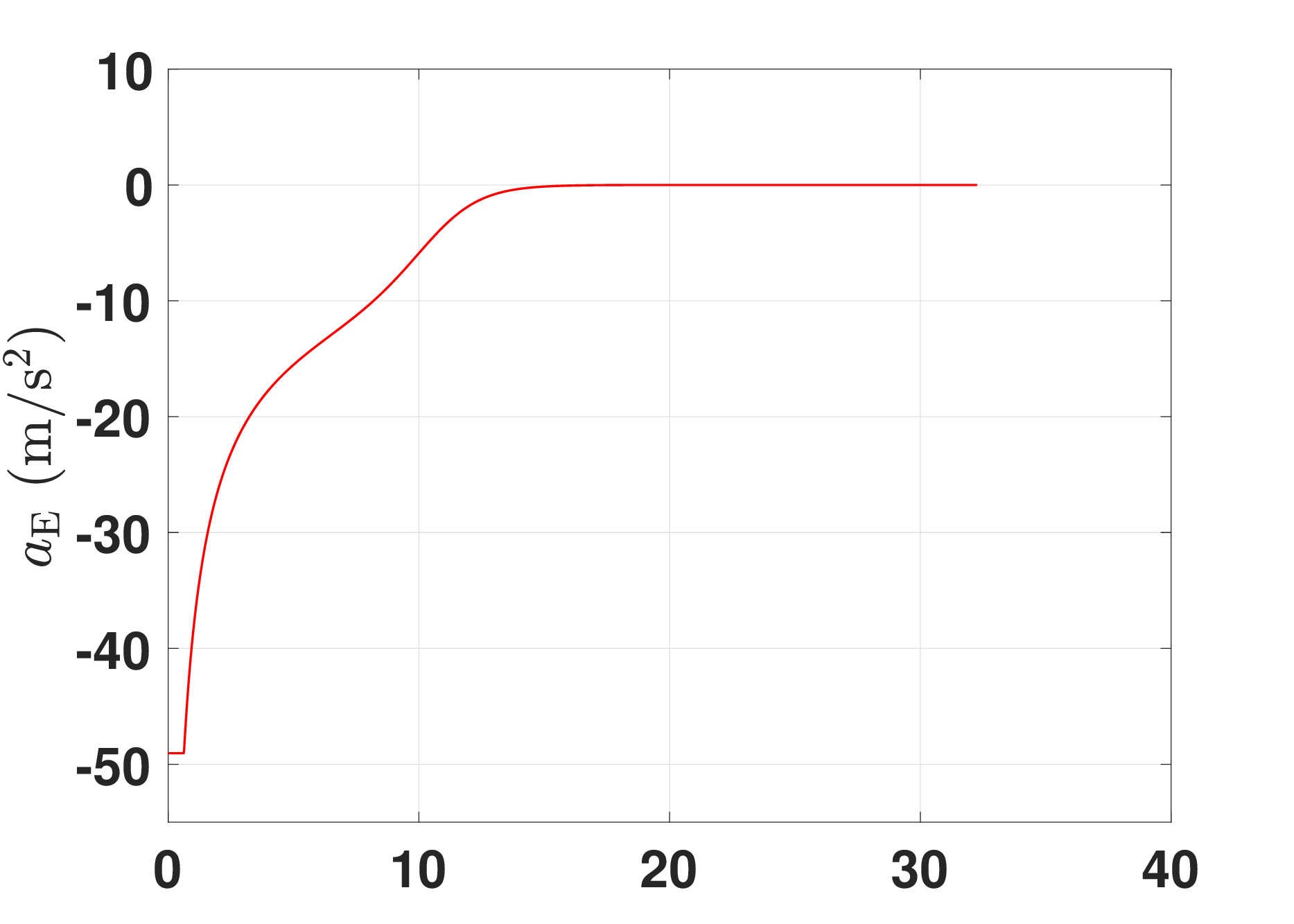}\vspace{0.3em}
			\includegraphics[width=1.1\linewidth]{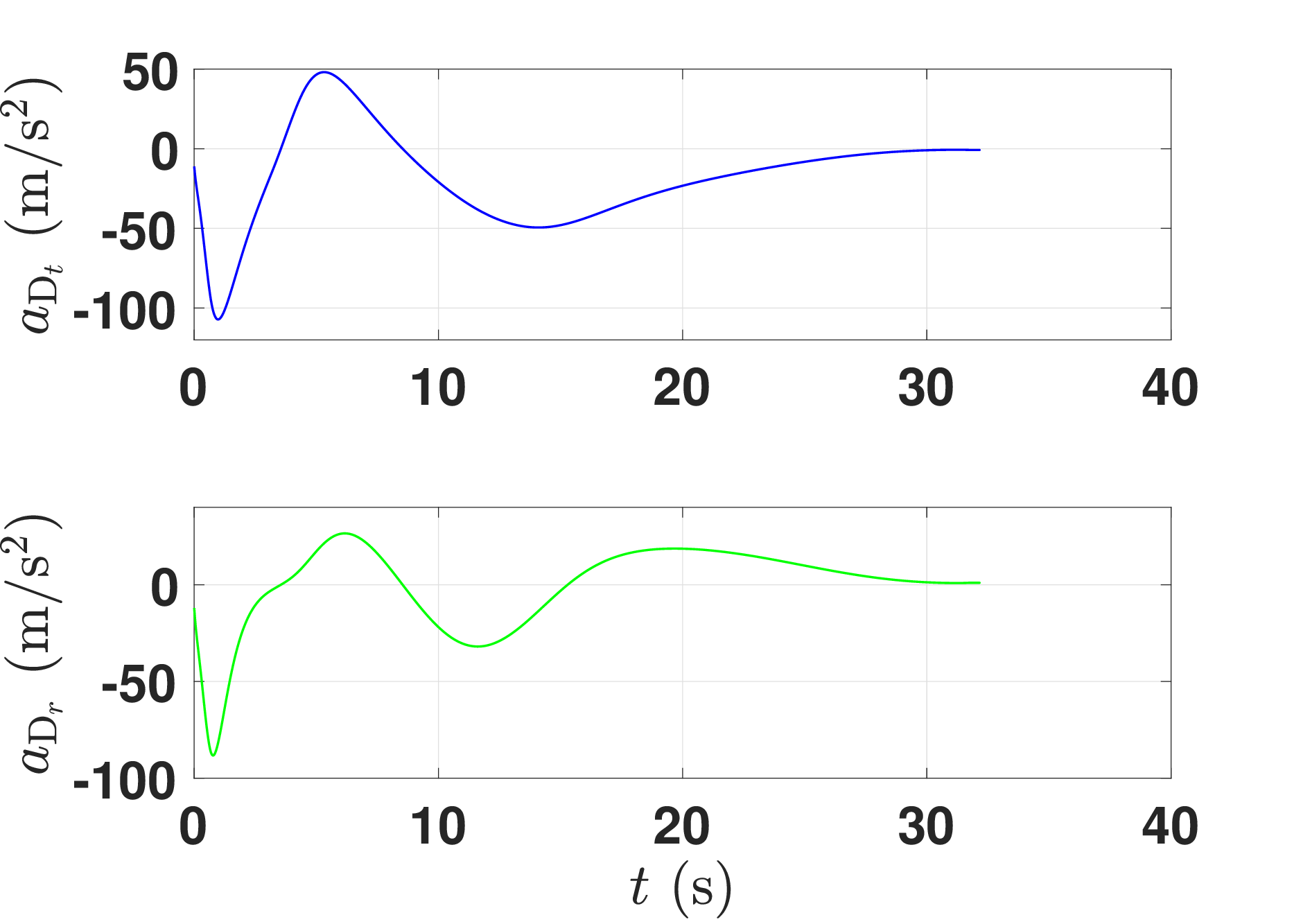}
			\caption{Lateral accelerations (steering controls).}
			\label{fig:acceleration_RTPN_without_aE_access}
		\end{subfigure}
		\hfill
		\begin{subfigure}[t]{.49\linewidth}
			\centering
			\includegraphics[width=1.1\linewidth]{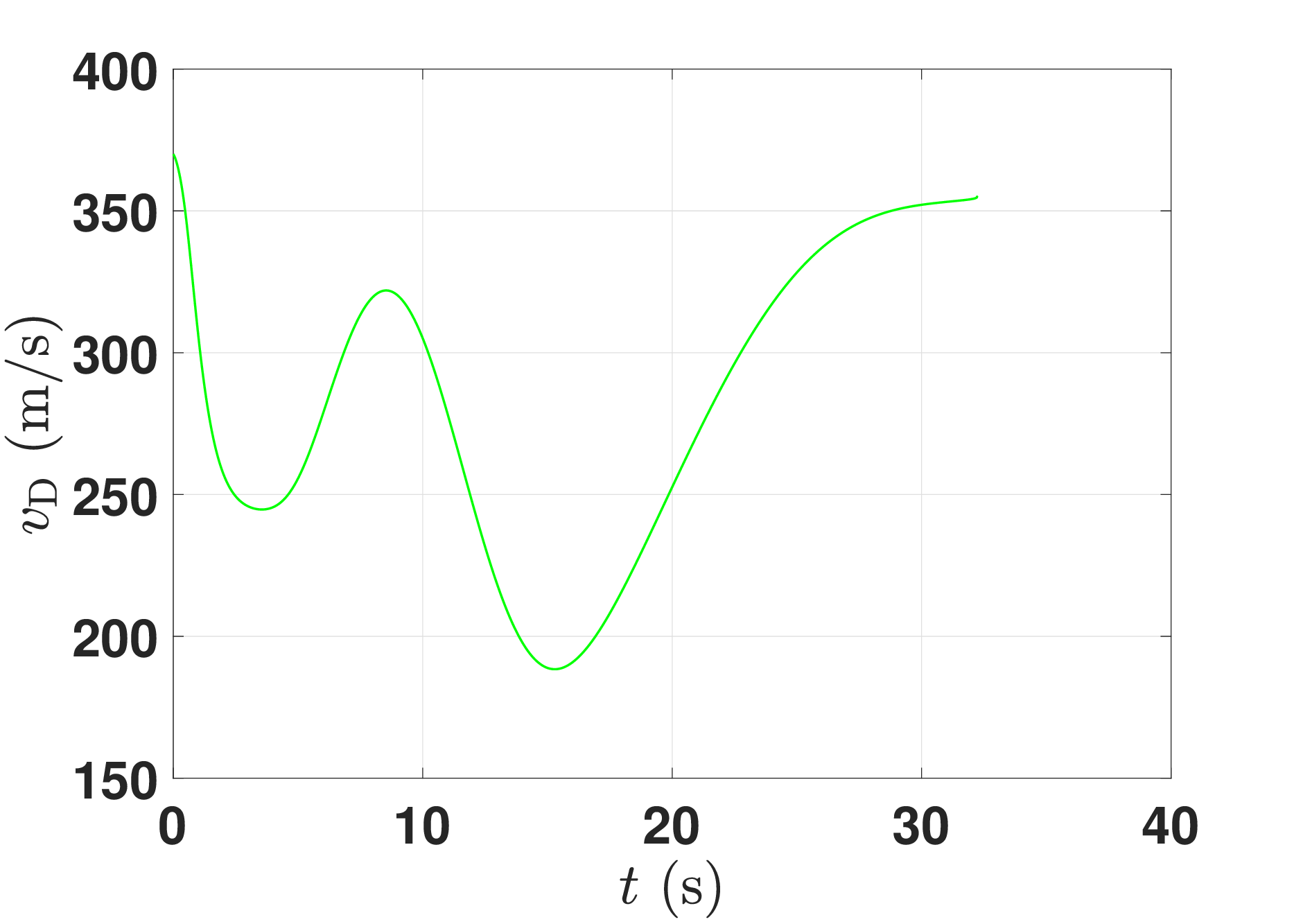}\vspace{0.3em}
			\includegraphics[width=1.1\linewidth]{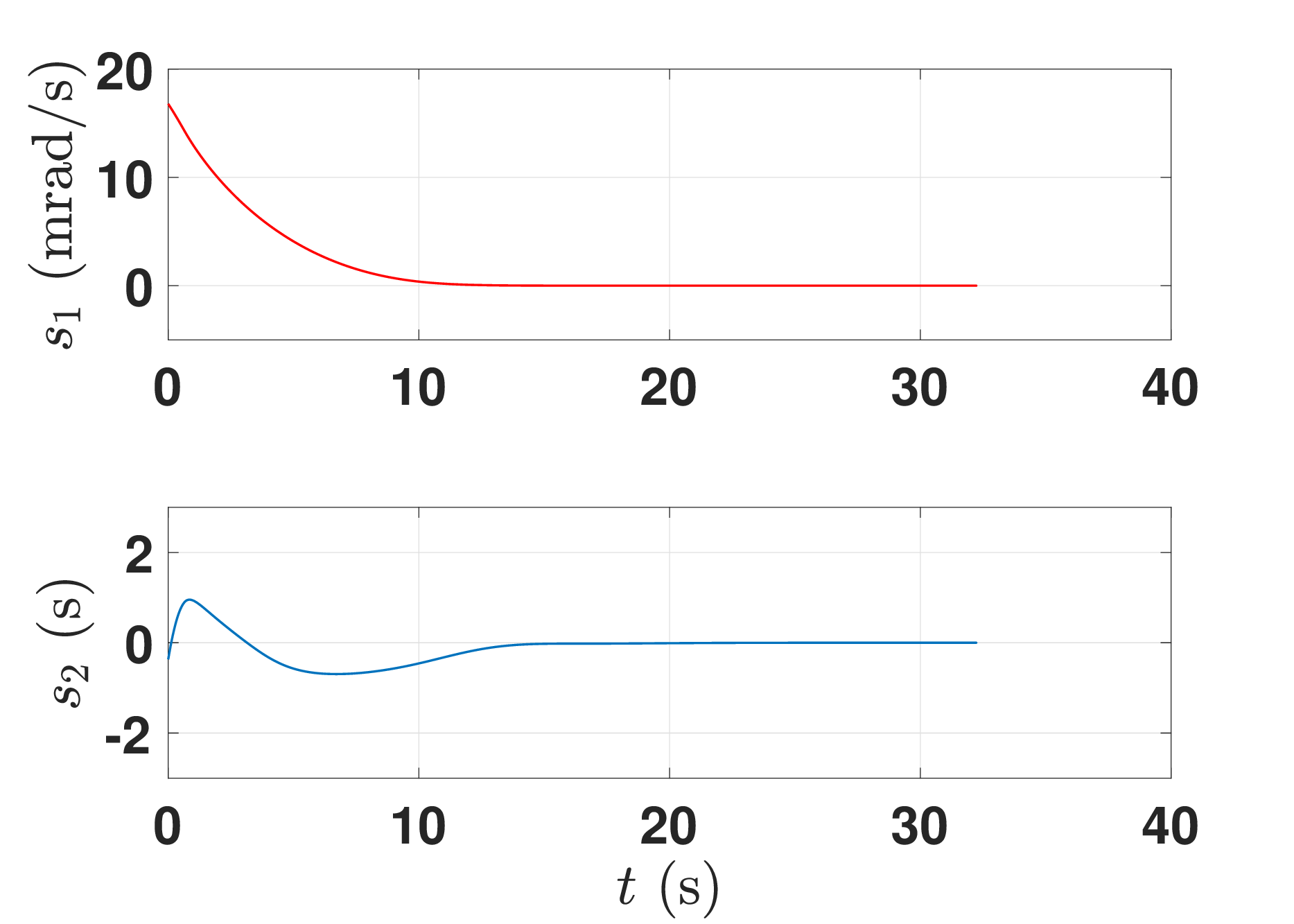}
			\caption{Defender's velocity and sliding manifolds.}
			\label{fig:errors_RTPN_without_aE_access}
		\end{subfigure}
		
		\caption{Performance evaluation under \(a_{\mathrm{P}}=-N \dot{r}_\mathrm{EP}\dot{\lambda}_{\mathrm{EP}}\)  and $a_\mathrm{E}$ is unavailable.}
		\label{fig:Results_RTPN_aE_without_access}
	\end{figure}

\begin{figure}[t]
		\centering
		\captionsetup[sub]{justification=centering}
		\begin{subfigure}[t]{.49\linewidth}
			\centering
			\includegraphics[width=1.1\linewidth]{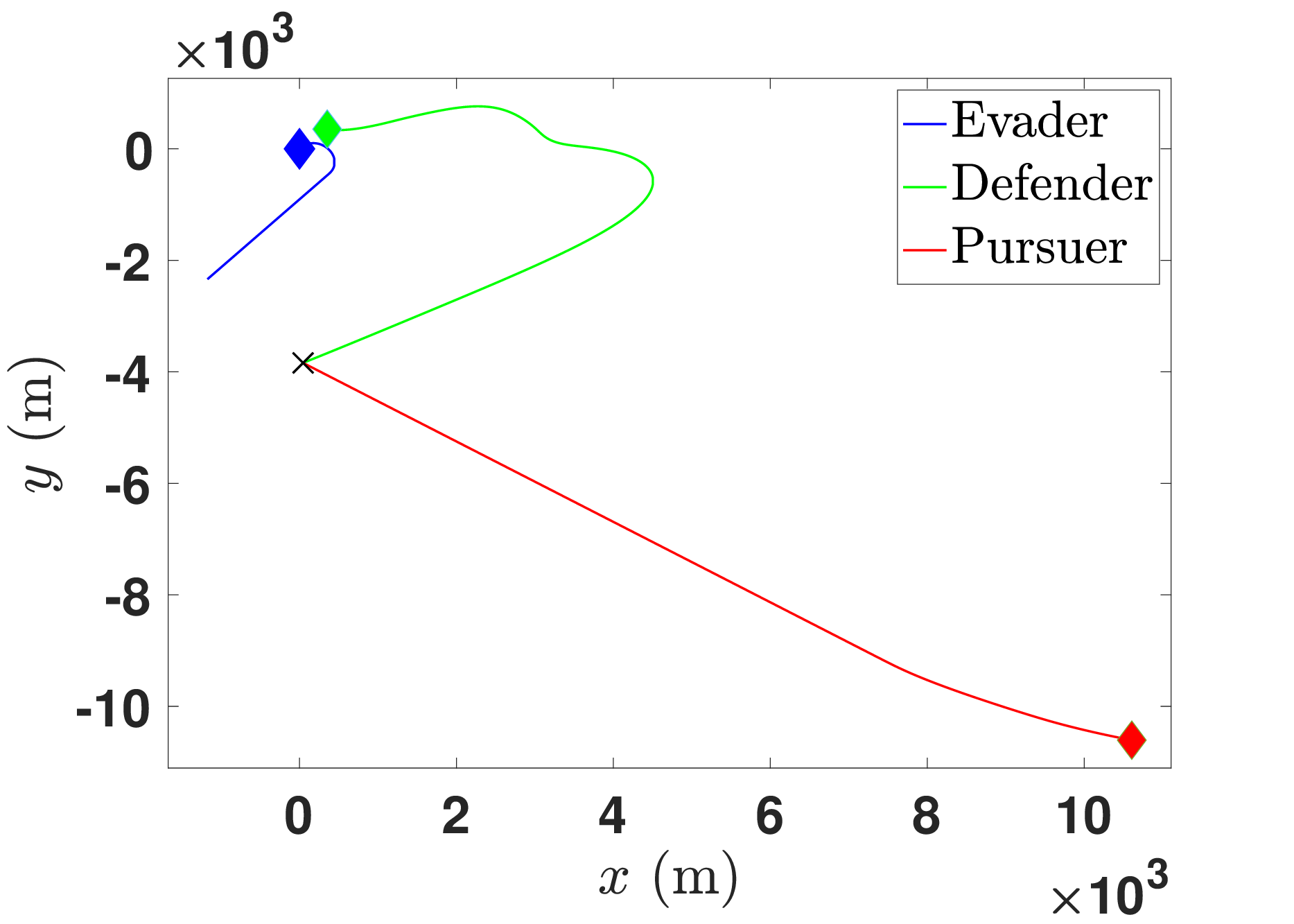}
			\caption{Trajectories.}
			\label{fig:traj_APN_without_aE_access}
		\end{subfigure}%
		\hfill
		\begin{subfigure}[t]{.49\linewidth}
			\centering
			\includegraphics[width= 1.1\linewidth]{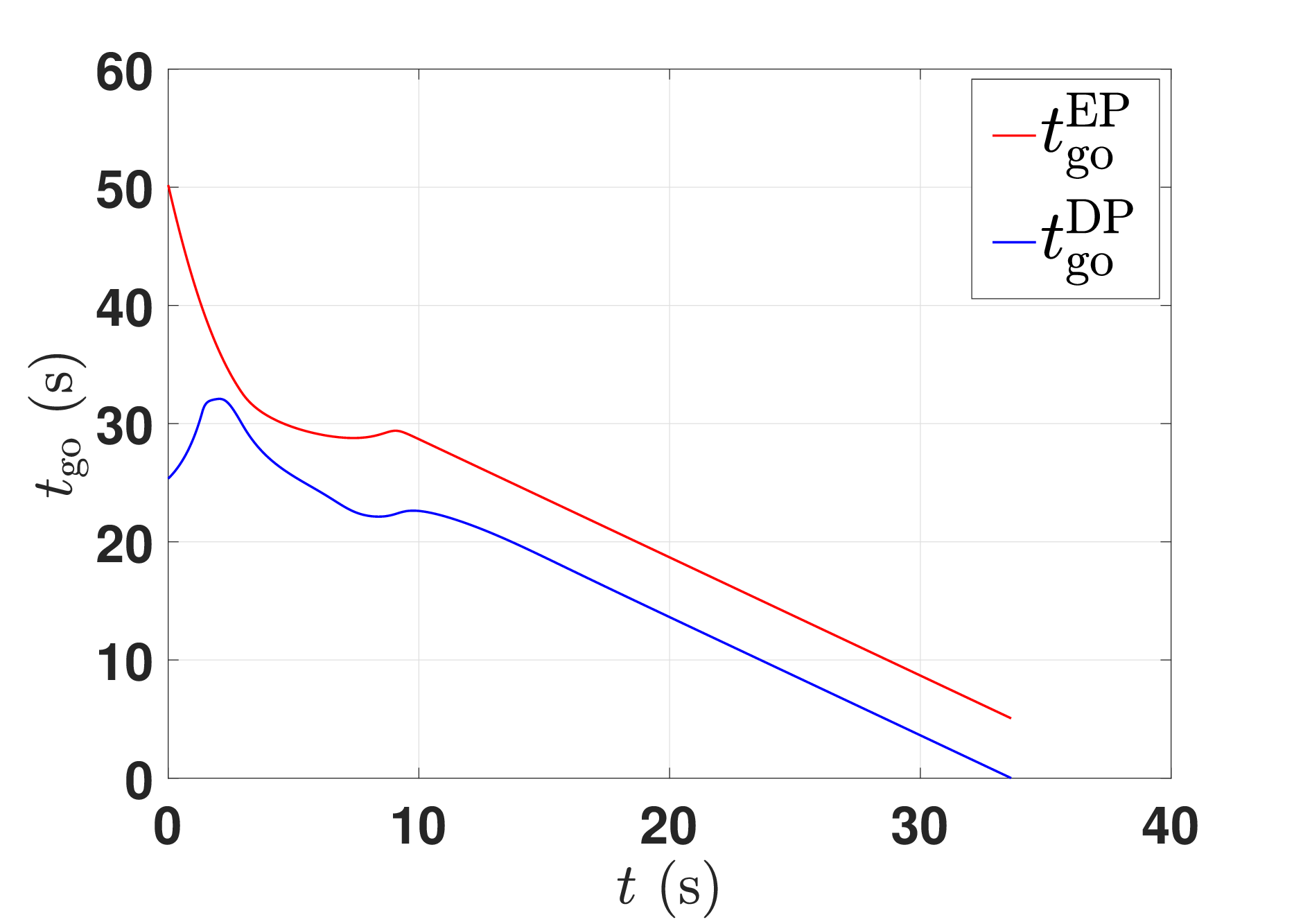}
			\caption{Time-to-go.}
			\label{fig:time_to_go_APN_without_aE_access}
		\end{subfigure}
		\vspace{0.6em}
		\begin{subfigure}[t]{.49\linewidth}
			\centering
			\includegraphics[width=1.1\linewidth]{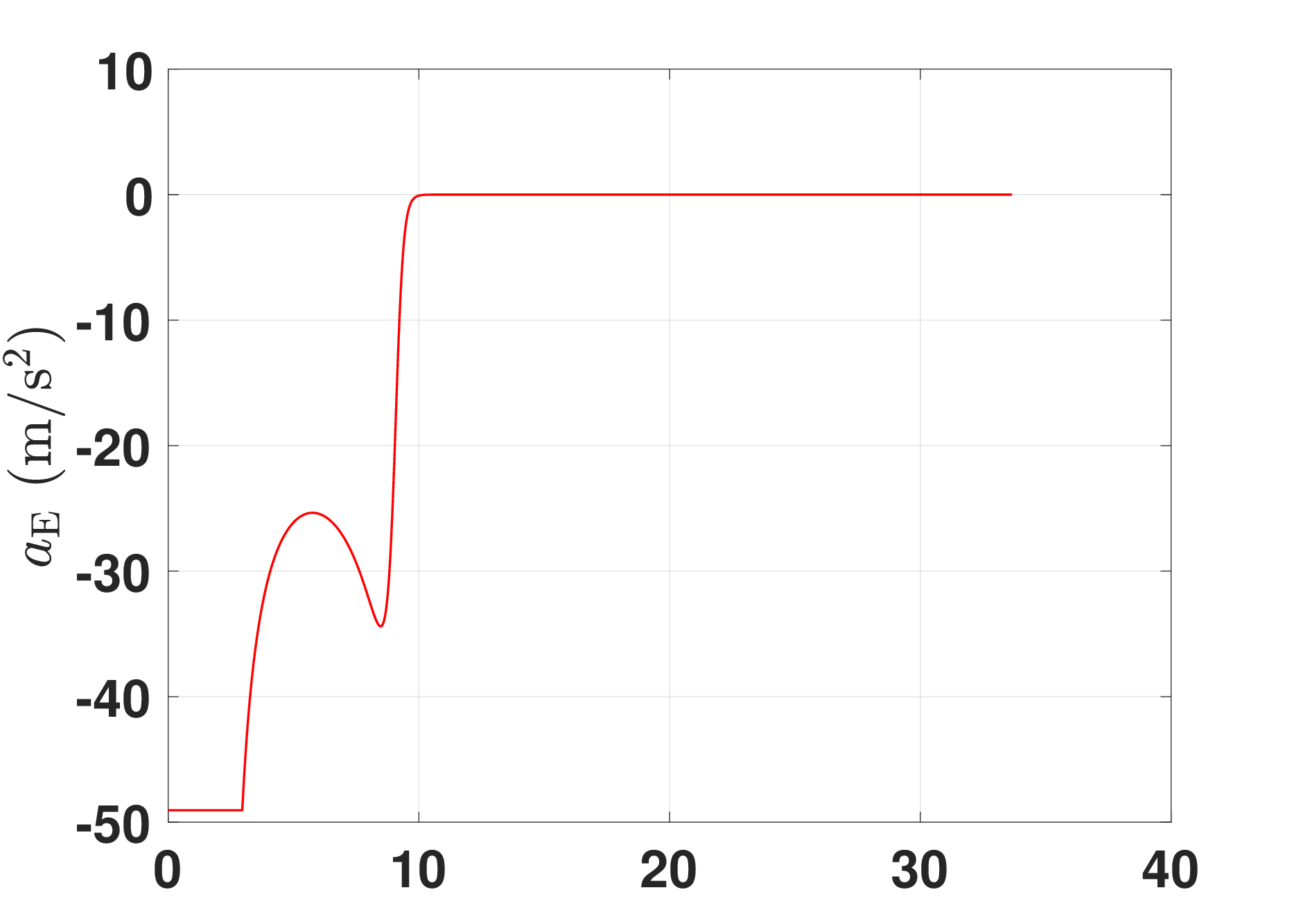}\vspace{0.3em}
			\includegraphics[width=1.1\linewidth]{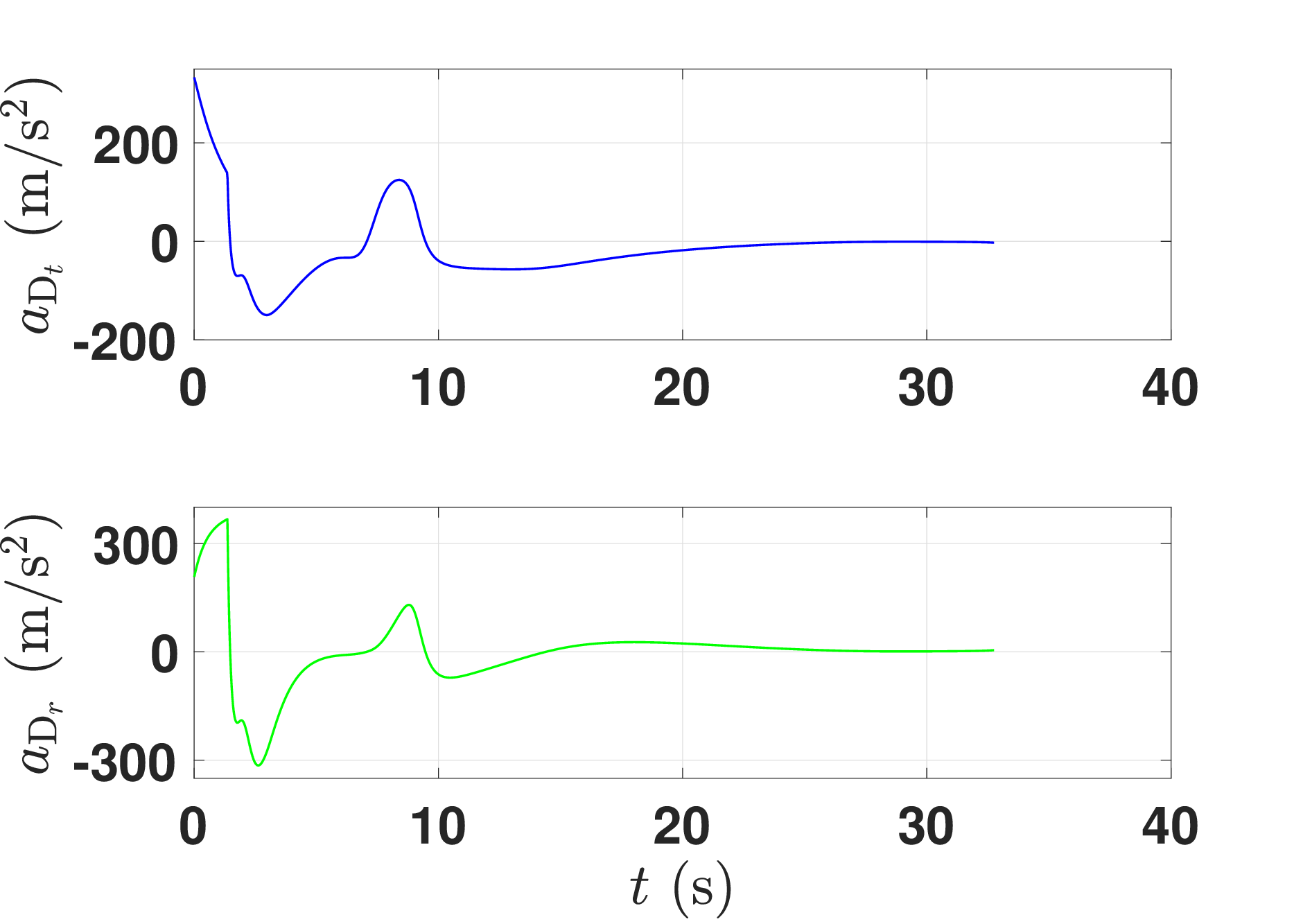}
			\caption{Lateral accelerations (steering controls).}
			\label{fig:acceleration_APN_without_aE_access}
		\end{subfigure}
		\hfill
		\begin{subfigure}[t]{.49\linewidth}
			\centering
			\includegraphics[width=1.1\linewidth]{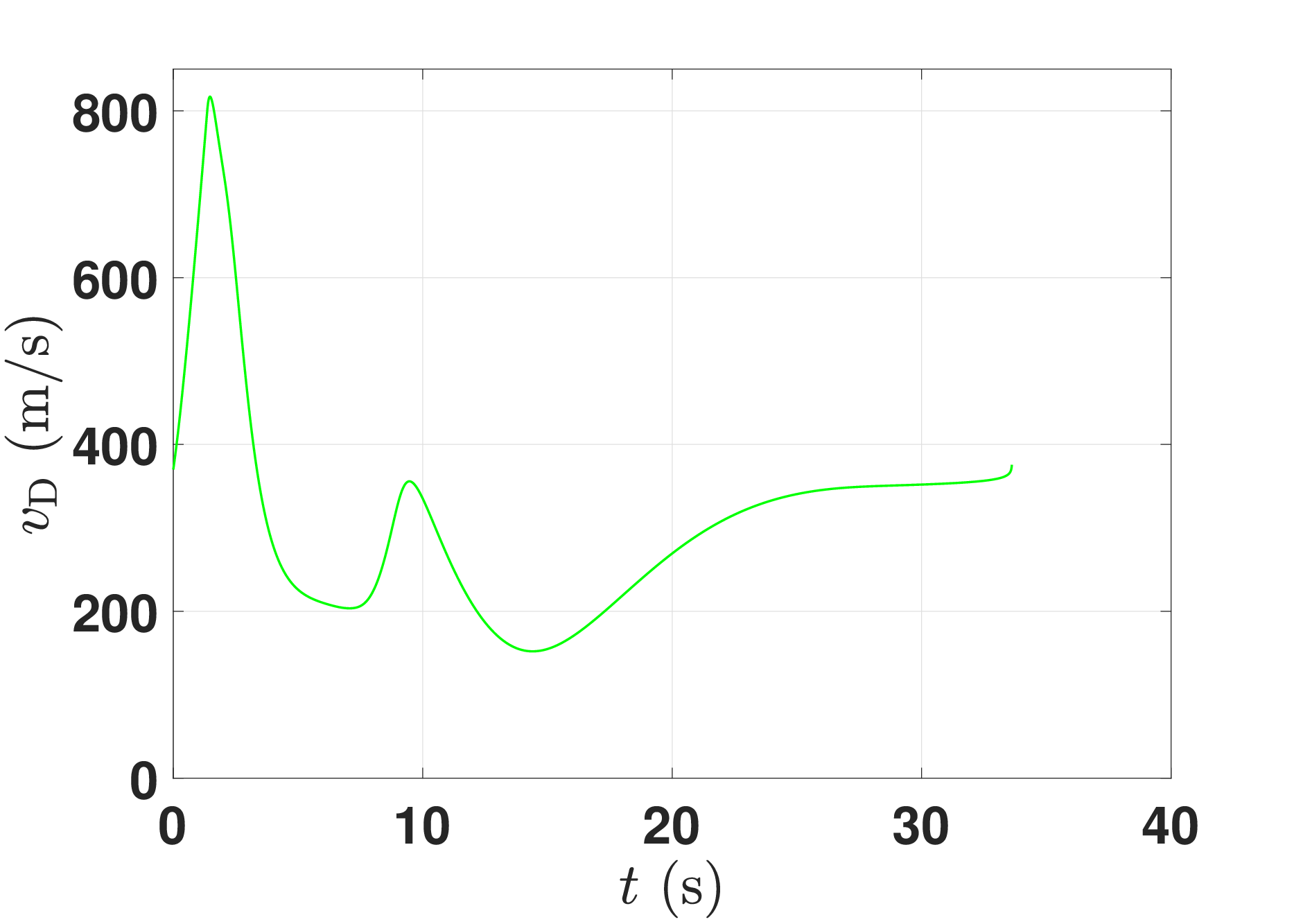}\vspace{0.3em}
			\includegraphics[width=1.1\linewidth]{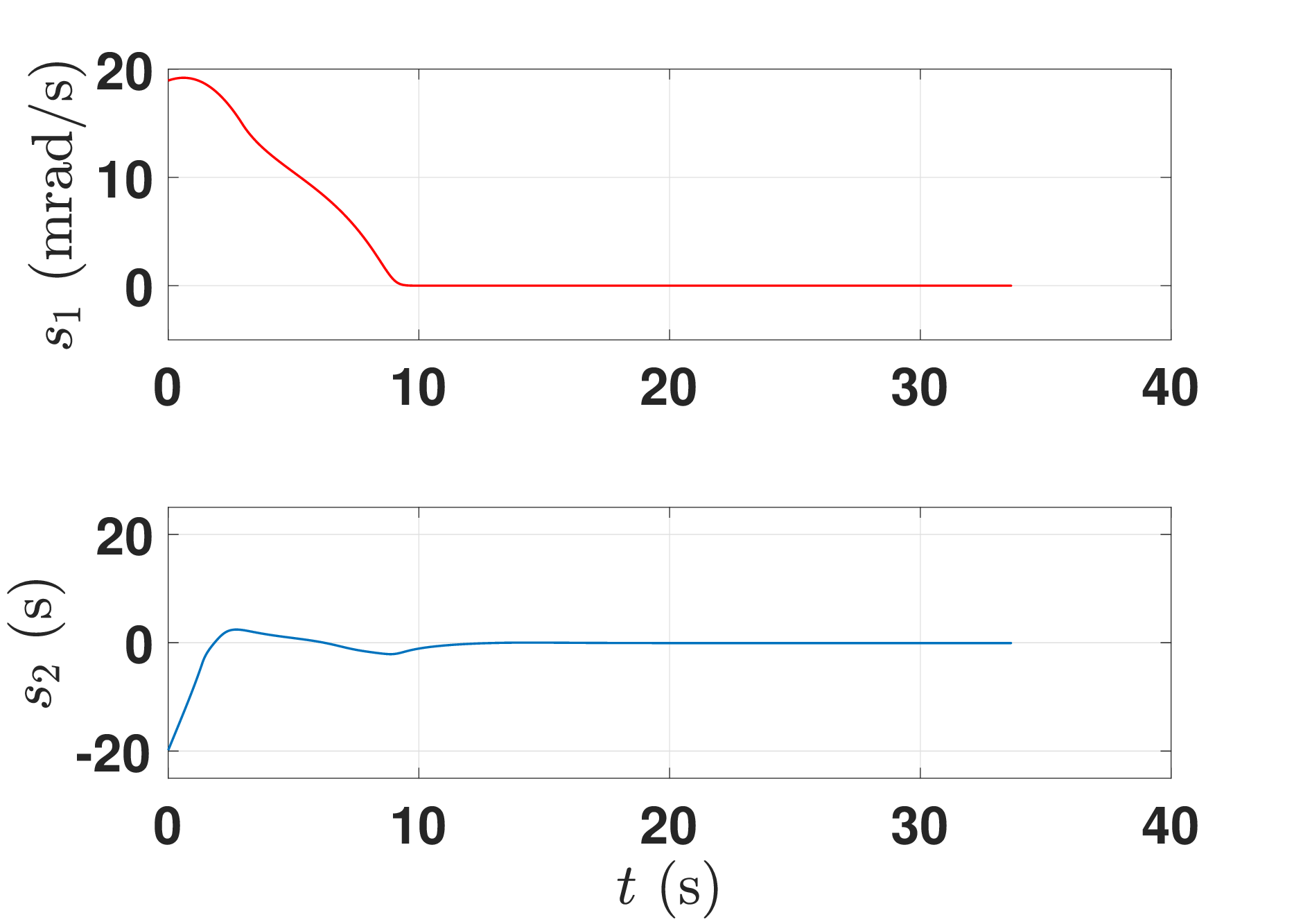}
			\caption{Defender's velocity and sliding manifolds.}
			\label{fig:errors_APN_without_aE_access}
		\end{subfigure}
		\caption{Performance evaluation under \(a_{\mathrm{P}}= -Nv_{\mathrm{P}}\dot{\lambda}_{\mathrm{EP}} +k_\mathrm{P}a_\mathrm{E}\)  and $a_\mathrm{E}$ is unavailable.}
		\label{fig:Results_APN_aE_without_access}
	\end{figure}

	Here, we conduct simulations for the scenario when the defender does not have access to the evader's maneuver. The pursuer employs different variants of the PN guidance strategy as before. The results of various cases are shown in \Crefrange{fig:Results_PPN_aE_without_access}{fig:Results_APN_aE_without_access}. For the case illustrated in \Cref{fig:Results_PPN_aE_without_access}, the initial range and the LOS angle between the evader and the defender are $r_\mathrm{EP}= 2$ km and $60^\circ$, respectively.The initial conditions for the evader-defender engagement are $r_\mathrm{DE}= 2$ km, and $\lambda_\mathrm{DE}= 60^\circ$, while the initial headings of the agents are $\gamma_\mathrm{E}=30^\circ$, and $\gamma_\mathrm{D}=-15^\circ$. The pursuer employs the pure PN guidance strategy, with tuning parameters selected as $\alpha_2 =0.3$, $\zeta_2= 0.05$, and $\xi_2=0.99$. The simulation results are portrayed in \Cref{fig:Results_PPN_aE_without_access}. The trajectories of the agents are depicted in \Cref{fig:traj_PPN_without_aE_access}. One may observe that the defender successfully intercepts the pursuer and safeguards the evader, even without access to the evader’s strategy, indicating that such information is not essential for ensuring successful interception. The time-to-go profiles in \Cref{fig:time_to_go_PPN_without_aE_access} exhibit trends consistent with the case where the defender has full access to the evader’s maneuvers. This indicates a degree of robustness in the defender's guidance system to any agent's maneuvers. The pursuer–evader control inputs are shown in \Cref{fig:acceleration_PPN_without_aE_access}, again following a similar pattern. \Cref{fig:errors_PPN_without_aE_access} depicts the defender’s velocity and sliding manifold profiles, showing that the LOS rate converges to zero within approximately $15$ s, while the time-to-go error vanishes in about $20$ s, soon after.
	
	
	Now, considering the second case, where the pursuer uses the realistic true PN guidance strategy, and the design parameters are chosen as $\alpha_2 = 0.3$, $\zeta_2=0.1275$, and $\xi_2 = 1.8$. The simulation results are presented in \Cref{fig:Results_RTPN_aE_without_access}. The initial conditions are $r_\mathrm{DE}=1.5$ km, $\lambda_\mathrm{DE}=110^\circ$, and headings $\gamma_\mathrm{E}=-5^\circ$, and $\gamma_\mathrm{D}=0^\circ$. The trajectories in \Cref{fig:traj_RTPN_without_aE_access} confirm that the defender intercepts the pursuer, with the time-to-go plot in \Cref{fig:time_to_go_RTPN_without_aE_access} indicating interception at approximately $32$ s. The control inputs of the evader–defender team are depicted in \Cref{fig:acceleration_RTPN_without_aE_access}, where the evader’s lateral acceleration remains comparable to the previous full-access case, while the defender experiences a different lateral acceleration demand relative to the scenario with full access to the evader’s maneuvering information \Cref{fig:acceleration_PPN_with_aE_access}.  The defender’s velocity and error profiles, shown in \Cref{fig:errors_RTPN_without_aE_access}, follow a trend similar to that observed when the evader’s control input is available to the defender.
	
	
	In this case, the pursuer is assumed to use the augmented PN strategy. The design parameters are kept as $\alpha_2 = 0.3$, $\zeta_2 = 0.01$, and $\xi_2=0.06$. The simulation results are presented in \Cref{fig:Results_APN_aE_without_access}. The initial conditions are $r_\mathrm{DE}=0.5$ km and $\lambda_\mathrm{DE}=45^\circ$, with agents heading $\gamma_\mathrm{E}=60^\circ$ and $\gamma_\mathrm{D}=-15^\circ$. The trajectories in \Cref{fig:traj_APN_without_aE_access} demonstrate that the defender intercepts the pursuer, despite the pursuer having access to the evader’s maneuvers while the defender does not. The time-to-go profiles in \Cref{fig:time_to_go_APN_without_aE_access} follow a similar trend to the case where the defender has access to the evader’s maneuvers. The lateral accelerations of the pursuer–evader team are shown in \Cref{fig:acceleration_APN_without_aE_access}, where the evader exhibits comparable behavior, while the defender’s lateral acceleration components appear smoother relative to the case with prior information access. \Cref{fig:errors_APN_without_aE_access} presents the defender’s velocity and error profiles, indicating that the LOS rate error converges to zero within approximately $10$ s, while the time-to-go error vanishes in about $15$ s.
	
	
	
	\subsection{Additional Statistical Analyses}
	
\begin{figure}[h!]
		\centering
		\includegraphics[width=1\linewidth]{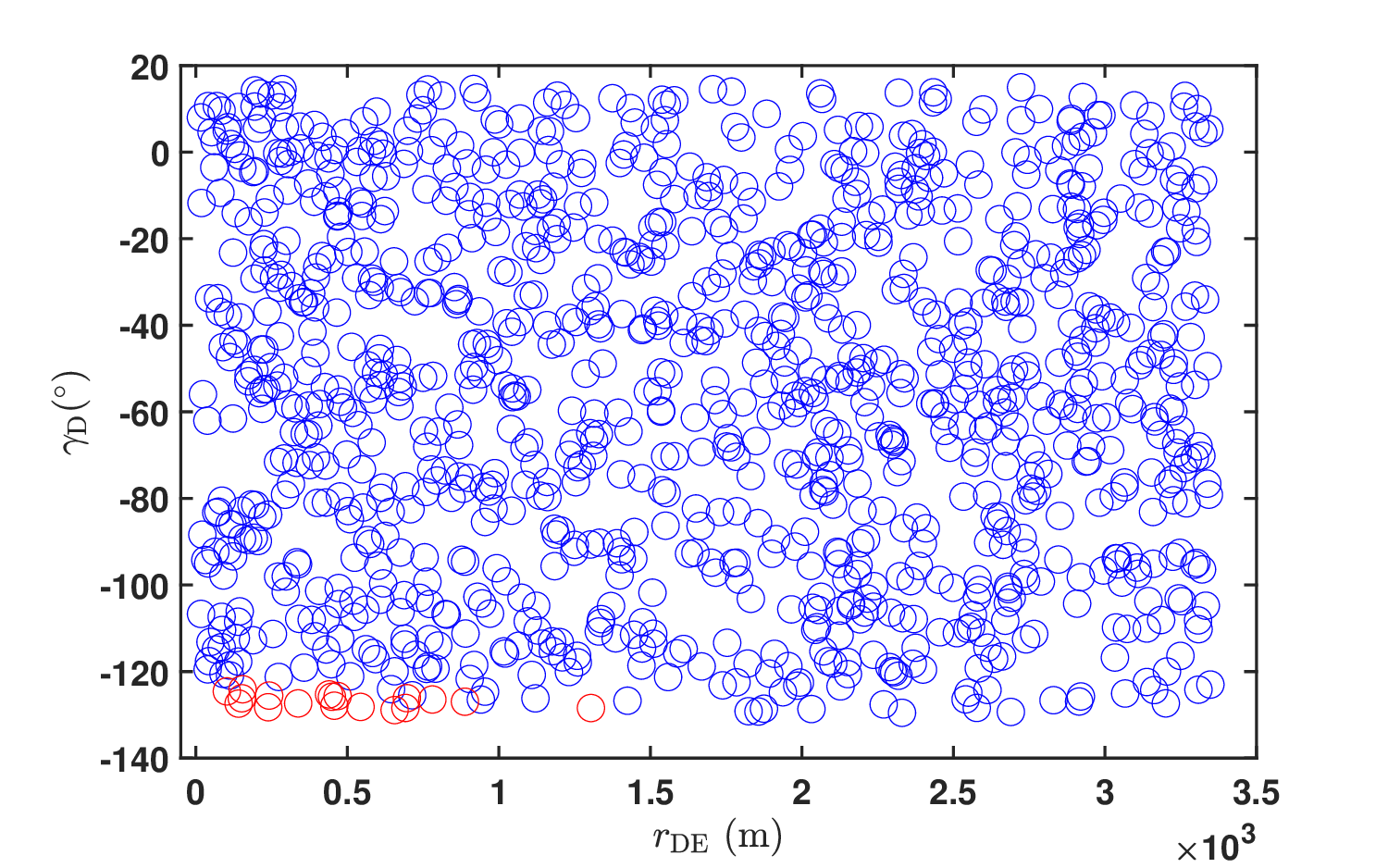}
		\caption{Performance evaluation under different initial evader–defender ranges and defender's headings. }
		\label{fig:r_DEandheading_D}
	\end{figure}

	Now, we perform simulations using the Monte Carlo method over $1100$ runs, with the agents’ initial conditions randomly sampled from a uniform distribution to demonstrate the robustness of the proposed guidance strategies for the evader–defender team across a wide range of initial geometric configurations. The initial range and LOS angle between the evader and pursuer are set to $r_\mathrm{EP}=15000$ m and $\lambda_\mathrm{EP}=-45^\circ$, respectively, while their initial headings are $\gamma_\mathrm{E}=-5^\circ$ and $\gamma_\mathrm{P}=165^\circ$. The winning criterion for the evader–defender team is defined such that interception is achieved if the defender–pursuer separation satisfies $r_\mathrm{DP}\leq3$, provided that the pursuer–evader separation remains $r_\mathrm{EP}>3$. In the first scenario, illustrated in \Cref{fig:r_DEandheading_D}, the initial evader–defender range is uniformly distributed over $[0, 3300]$ m, while the defender’s initial heading is uniformly sampled between $-120^\circ$ and $15^\circ$. The heading interval is determined by the agents’ geometric configuration and is consistent with the proposed design objective of ensuring interception of the pursuer at a $5$-s time margin. Blue markers indicate initial configurations in which the defender intercepts the pursuer, while red markers denote scenarios where the pursuer captures the evader before interception occurs. Overall, the results indicate a winning rate of $98.36\%$ for the evader-defender team. 
	
\begin{figure}[h!]
		\centering
		\includegraphics[width=1\linewidth]{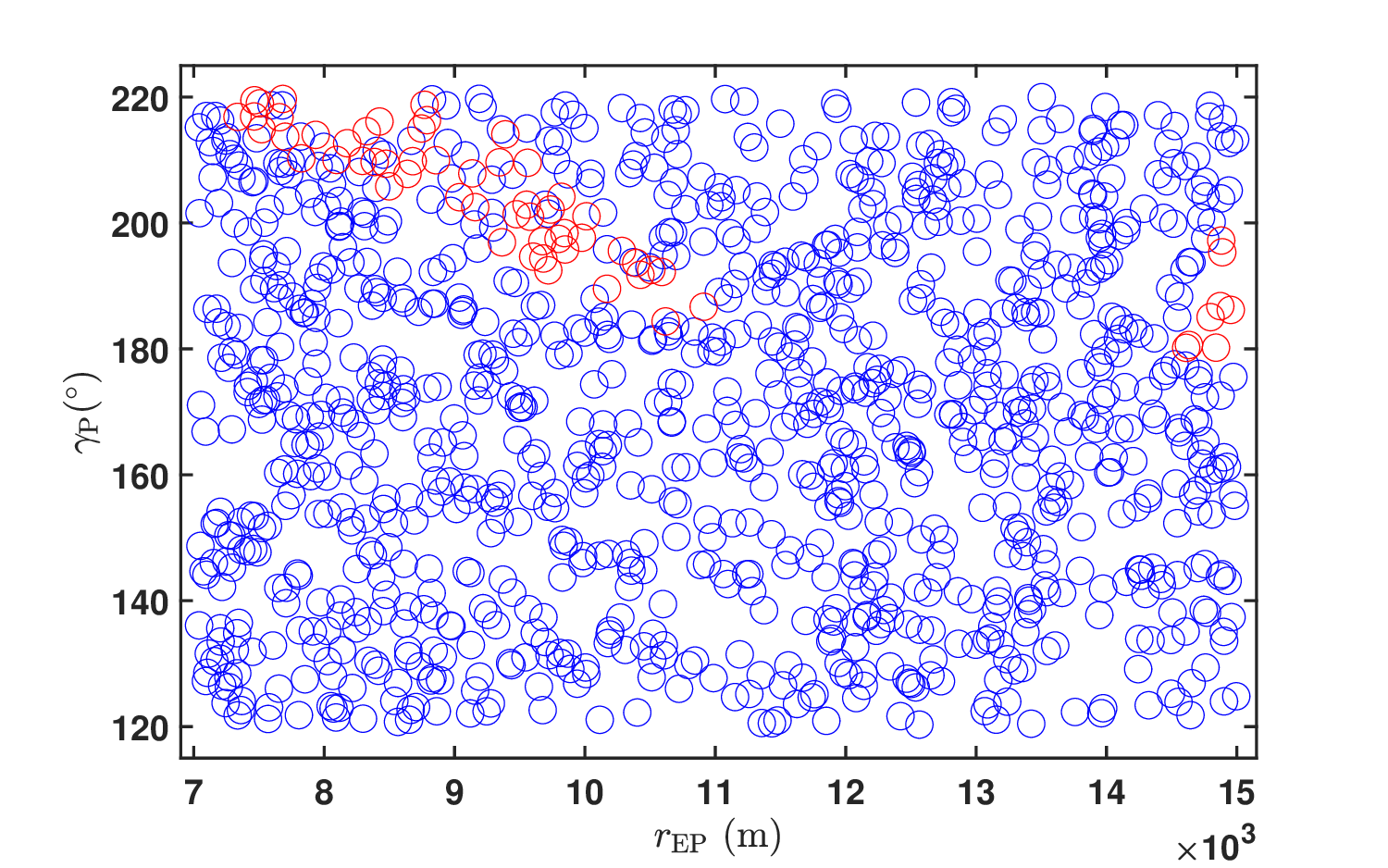}
		\caption{Performance evaluation under different initial evader–pursuer ranges and pursuer's headings. }
		\label{fig:r_EPandheading_P}
	\end{figure}

	\Cref{fig:r_EPandheading_P} represents the second scenario for varying initial evader–pursuer ranges and pursuer's headings. The initial range, $r_\mathrm{EP}$, is uniformly sampled within $[7000,\,15000]$ m, while the pursuer’s heading $\gamma_\mathrm{P}$ is sampled over $(120^\circ,\,220^\circ)$. Across $1100$ simulation runs, the evader–defender team achieves a winning rate of $94.45\%$, demonstrating consistent performance under wide initial conditions.

\begin{figure}[h!]
		\centering
		\includegraphics[width=1\linewidth]{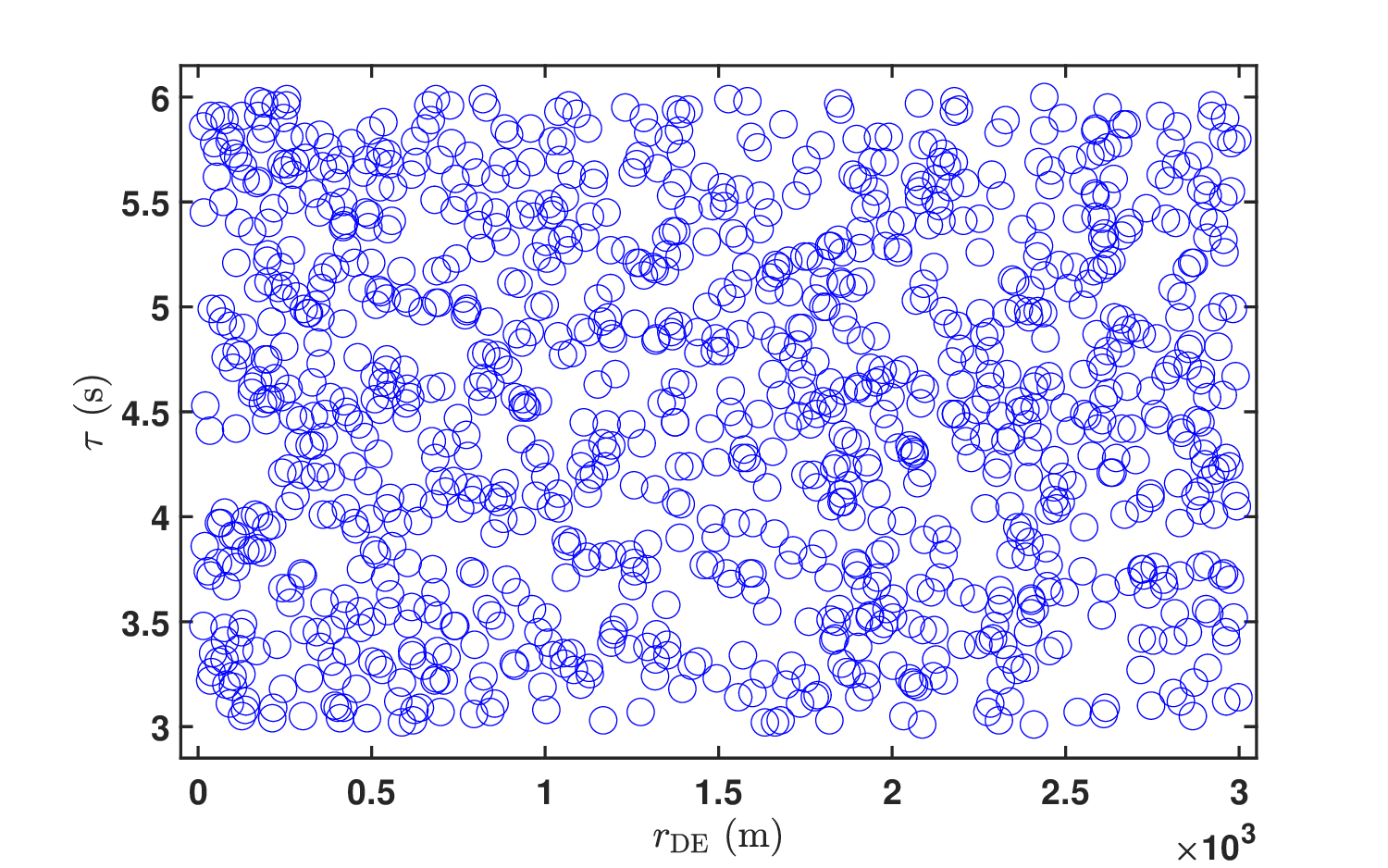}
		\caption{Performance evaluation under different initial evader–pursuer ranges and time margins. }
		\label{fig:r_DETimemargin}
	\end{figure}

	In the third case, shown in \Cref{fig:r_DETimemargin}, simulations are conducted for varying initial evader–defender ranges and time margins. The initial range $r_\mathrm{DE}$ is uniformly sampled within $[0,\,3000]$ m, while the time margin is drawn from $(3,\,6)$ seconds. Across all runs, the evader–defender team achieves a $100\%$ winning rate, demonstrating the robustness of the strategy over wide variations in range and time margin.

	\section{Conclusions}\label{sec:conclusions}
	
	In this work, a cooperative approach was proposed for the evader-defender team to safeguard the evader from the attacking pursuer. The evader deploys a defender with similar capabilities as those of the pursuer. The cooperative strategies were designed using the concepts of impact time guidance strategies. The cooperative strategies design involved a control scheme that forced the errors to vanish in a fixed-time, independent of the initial error value. The evader's strategy was to nullify its line-of-sight rate with respect to the pursuer to attract the latter on the collision path, and render its non-maneuvering. This was used as an advantage by the defender, which then intercepted the non-maneuvering pursuer by means of a true-proportional navigation guidance law. The numerical simulation results demonstrate the effectiveness of the proposed cooperative approach for the evader-defender team, irrespective of the pursuer's guidance strategy. Moreover, the defender was successfully intercepted the non-maneuvering pursuer even when the former did not have access to the evader's maneuvers.

	\bibliographystyle{IEEEtaes}
	\bibliography{references}

\end{document}